\documentclass[12pt]{article}

\usepackage{jheppub, bm, wrapfig,float,array}
\usepackage[utf8]{inputenc}
\numberwithin{equation}{section}
\renewcommand\afterTocRuleSpace{\bigskip}

\usepackage{amsfonts}
\usepackage{amsmath}
\usepackage{color}
\usepackage{graphicx}
\usepackage{float}
\usepackage[T1]{fontenc}
\usepackage[utf8]{inputenc}
\usepackage{alphabeta}
\usepackage{fancyvrb}
\usepackage[dvipsnames]{xcolor}
\usepackage{bold-extra}
\usepackage{lmodern}

\usepackage{tikz}
\usetikzlibrary{arrows}
\usetikzlibrary{shapes.geometric,calc,arrows, positioning,shapes.misc,decorations.markings}
\tikzset{
  big arrow/.style={
    decoration={markings,mark=at position 1 with {\arrow[scale=2,#1]{>}}},
    postaction={decorate},
    shorten >=0.4pt},
  big arrow/.default=black}

\newcommand{\bea}{\begin{eqnarray}}
\newcommand{\eea}{\end{eqnarray}}
\newcommand{\be}{\begin{equation}}
\newcommand{\ee}{\end{equation}}
\newcommand{\bit}{\begin{itemize}}
\newcommand{\eit}{\end{itemize}}
\newcommand{\ben}{\begin{enumerate}}
\newcommand{\een}{\end{enumerate}}

\newcommand{\half}{\frac{1}{2}}

\newcommand{\Z}{{\mathbb Z}}
\newcommand{\R}{{\mathbb R}}
\newcommand{\C}{{\mathbb C}}

\renewcommand{\P}{{\mathbb P}}

\newcommand{\cF}{\mathcal{F}}

\newcommand{\cI}{\mathcal{I}}

\newcommand{\cN}{\mathcal{N}}

\newcommand{\cP}{\mathcal{P}}

\newcommand{\cR}{\mathcal{R}}
\newcommand{\cS}{\mathcal{S}}

\newcommand{\F}{\mathsf{F}}
\renewcommand{\S}{\mathsf{S}}
\newcommand{\Asym}{\mathsf{\Lambda}^2}

\renewcommand{\C}{\mathsf{C}}
\newcommand{\A}{\mathsf{A}}

\renewcommand{\C}{\mathsf{C}}

\renewcommand{\L}{\mathsf{\Lambda}}

\newcommand{\fe}{\mathfrak{e}}
\newcommand{\ff}{\mathfrak{f}}
\newcommand{\fg}{\mathfrak{g}}
\newcommand{\fh}{\mathfrak{h}}
\newcommand{\su}{\mathfrak{su}}
\renewcommand{\sp}{\mathfrak{sp}}
\newcommand{\so}{\mathfrak{so}}
\renewcommand{\u}{\mathfrak{u}}

\newcommand{\ubf}[1]{\underline{\bf #1}}

\newcommand{\lra}{\leftrightarrow}

\newcommand{\bF}{{\mathbb F}}

\title{Dualities of $5d$ gauge theories from S-duality}

\author{Lakshya Bhardwaj}

\affiliation{Department of Physics, Harvard University\\
17 Oxford St, Cambridge, MA 02138, USA}

\abstract{We describe a general method to determine dualities between supersymmetric $5d$ gauge theories. The method is based on performing local S-dualities in the geometry associated to the gauge theory. We find that often a duality can be obtained by adding matter to both sides of a more primitive duality. This allows us to define the notion of irreducible dualities which cannot be obtained from more primitive dualities. More general dualities then are obtained by adding matter to both sides of an irreducible duality. The geometric method described in this paper allows us to systematically construct irreducible dualities. As an application, we explicitly determine a special class of irreducible dualities classified by removal and addition of edges into a Dynkin diagram. This class of dualities vastly generalizes many of the known $5d$ dualities in the literature.
}

\begin{document}

\maketitle

\section{Introduction} \label{intro}
It is well-known by now that different supersymmetric\footnote{Throughout this paper, a $5d$ gauge theory will refer to a supersymmetric $5d$ gauge theory with eight supercharges, or in other words, $\cN=1$ supersymmetry in $5d$. Theories admitting more supersymmetry will also be treated in an $\cN=1$ language.} $5d$ gauge theories can be dual to each other\footnote{See \cite{Bergman:2013aca,Zafrir:2014ywa,Bergman:2014kza,Zafrir:2015rga,Zafrir:2015ftn,Hayashi:2018lyv,Hayashi:2016abm,Hayashi:2015vhy,Hayashi:2015zka,Hayashi:2015fsa,Jefferson:2018irk,Bergman:2015dpa,Closset:2018bjz} for a sampling of recent work on this topic.}. A prime example is the duality between $SU(2)_\pi\times SU(2)_\pi$ with a bifundamental and\footnote{The notation $SU(n)_k$ for $n\ge3$ denotes $SU(n)$ with Chern-Simons level $k$. The notation $SU(2)_\theta$ denotes $SU(2)$ with theta angle $\theta$ which is $\Z_2$ valued and can take values $0,\pi$. The same holds for $Sp(n)_\theta$.} $SU(3)_0$ with two flavors \cite{Aharony:1997ju,Jefferson:2017ahm}. This paper is devoted to studying such dualities abstractly.

A $5d$ gauge theory is non-renormalizable, thus its UV completion is not guaranteed. Three kinds of UV completions are especially interesting:
\ben
\item The $5d$ gauge theory describes a region (i.e. an open set) of the Coulomb branch of a mass-deformed $5d$ SCFT.
\item The $5d$ gauge theory describes a region of the Coulomb branch of a (possibly) mass-deformed $6d$ SCFT compactified on circle.
\item The $5d$ gauge theory arises at low-energies in a $5d$ supergravity theory admitting a consistent UV completion.
\een
As discussed later in the paper, it is possible to associate a \emph{local piece} of a Calabi-Yau threefold to any arbitrary gauge theory. The correspondence between the local piece and the gauge theory is established by imagining that we are compactifying M-theory on a Calabi-Yau threefold\footnote{It is important to keep in mind that the existence of a Calabi-Yau threefold carrying this local piece is not guaranteed. In the cases where such a Calabi-Yau threefold does not exist, we believe that the corresponding gauge theory cannot be UV completed in any way and hence should be discarded.} which carries this local piece. In terms of this local piece, the above-mentioned three kinds of UV completions take the following form:
\ben
\item If this local piece can be simultaneously shrunk to a point at a finite distance in the Calabi-Yau moduli space, then this piece describes all the finite volume structure of a non-compact Calabi-Yau threefold. The corresponding $5d$ gauge theory then describes a region of the Coulomb branch of a mass-deformed $5d$ SCFT.
\item If this local piece can be expressed as an elliptic fibration\footnote{The existence of a section is not necessary. If there is no section, the fibration is referred to as a \emph{genus-one fibration} rather than an elliptic fibration.} whose base can be shrunk to a point, then this piece describes all the finite volume structure of a non-compact elliptically fibered Calabi-Yau threefold. The corresponding $5d$ gauge theory then describes a region of the Coulomb branch of a (possibly) mass-deformed $6d$ SCFT compactified on circle.
\item If the local piece embeds into a compact Calabi-Yau threefold, then the corresponding $5d$ gauge theory describes a low-energy sector of a $5d$ supergravity theory admitting a consistent UV completion.
\een

Now, it is often possible to change the parameters and moduli of the UV complete theory to transition from one $5d$ gauge theory description to another $5d$ gauge theory description. In such a case, the two $5d$ gauge theories are referred as being \emph{dual} to each other\footnote{Extra care must be taken in the case when UV completion is a supergravity theory. In these cases, the changes in parameters and moduli should be smaller than the energy scale below which the $5d$ gauge theory description is valid.}. From the point of view of the local pieces associated to the two $5d$ gauge theories, such changes correspond to a geometric transition from the local piece associated to the first gauge theory to the local piece associated to the second gauge theory.

It is thus possible to forget about the underlying UV completion of $5d$ gauge theories, and define a duality between two $5d$ gauge theories in terms of the existence of a geometric transition between local pieces of Calabi-Yau threefolds associated to the two gauge theories. If the local pieces are such that they can be embedded in a compact or a non-compact Calabi-Yau threefold, then the corresponding duality is physical and interesting. If the local pieces do not admit a consistent completion into a Calabi-Yau threefold, then the duality is unphysical and can be discarded.

In this paper, we study dualities from this abstract, geometric point of view, without regard for whether or not the gauge theories on two sides of the duality actually admit a UV completion. The exploration of latter issue is left to future papers on this topic. However, on the basis of the results of \cite{Jefferson:2017ahm}, we do collect all the dualities\footnote{This does not mean that the other dualities appearing in this paper can be discarded, as the gauge theories participating in these dualities may UV complete into a consistent supergravity theory. The reader might suspect that only the gauge theories which can UV complete into a theory without gravity may admit UV completion into a theory with gravity. However, this is not true. For example, the base in the F-theory construction of $6d$ supergravity theories can admit curves of positive self-intersection, but such curves are forbidden in the F-theory constructions for $6d$ SCFTs and little string theories.} appearing in this paper which can correspond to dualities between $5d$ gauge theories which describe regions of Coulomb branch of some mass deformed $5d$ or $6d$ SCFT (the latter being compactified on circle). See Section \ref{list} for the list of these dualities.

Moving onto the Coulomb branch of a $5d$ gauge theory corresponds to resolving the associated local piece. The resolved local piece\footnote{For ease of communication, we refer to a resolved local piece as a ``geometry'' from hereafter.} then consists of a collection of Kahler surfaces intersecting each other in a way consistent with the Calabi-Yau condition. A duality between two $5d$ gauge theories means that the two associated geometries are isomorphic to each other upto flop transitions associated to compact $-1$ curves living inside the Kahler surfaces. In this paper, we will focus our attention on a set of dualities which can be thought of as implementing S-duality of Type IIB string theory on some of the surfaces inside the geometry associated to one of the gauge theories. See Section \ref{S} for more details.

We will also impose some further technical restrictions on the geometries to simplify our analysis. These restrictions are described at the beginning of Section \ref{ID}. We discuss a (partial) list of generalizations of the methods of this paper in Section \ref{FG}.

Rest of this paper is organized as follows. Section \ref{list} contains a list of dualities appearing in this paper which are relevant in the context of $5d$ gauge theory descriptions of $5d$ and $6d$ SCFTs. In Section \ref{review}, we describe how a geometry is associated to any $5d$ gauge theory. In this section, we also discuss how the S-duality of Type IIB string theory appears as a geometric isomorphism $\cS$. We discuss how implementing $\cS$ leads to dualities between $5d$ gauge theories. Another interesting point is that if two gauge theories are found to be dual, then adding suitable matter on both sides of the duality preserves the duality. Thus, each duality lives in a tower of dualities built on top of what we call an \emph{irreducible duality} which captures the minimum amount of matter necessary to implement the duality. In Section \ref{ID}, we define a special class of irreducible dualities which are classified by adding and removing edges from the Dynkin diagram of a simple gauge algebra. We explicitly work out all the irreducible dualities in this class and display them inside boxes in Section \ref{ID} for easy identification. We also describe the map between the Dynkin diagrams on two sides of the dualities, thus completely specifying the full tower of dualities that can be built on top of the irreducible dualities in this special class. In Section \ref{FG}, we close the paper by discussing many ways in which the considerations of this paper can be generalized.

\subsection{Dualities for $5d$ gauge theories describing $5d$ and $6d$ SCFTs}\label{list}
Here we collect all those dualities appearing in this paper which involve $5d$ gauge theories that can describe regions of Coulomb branches of $5d$ and $6d$ SCFTs (with the latter being compactified on circle). All the dualities we study in this paper involve a gauge theory with a simple gauge algebra on one side of the duality, and $5d$ gauge theories with simple gauge algebra which can describe $5d$ and $6d$ SCFTs were constrained in the work of \cite{Jefferson:2017ahm}. In other words, in this section, we collect all those dualities which involve a gauge theory appearing in \cite{Jefferson:2017ahm}. It should be noted that \cite{Jefferson:2017ahm} only proposed necessary conditions for these gauge theories, thus the theories appearing here may not have a UV completion into a $5d$ or $6d$ SCFT. In fact, as we will see, we find that some of the theories appearing in \cite{Jefferson:2017ahm} should be dual to theories not appearing in \cite{Jefferson:2017ahm}, implying that the former cannot be UV completed into a $5d$ or $6d$ SCFT even though they satisfy the conditions of \cite{Jefferson:2017ahm}. See below for such examples.

Some of the dualities appearing below have appeared in the literature before, particularly in \cite{Bergman:2013aca,Zafrir:2014ywa,Bergman:2014kza,Zafrir:2015rga,Zafrir:2015ftn,Hayashi:2018lyv,Hayashi:2016abm,Hayashi:2015vhy,Hayashi:2015zka,Hayashi:2015fsa,Jefferson:2018irk,Bergman:2015dpa}.

In the following, $\F$ will denote the fundamental representation, $\L^n$ will denote the $n$-index antisymmetric irrep, $\S^2$ will denote the 2-index symmetric irrep, $\S$ and $\C$ will denote the two spinor irreps, and $\A$ will denote the adjoint representation.

(\ref{su1}) and (\ref{su3}) contain the following dualities:
\be
\begin{tikzpicture} [scale=1.9]
\node (v1) at (-5.4,0.4) {$\su(2)_0$};
\node at (-5.4,0.9) {$\su(n)_{\frac{m-p}2}+(2n+4-m-p)\F$};
\node at (-7,0.4) {$=$};
\node (v2) at (-6.4,0.4) {$\su(2)_{\pi}$};
\draw  (v2) edge (v1);
\node (v3) at (-4.4,0.4) {$\su(2)_0$};
\draw  (v1) edge (v3);
\node (v4) at (-3.7,0.4) {$\cdots$};
\draw  (v3) edge (v4);
\node (v5) at (-3,0.4) {$\su(2)_0$};
\draw  (v4) edge (v5);
\node (v6) at (-2,0.4) {$\su(2)_{\pi}$};
\draw  (v5) edge (v6);
\begin{scope}[shift={(-2.5,0.05)}]
\node at (-1.8,-0.1) {$n-3$};
\draw (-3.2,0.15) .. controls (-3.2,0.1) and (-3.2,0.05) .. (-3.1,0.05);
\draw (-3.1,0.05) -- (-1.9,0.05);
\draw (-1.8,0) .. controls (-1.8,0.05) and (-1.85,0.05) .. (-1.9,0.05);
\draw (-1.8,0) .. controls (-1.8,0.05) and (-1.75,0.05) .. (-1.7,0.05);
\draw (-1.7,0.05) -- (-0.35,0.05);
\draw (-0.25,0.15) .. controls (-0.25,0.1) and (-0.25,0.05) .. (-0.35,0.05);
\end{scope}
\node (v7) at (-6.4,-0.2) {$(4-p)\F$};
\draw  (v2) edge (v7);
\node (v8) at (-2,-0.2) {$(4-m)\F$};
\draw  (v6) edge (v8);
\end{tikzpicture}
\ee
Every integer $m,n,p$ will always be assumed to be non-negative. An edge between two gauge algebras denotes a bifundamental hyper and an edge between a gauge algebra and $n\F$ denotes that the gauge algebra carries additional $n$ number of hypers in fundamental representation. The theta angle for $\su(2)$ is relevant only if it carries no fundamental hypers which are uncharged under any other gauge algebras. Thus, the theta angle for left-most $\su(2)$ is relevant only when $p=4$, in which case it is $\pi$ as displayed above. Similarly, the theta angle for right-most $\su(2)$ is relevant only when $m=4$, in which case it is $\pi$.
\be

\ee
where we remind the reader that $n$ is non-negative, that is $n=0$ is allowed.
\be
%
\ee
where $\su(4)$ contains a hyper in 2-index antisymmetric representation.
\be
%
\ee

(\ref{su2}) includes the following dualities:
\be
%
\ee

(\ref{suu4}) includes the following dualities:
\be
%
\ee
Like $\su(2)$, the theta angle for $\sp(n)$ is relevant only when it carries no fundamental hypers that are not charged under some other gauge algebra. That is, the theta angle of $\sp(2)$ in above duality is relevant only when $p=14-2n$.
\be
%
\ee

(\ref{suu5}) includes the following dualities:
\be
%
\ee

(\ref{soo1}) and (\ref{soo2}) contain the following dualities:
\be
%
\ee

(\ref{soo3}) and (\ref{soo4}) contain the following dualities:
\be
%
\ee

(\ref{soo5}) contains the following dualities:
\be
\so(7)+(7-m)\S\:=\:\su(4)_{1-\frac m2}+3\L^2+(4-m)\F
\ee
where $m\le4$.
\be
\so(7)+\F+(6-m)\S\:=\:\su(4)_{-\frac m2}+3\L^2+(4-m)\F
\ee
where $m\le3$.
\be
\so(7)+2\F+(5-m)\S\:=\:\su(4)_{-1-\frac m2}+3\L^2+(4-m)\F
\ee
where $m\le2$.
\be
\so(7)+3\F+(4-m)\S\:=\:\su(4)_{-2-\frac m2}+3\L^2+(4-m)\F
\ee
where $m\le1$.
\be
\su(4)_{-1}+4\L^2\:=\:\so(7)+3\S+\A
\ee
Since $\so(7)+\A$ already lifts to a $6d$ SCFT, adding extra $3\F$ should cause the theory to have no UV completion. Thus, according to the above duality, $\su(4)_{-1}+4\L^2$ should not lift to a $5d$ or $6d$ SCFT despite it satisfying the conditions of \cite{Jefferson:2017ahm}.
\be
\so(9)+3\S+(3-m)\F\:=\:\su(5)_{\frac m2}+3\L^2+(3-m)\F
\ee

\be
\so(9)+4\S+(1-m)\F\:=\:\su(5)_{\frac {3+m}2}+3\L^2+(2-m)\F
\ee

(\ref{soo6}) contains the following two dualities:
\be
\so(7)+7\S\:=\:\so(7)+5\S+2\F
\ee

\be
\so(7)+6\S\:=\:\so(7)+5\S+\F
\ee

(\ref{spp1}) contains the following duality:
\be
\begin{tikzpicture} [scale=1.9]
\node (v1) at (-6.4,0.4) {$\su(2)_\pi$};
\node at (-8.1,0.4) {$\sp(3)+\L^2+8\F$};
\node at (-7.1,0.4) {$=$};
\node (v3) at (-5.4,0.4) {$\sp(2)$};
\draw  (v1) edge (v3);
\node (v4) at (-5.4,-0.2) {$7\F$};
\draw  (v3) edge (v4);
\end{tikzpicture}
\ee

(\ref{spsu}) contains the following dualities where $n\ge2$:
\be
\sp(n)_{(n+1)\pi}+(2n+6-m)\F\:=\:\su(n+1)_{-\frac m2}+(2n+6-m)\F
\ee

\be
\sp(n)_{(n+1)\pi}+\L^2+(8-m)\F\:=\:\su(n+1)_{-\frac {n+1+m}2}+\L^2+(8-m)\F
\ee

\be
\sp(n)_{(n+1)\pi}+\A\:=\:\su(n+1)_{-\frac {n+1}2}+\S^2
\ee

\be
\sp(2)_{\pi}+2\L^2+(4-m)\F\:=\:\su(3)_{-4-\frac m2}+(6-m)\F
\ee

\be
\sp(2)_{\pi}+3\L^2\:=\:\su(3)_{-\frac{13}2}+3\F
\ee

\be
\sp(3)_{0}+2\L^2\:=\:\su(4)_{-6}+2\L^2
\ee

\be
\sp(3)+\L^3+(5-m)\F\:=\:\su(4)_{-4-\frac m2}+(6-m)\F
\ee

\be
\su(4)_{-3-\frac m2}+(8-m)\F\:=\:\sp(3)+\L^3+(7-m)\F
\ee
Since the right hand side does not satisfy the conditions of \cite{Jefferson:2017ahm} for $m=1,2$, the left hand side should not admit a UV completion into $5d$ or $6d$ SCFT for $m=1,2$ even though it satisfies the conditions of \cite{Jefferson:2017ahm}.
\be
\su(5)_{-\frac {11+m}2}+(5-m)\F\:=\:\sp(4)_\pi+\L^4+(4-m)\F
\ee
Since the right hand side does not satisfy the conditions of \cite{Jefferson:2017ahm}, the left hand side should not admit a UV completion into $5d$ or $6d$ SCFT even though it satisfies the conditions of \cite{Jefferson:2017ahm}.

(\ref{soe}) and (\ref{soe2}) contain the following dualities:
\be
\begin{tikzpicture} [scale=1.9]
\node (v1) at (-6.4,0.4) {$\su(2)_0$};
\node at (-8.5,0.4) {$\so(8)+(5-m)\F+\S+\C$};
\node at (-7.1,0.4) {$=$};
\node (v3) at (-5.4,0.4) {$\su(4)_0$};
\draw  (v1) edge (v3);
\node (v4) at (-5.4,-0.2) {$2\L^2$};
\draw  (v3) edge (v4);
\node (v2) at (-6.4,-0.2) {$(3-m)\F$};
\draw  (v1) edge (v2);
\end{tikzpicture}
\ee
For $\so(8)$ we are only going to display results upto the action of triality.
\be
\begin{tikzpicture} [scale=1.9]
\node (v1) at (-6.4,0.4) {$\su(2)_0$};
\node at (-8.8,0.4) {$\so(8)+(4-m)\F+(3-n)\S+\C$};
\node at (-7.1,0.4) {$=$};
\node (v3) at (-5.2,0.4) {$\su(4)_{1-\frac n2}$};
\draw  (v1) edge (v3);
\node (v4) at (-5.2,-0.2) {$2\L^2$};
\draw  (v3) edge (v4);
\node (v2) at (-6.4,-0.2) {$(2-m)\F$};
\draw  (v1) edge (v2);
\node (v5) at (-4,0.4) {$(2-n)\F$};
\draw  (v3) edge (v5);
\end{tikzpicture}
\ee
\be
\begin{tikzpicture} [scale=1.9]
\node (v1) at (-6.4,0.4) {$\su(2)_0$};
\node at (-8.6,0.4) {$\so(8)+(3-m)\F+4\S+\C$};
\node at (-7.1,0.4) {$=$};
\node (v3) at (-5.3,0.4) {$\su(4)_{\frac 32}$};
\draw  (v1) edge (v3);
\node (v4) at (-5.3,-0.2) {$2\L^2$};
\draw  (v3) edge (v4);
\node (v2) at (-6.4,-0.2) {$(1-m)\F$};
\draw  (v1) edge (v2);
\node (v5) at (-4.4,0.4) {$3\F$};
\draw  (v3) edge (v5);
\end{tikzpicture}
\ee
\be
\begin{tikzpicture} [scale=1.9]
\node (v1) at (-6.4,0.4) {$\su(2)_0$};
\node at (-9.1,0.4) {$\so(8)+(4-m)\F+(2-n)\S+(2-p)\C$};
\node at (-7.1,0.4) {$=$};
\node (v3) at (-5.3,0.4) {$\su(4)_{\frac{p-n}2}$};
\draw  (v1) edge (v3);
\node (v4) at (-5.3,-0.2) {$2\L^2$};
\draw  (v3) edge (v4);
\node (v2) at (-6.4,-0.2) {$(1-m)\F$};
\draw  (v1) edge (v2);
\node (v5) at (-4.1,0.4) {$(1-n)\F$};
\draw  (v3) edge (v5);
\node (v6) at (-5.3,1) {$(1-p)\F$};
\draw  (v6) edge (v3);
\end{tikzpicture}
\ee
where we have not combined the fundamentals for $\su(4)$ to manifest the possible values for $n$ and $p$, which can at most be 1.
\be
\begin{tikzpicture} [scale=1.9]
\node (v1) at (-6.4,0.4) {$\su(2)_0$};
\node at (-8.9,0.4) {$\so(8)+2\F+(4-n)\S+(2-p)\C$};
\node at (-7.1,0.4) {$=$};
\node (v3) at (-5.2,0.4) {$\su(4)_{1+\frac{p-n}2}$};
\draw  (v1) edge (v3);
\node (v4) at (-5.2,-0.2) {$2\L^2$};
\draw  (v3) edge (v4);
\node (v5) at (-3.9,0.4) {$(3-n)\F$};
\draw  (v3) edge (v5);
\node (v6) at (-5.2,1) {$(1-p)\F$};
\draw  (v6) edge (v3);
\end{tikzpicture}
\ee
\be
\begin{tikzpicture} [scale=1.9]
\node (v1) at (-6.4,0.4) {$\su(2)_0$};
\node at (-9.1,0.4) {$\so(8)+(3-m)\F+(3-n)\S+(2-p)\C$};
\node at (-7.1,0.4) {$=$};
\node (v3) at (-5.2,0.4) {$\su(4)_{\frac{1+p-n}2}$};
\draw  (v1) edge (v3);
\node (v4) at (-5.2,-0.2) {$2\L^2$};
\draw  (v3) edge (v4);
\node (v2) at (-6.4,-0.2) {$(1-m)\F$};
\draw  (v1) edge (v2);
\node (v5) at (-3.9,0.4) {$(2-n)\F$};
\draw  (v3) edge (v5);
\node (v6) at (-5.2,1) {$(1-p)\F$};
\draw  (v6) edge (v3);
\end{tikzpicture}
\ee
\be
\begin{tikzpicture} [scale=1.9]
\node (v1) at (-6.4,0.4) {$\su(2)_0$};
\node at (-8.9,0.4) {$\so(8)+2\F+(3-n)\S+(3-p)\C$};
\node at (-7.1,0.4) {$=$};
\node (v3) at (-5.3,0.4) {$\su(4)_{\frac{p-n}2}$};
\draw  (v1) edge (v3);
\node (v4) at (-5.3,-0.2) {$2\L^2$};
\draw  (v3) edge (v4);
\node (v5) at (-4.1,0.4) {$(2-n)\F$};
\draw  (v3) edge (v5);
\node (v6) at (-5.3,1) {$(2-p)\F$};
\draw  (v6) edge (v3);
\end{tikzpicture}
\ee
\be
\begin{tikzpicture} [scale=1.9]
\node (v1) at (-6.1,0.4) {$\su(2)_\pi$};
\node at (-8.9,0.4) {$\so(10)+4\F+3\S$};
\node at (-7.8,0.4) {$=$};
\node (v3) at (-5.1,0.4) {$\su(4)_{\frac12}$};
\draw  (v1) edge (v3);
\node (v4) at (-5.1,-0.2) {$2\L^2$};
\draw  (v3) edge (v4);
\node (v5) at (-4.3,0.4) {$\F$};
\draw  (v3) edge (v5);
\node (v2) at (-7.1,0.4) {$\su(2)_\pi$};
\draw  (v2) edge (v1);
\end{tikzpicture}
\ee
\be
\begin{tikzpicture} [scale=1.9]
\node (v1) at (-6.1,0.4) {$\su(2)_\pi$};
\node at (-9.2,0.4) {$\so(10)+(6-m)\F+2\S$};
\node at (-7.8,0.4) {$=$};
\node (v3) at (-5.1,0.4) {$\su(4)_{0}$};
\draw  (v1) edge (v3);
\node (v4) at (-5.1,-0.2) {$2\L^2$};
\draw  (v3) edge (v4);
\node (v5) at (-7.1,-0.2) {$(2-m)\F$};
\node (v2) at (-7.1,0.4) {$\su(2)_\pi$};
\draw  (v2) edge (v1);
\draw  (v2) edge (v5);
\end{tikzpicture}
\ee

(\ref{soe3}) and (\ref{soe4}) contain the following dualities:
\be
\begin{tikzpicture} [scale=1.9]
\node (v1) at (-6.1,0.4) {$\su(3)_3$};
\node at (-9.1,0.4) {$\so(8)+(5-m)\F+2\S$};
\node at (-7.8,0.4) {$=$};
\node (v3) at (-5.1,0.4) {$\su(2)_{\pi}$};
\draw  (v1) edge (v3);
\node (v5) at (-7.1,-0.2) {$(3-m)\F$};
\node (v2) at (-7.1,0.4) {$\su(2)_\pi$};
\draw  (v2) edge (v1);
\draw  (v2) edge (v5);
\end{tikzpicture}
\ee
\be
\begin{tikzpicture} [scale=1.9]
\node (v1) at (-6.1,0.4) {$\su(3)_3$};
\node at (-9.4,0.4) {$\so(8)+(4-m)\F+(4-n)\S$};
\node at (-7.8,0.4) {$=$};
\node (v3) at (-5.1,0.4) {$\su(2)_{\pi}$};
\draw  (v1) edge (v3);
\node (v4) at (-5.1,-0.2) {$(2-n)\F$};
\draw  (v3) edge (v4);
\node (v5) at (-7.1,-0.2) {$(2-m)\F$};
\node (v2) at (-7.1,0.4) {$\su(2)_\pi$};
\draw  (v2) edge (v1);
\draw  (v2) edge (v5);
\end{tikzpicture}
\ee
\be
\begin{tikzpicture} [scale=1.9]
\node (v1) at (-6.1,0.4) {$\su(3)_\frac52$};
\node at (-9.6,0.4) {$\so(8)+(4-m)\F+(3-n)\S+\C$};
\node at (-7.8,0.4) {$=$};
\node (v3) at (-5.1,0.4) {$\su(2)_{\pi}$};
\draw  (v1) edge (v3);
\node (v4) at (-5.1,-0.2) {$(1-n)\F$};
\draw  (v3) edge (v4);
\node (v5) at (-7.1,-0.2) {$(2-m)\F$};
\node (v2) at (-7.1,0.4) {$\su(2)_\pi$};
\draw  (v2) edge (v1);
\draw  (v2) edge (v5);
\node (v6) at (-6.1,-0.2) {$\F$};
\draw  (v1) edge (v6);
\end{tikzpicture}
\ee
\be
\begin{tikzpicture} [scale=1.9]
\node (v1) at (-6.1,0.4) {$\su(3)_2$};
\node at (-9.4,0.4) {$\so(8)+(4-m)\F+2\S+2\C$};
\node at (-7.8,0.4) {$=$};
\node (v3) at (-5.1,0.4) {$\su(2)_{\pi}$};
\draw  (v1) edge (v3);
\node (v5) at (-7.1,-0.2) {$(2-m)\F$};
\node (v2) at (-7.1,0.4) {$\su(2)_\pi$};
\draw  (v2) edge (v1);
\draw  (v2) edge (v5);
\node (v6) at (-6.1,-0.2) {$2\F$};
\draw  (v1) edge (v6);
\end{tikzpicture}
\ee
\be
\begin{tikzpicture} [scale=1.9]
\node (v1) at (-6,0.4) {$\su(3)_{1+\frac p2}$};
\node at (-9.4,0.4) {$\so(8)+2\F+2\S+(4-p)\C$};
\node at (-7.8,0.4) {$=$};
\node (v3) at (-4.9,0.4) {$\su(2)_{\pi}$};
\draw  (v1) edge (v3);
\node (v2) at (-7.1,0.4) {$\su(2)_\pi$};
\draw  (v2) edge (v1);
\node (v6) at (-6,-0.2) {$(4-p)\F$};
\draw  (v1) edge (v6);
\end{tikzpicture}
\ee
\be
\begin{tikzpicture} [scale=1.9]
\node (v1) at (-6.1,0.4) {$\su(3)_{2}$};
\node at (-9.7,0.4) {$\so(8)+(3-m)\F+(3-n)\S+2\C$};
\node at (-7.8,0.4) {$=$};
\node (v3) at (-5.1,0.4) {$\su(2)_{\pi}$};
\draw  (v1) edge (v3);
\node (v2) at (-7.1,0.4) {$\su(2)_\pi$};
\node (v4) at (-5.1,-0.2) {$(1-n)\F$};
\draw  (v3) edge (v4);
\node (v5) at (-7.1,-0.2) {$(1-m)\F$};
\draw  (v2) edge (v5);
\draw  (v2) edge (v1);
\node (v6) at (-6.1,-0.2) {$2\F$};
\draw  (v1) edge (v6);
\end{tikzpicture}
\ee
\be
\begin{tikzpicture} [scale=1.9]
\node (v1) at (-6.1,0.4) {$\su(3)_{\frac32}$};
\node at (-9.4,0.4) {$\so(8)+(3-m)\F+2\S+3\C$};
\node at (-7.8,0.4) {$=$};
\node (v3) at (-5.1,0.4) {$\su(2)_{\pi}$};
\draw  (v1) edge (v3);
\node (v2) at (-7.1,0.4) {$\su(2)_\pi$};
\node (v5) at (-7.1,-0.2) {$(1-m)\F$};
\draw  (v2) edge (v5);
\draw  (v2) edge (v1);
\node (v6) at (-6.1,-0.2) {$3\F$};
\draw  (v1) edge (v6);
\end{tikzpicture}
\ee
\be
\begin{tikzpicture} [scale=1.9]
\node (v1) at (-6.1,0.4) {$\su(2)_{0}$};
\node at (-9.2,0.4) {$\so(10)+(6-m)\F+2\S$};
\node at (-7.8,0.4) {$=$};
\node (v3) at (-5.1,0.4) {$\su(3)_{3}$};
\draw  (v1) edge (v3);
\node (v2) at (-7.1,0.4) {$\su(2)_\pi$};
\node (v5) at (-7.1,-0.2) {$(2-m)\F$};
\draw  (v2) edge (v5);
\draw  (v2) edge (v1);
\node (v4) at (-4.1,0.4) {$\su(2)_\pi$};
\draw  (v3) edge (v4);
\end{tikzpicture}
\ee
\be
\begin{tikzpicture} [scale=1.9]
\node (v1) at (-6.1,0.4) {$\su(2)_{0}$};
\node at (-8.9,0.4) {$\so(10)+4\F+3\S$};
\node at (-7.8,0.4) {$=$};
\node (v3) at (-5.1,0.4) {$\su(3)_{3}$};
\draw  (v1) edge (v3);
\node (v2) at (-7.1,0.4) {$\su(2)_\pi$};
\draw  (v2) edge (v1);
\node (v4) at (-4.1,0.4) {$\su(2)$};
\draw  (v3) edge (v4);
\node (v5) at (-3.3,0.4) {$\F$};
\draw  (v4) edge (v5);
\begin{scope}[shift={(0,-0.6)}]
\node (v1) at (-6.1,0.4) {$\su(2)_{0}$};
\node at (-7.8,0.4) {$=$};
\node (v3) at (-5.1,0.4) {$\su(3)_{\frac52}$};
\draw  (v1) edge (v3);
\node (v2) at (-7.1,0.4) {$\su(2)_\pi$};
\node (v4) at (-5.1,-0.2) {$\F$};
\draw  (v3) edge (v4);
\draw  (v2) edge (v1);
\node (v4) at (-4.1,0.4) {$\su(2)_\pi$};
\draw  (v3) edge (v4);
\end{scope}
\end{tikzpicture}
\ee

(\ref{soe5}) contains the following dualities:
\be
\begin{tikzpicture} [scale=1.9]
\node (v1) at (-6.4,0.4) {$\su(2)_0$};
\node at (-8.5,0.4) {$\so(10)+(4-m)\F+3\S$};
\node at (-7.1,0.4) {$=$};
\node (v3) at (-5.2,0.4) {$\su(5)_{\frac{1+m}2}$};
\draw  (v1) edge (v3);
\node (v4) at (-5.2,-0.2) {$2\L^2$};
\draw  (v3) edge (v4);
\node (v5) at (-4,0.4) {$(3-m)\F$};
\draw  (v3) edge (v5);
\end{tikzpicture}
\ee
\be
\begin{tikzpicture} [scale=1.9]
\node (v1) at (-6.3,0.4) {$\su(2)$};
\node at (-8.3,0.4) {$\so(10)+(2-m)\F+4\S$};
\node at (-6.9,0.4) {$=$};
\node (v3) at (-5.2,0.4) {$\su(5)_{\frac{3+m}2}$};
\draw  (v1) edge (v3);
\node (v4) at (-5.2,-0.2) {$2\L^2$};
\draw  (v3) edge (v4);
\node (v2) at (-6.3,-0.2) {$\F$};
\draw  (v1) edge (v2);
\node (v5) at (-4,0.4) {$(1-m)\F$};
\draw  (v3) edge (v5);
\begin{scope}[shift={(0,-1.7)}]
\node (v1) at (-6.3,0.4) {$\su(2)_0$};
\node at (-6.9,0.4) {$=$};
\node (v3) at (-5.2,0.4) {$\su(5)_{\frac{3+m}2}$};
\draw  (v1) edge (v3);
\node (v4) at (-5.2,-0.2) {$2\L^2$};
\draw  (v3) edge (v4);
\node (v5) at (-4,0.4) {$(1-m)\F$};
\draw  (v3) edge (v5);
\end{scope}
\node (v6) at (-5.2,-0.7) {$\F$};
\draw  (v6) edge (v3);
\end{tikzpicture}
\ee

(\ref{soe6}) contains the following two dualities:
\be
\begin{tikzpicture} [scale=1.9]
\node (v1) at (-6.3,0.4) {$\su(2)_\pi$};
\node at (-8.6,0.4) {$\so(8)+(4-m)\F+2\S+2\C$};
\node at (-7,0.4) {$=$};
\node (v3) at (-5.3,0.4) {$\su(2)_{\pi}$};
\draw  (v1) edge (v3);
\node (v4) at (-5.3,-0.2) {$\su(2)_\pi$};
\draw  (v3) edge (v4);
\node (v2) at (-6.3,-0.2) {$(2-m)\F$};
\draw  (v1) edge (v2);
\node (v5) at (-4.3,0.4) {$\su(2)_\pi$};
\draw  (v3) edge (v5);
\end{tikzpicture}
\ee
\be
\begin{tikzpicture} [scale=1.9]
\node (v1) at (-6.3,0.4) {$\su(2)_\pi$};
\node at (-8.9,0.4) {$\so(8)+(3-m)\F+(3-n)\S+2\C$};
\node at (-7,0.4) {$=$};
\node (v3) at (-5.3,0.4) {$\su(2)_{\pi}$};
\draw  (v1) edge (v3);
\node (v4) at (-5.3,-0.2) {$\su(2)_\pi$};
\draw  (v3) edge (v4);
\node (v2) at (-6.3,-0.2) {$(1-m)\F$};
\draw  (v1) edge (v2);
\node (v5) at (-4.3,0.4) {$\su(2)_\pi$};
\draw  (v3) edge (v5);
\node (v6) at (-3.2,0.4) {$(1-n)\F$};
\draw  (v5) edge (v6);
\end{tikzpicture}
\ee

(\ref{soe7}) contains the following duality:
\be
\begin{tikzpicture} [scale=1.9]
\node (v1) at (-6.3,0.4) {$\su(2)_\pi$};
\node at (-8.4,0.4) {$\so(10)+(2-m)\F+4\S$};
\node at (-7,0.4) {$=$};
\node (v3) at (-5.1,0.4) {$\su(4)_{3+\frac m2}$};
\draw  (v1) edge (v3);
\node (v4) at (-5.1,-0.2) {$(2-m)\F$};
\draw  (v3) edge (v4);
\node (v5) at (-3.9,0.4) {$\su(2)_\pi$};
\draw  (v3) edge (v5);
\end{tikzpicture}
\ee

(\ref{soe8}) and (\ref{soe9}) contain the following dualities:
\be
\begin{tikzpicture} [scale=1.9]
\node (v1) at (-6.1,0.4) {$\sp(2)_\pi$};
\node at (-9.1,0.4) {$\so(8)+(5-m)\F+2\S$};
\node at (-7.8,0.4) {$=$};
\node (v3) at (-5.1,0.4) {$\su(2)_{0}$};
\draw  (v1) edge (v3);
\node (v5) at (-7.1,-0.2) {$(3-m)\F$};
\node (v2) at (-7.1,0.4) {$\su(2)_0$};
\draw  (v2) edge (v1);
\draw  (v2) edge (v5);
\end{tikzpicture}
\ee
\be
\begin{tikzpicture} [scale=1.9]
\node (v1) at (-6.1,0.4) {$\sp(2)_\pi$};
\node at (-9.4,0.4) {$\so(8)+(4-m)\F+(4-n)\S$};
\node at (-7.8,0.4) {$=$};
\node (v3) at (-5.1,0.4) {$\su(2)_{0}$};
\draw  (v1) edge (v3);
\node (v4) at (-5.1,-0.2) {$(2-n)\F$};
\draw  (v3) edge (v4);
\node (v5) at (-7.1,-0.2) {$(2-m)\F$};
\node (v2) at (-7.1,0.4) {$\su(2)_0$};
\draw  (v2) edge (v1);
\draw  (v2) edge (v5);
\end{tikzpicture}
\ee
\be
\begin{tikzpicture} [scale=1.9]
\node (v1) at (-6.1,0.4) {$\sp(2)$};
\node at (-9.6,0.4) {$\so(8)+(4-m)\F+(3-n)\S+\C$};
\node at (-7.8,0.4) {$=$};
\node (v3) at (-5.1,0.4) {$\su(2)_{0}$};
\draw  (v1) edge (v3);
\node (v4) at (-5.1,-0.2) {$(1-n)\F$};
\draw  (v3) edge (v4);
\node (v5) at (-7.1,-0.2) {$(2-m)\F$};
\node (v2) at (-7.1,0.4) {$\su(2)_0$};
\draw  (v2) edge (v1);
\draw  (v2) edge (v5);
\node (v6) at (-6.1,-0.2) {$\F$};
\draw  (v1) edge (v6);
\end{tikzpicture}
\ee
\be
\begin{tikzpicture} [scale=1.9]
\node (v1) at (-6.1,0.4) {$\sp(2)$};
\node at (-9.4,0.4) {$\so(8)+(4-m)\F+2\S+2\C$};
\node at (-7.8,0.4) {$=$};
\node (v3) at (-5.1,0.4) {$\su(2)_{0}$};
\draw  (v1) edge (v3);
\node (v5) at (-7.1,-0.2) {$(2-m)\F$};
\node (v2) at (-7.1,0.4) {$\su(2)_0$};
\draw  (v2) edge (v1);
\draw  (v2) edge (v5);
\node (v6) at (-6.1,-0.2) {$2\F$};
\draw  (v1) edge (v6);
\end{tikzpicture}
\ee
\be
\begin{tikzpicture} [scale=1.9]
\node (v1) at (-6,0.4) {$\sp(2)_{\pi}$};
\node at (-9.4,0.4) {$\so(8)+2\F+2\S+(4-p)\C$};
\node at (-7.8,0.4) {$=$};
\node (v3) at (-4.9,0.4) {$\su(2)_{0}$};
\draw  (v1) edge (v3);
\node (v2) at (-7.1,0.4) {$\su(2)_0$};
\draw  (v2) edge (v1);
\node (v6) at (-6,-0.2) {$(4-p)\F$};
\draw  (v1) edge (v6);
\end{tikzpicture}
\ee
\be
\begin{tikzpicture} [scale=1.9]
\node (v1) at (-6.1,0.4) {$\sp(2)$};
\node at (-9.7,0.4) {$\so(8)+(3-m)\F+(3-n)\S+2\C$};
\node at (-7.8,0.4) {$=$};
\node (v3) at (-5.1,0.4) {$\su(2)_{0}$};
\draw  (v1) edge (v3);
\node (v2) at (-7.1,0.4) {$\su(2)_0$};
\node (v4) at (-5.1,-0.2) {$(1-n)\F$};
\draw  (v3) edge (v4);
\node (v5) at (-7.1,-0.2) {$(1-m)\F$};
\draw  (v2) edge (v5);
\draw  (v2) edge (v1);
\node (v6) at (-6.1,-0.2) {$2\F$};
\draw  (v1) edge (v6);
\end{tikzpicture}
\ee
\be
\begin{tikzpicture} [scale=1.9]
\node (v1) at (-6.1,0.4) {$\sp(2)$};
\node at (-9.4,0.4) {$\so(8)+(3-m)\F+2\S+3\C$};
\node at (-7.8,0.4) {$=$};
\node (v3) at (-5.1,0.4) {$\su(2)_{0}$};
\draw  (v1) edge (v3);
\node (v2) at (-7.1,0.4) {$\su(2)_0$};
\node (v5) at (-7.1,-0.2) {$(1-m)\F$};
\draw  (v2) edge (v5);
\draw  (v2) edge (v1);
\node (v6) at (-6.1,-0.2) {$3\F$};
\draw  (v1) edge (v6);
\end{tikzpicture}
\ee
\be
\begin{tikzpicture} [scale=1.9]
\node (v1) at (-6.1,0.4) {$\su(2)_{\pi}$};
\node at (-9.2,0.4) {$\so(10)+(6-m)\F+2\S$};
\node at (-7.8,0.4) {$=$};
\node (v3) at (-5.1,0.4) {$\sp(2)_{\pi}$};
\draw  (v1) edge (v3);
\node (v2) at (-7.1,0.4) {$\su(2)_0$};
\node (v5) at (-7.1,-0.2) {$(2-m)\F$};
\draw  (v2) edge (v5);
\draw  (v2) edge (v1);
\node (v4) at (-4.1,0.4) {$\su(2)_0$};
\draw  (v3) edge (v4);
\end{tikzpicture}
\ee
\be
\begin{tikzpicture} [scale=1.9]
\node (v1) at (-6.1,0.4) {$\su(2)_{\pi}$};
\node at (-8.9,0.4) {$\so(10)+4\F+3\S$};
\node at (-7.8,0.4) {$=$};
\node (v3) at (-5.1,0.4) {$\sp(2)_{\pi}$};
\draw  (v1) edge (v3);
\node (v2) at (-7.1,0.4) {$\su(2)_0$};
\draw  (v2) edge (v1);
\node (v4) at (-4.1,0.4) {$\su(2)$};
\draw  (v3) edge (v4);
\node (v5) at (-3.3,0.4) {$\F$};
\draw  (v4) edge (v5);
\begin{scope}[shift={(0,-0.6)}]
\node (v1) at (-6.1,0.4) {$\su(2)_{\pi}$};
\node at (-7.8,0.4) {$=$};
\node (v3) at (-5.1,0.4) {$\sp(2)$};
\draw  (v1) edge (v3);
\node (v2) at (-7.1,0.4) {$\su(2)_0$};
\node (v4) at (-5.1,-0.2) {$\F$};
\draw  (v3) edge (v4);
\draw  (v2) edge (v1);
\node (v4) at (-4.1,0.4) {$\su(2)_0$};
\draw  (v3) edge (v4);
\end{scope}
\end{tikzpicture}
\ee

(\ref{soe10}) contains the following duality:
\be
\begin{tikzpicture} [scale=1.9]
\node (v1) at (-7.1,0.4) {$\su(2)_{\pi}$};
\node at (-9.2,0.4) {$\so(10)+(2-m)\F+4\S$};
\node at (-7.8,0.4) {$=$};
\node (v3) at (-6.1,0.4) {$\sp(3)_{0}$};
\draw  (v1) edge (v3);
\node (v5) at (-6.1,-0.2) {$(2-m)\F$};
\node (v4) at (-5.1,0.4) {$\su(2)_\pi$};
\draw  (v3) edge (v4);
\draw  (v3) edge (v5);
\end{tikzpicture}
\ee

(\ref{f4}) contains the following dualities:
\be
\ff_4+3\F\:=\:\su(5)_{-\frac72}+3\L^2
\ee

\be
\su(5)_{-3}+3\L^2+\F\:=\:\ff_4+4\F
\ee
Since $\ff_4+4\F$ does not satisfy the conditions of \cite{Jefferson:2017ahm}, it follows from the above duality that $\su(5)_{-3}+3\L^2+\F$ cannot have a UV completion into $5d$ or $6d$ SCFT despite it satisfying the conditions of \cite{Jefferson:2017ahm}.

(\ref{g2sp2}) contains the following dualities:
\be
\fg_2+(6-m)\F\:=\:\sp(2)_{\pi}+2\L^2+(4-m)\F
\ee

\be
\sp(2)_\pi+3\L^2\:=\:\fg_2+\A+2\F
\ee
The above duality implies that $\sp(2)_\pi+3\L^2$ cannot UV complete into a $5d$ or $6d$ SCFT even though it satisfies the conditions of \cite{Jefferson:2017ahm}.

(\ref{g2su}) contains the following dualities:
\be
\fg_2+(6-m)\F\:=\:\su(3)_{-4-\frac m2}+(6-m)\F
\ee

\be
\su(3)_{-\frac{13+m}2}+(3-m)\F\:=\:\fg_2+\A+(2-m)\F
\ee
Thus, $\su(3)_{-\frac{13+m}2}+(3-m)\F$ for $m=0,1$ cannot be UV completed into a $5d$ or $6d$ SCFT even though they satisfy the conditions of \cite{Jefferson:2017ahm}.

\section{The geometry associated to a $5d$ gauge theory}\label{review}
In this section, we describe how one can associate a local resolved piece of a Calabi-Yau threefold to a $5d$ gauge theory in a particular phase. Many of the aspects discussed here are taken from \cite{Seiberg:1996bd,Morrison:1996xf,Intriligator:1997pq,Witten:1996qb,Diaconescu:1998cn,Jefferson:2018irk,Bhardwaj:2018yhy,Bhardwaj:2018vuu,DelZotto:2017pti,Bhardwaj:2019fzv} (see also \cite{Apruzzi:2018nre,Apruzzi:2019vpe,Apruzzi:2019opn,Apruzzi:2019enx}).

\subsection{Some geometric background}
This local piece is described in terms of a collection of complex surfaces intersecting each other. The complex surfaces relevant to the construction of gauge theories are Hirzebruch surfaces which are defined by specifying a $\P^1$ fibration over a base $\P^1$. If the fibration has degree $n$ (valued in non-negative integers), the corresponding Hirzebruch surface is denoted by $\bF_n$. Of particular interest to us are the holomorphic curve classes in $\bF_n$ and their intersection numbers. These classes are generated by two classes $e$ and $f$ with the  following intersection numbers\footnote{Note that intersection numbers of complex curves inside complex surfaces are symmetric.}
\begin{align}
e^2&=-n\\
f^2&=0\\
e\cdot f&=1
\end{align}
where $e$ is associated to the base $\P^1$ and $f$ is associated to the fiber $\P^1$. 

More general curves can be obtained by positive linear combinations of $e$ and $f$. An important curve class in $\bF_n$ is
\be
h:=e+nf
\ee
whose intersection numbers are
\begin{align}
h^2&=n\\
h\cdot e&=0\\
h\cdot f&=1
\end{align}
The curve $h$ has genus zero. The genus $g$ of a general curve 
\be\label{C}
C=\alpha e+\beta f
\ee
can be obtained using the adjunction formula
\be\label{adj}
(K+C)\cdot C=2g-2
\ee
where K is the canonical class of the surface and does not depend on the identity of $C$. The fact that $e,f$ have genus zero determines the values of $K\cdot e$ and $K\cdot f$ to be
\begin{align}
K\cdot e&=n-2\\
K\cdot f&=-2
\end{align}
in terms of which one can determine the genus of any $C$ of the form (\ref{C}).

We also consider blowups of Hirzebruch surfaces. If we perform $b$ blowups of $\bF_n$, we denote\footnote{It is possible to obtain many different surfaces by performing $b$ blowups on $\bF_n$. The difference between these surfaces is being tracked implicitly in this paper. Effectively, the gluing curves track how non-generic a blowup is, since every blowup is required to be as generic as is allowed by the existence of gluing curves. See Section 2.5 of \cite{Bhardwaj:2018vuu} to see how the full Mori cone of the surface can be determined from the data of gluing curves by applying the above genericity criterion.} the resulting surface as $\bF_n^b$. Let us denote the exceptional curves created by the blowups by $x_i$ with $i=1,\cdots,b$. We will use the convention that the total transforms\footnote{If $B:\tilde S\to S$ is a blowup of a surface $S$, then the total transform of a curve $C$ in $S$ is the curve $f^{-1}(C)$ in $\tilde S$.} of the curves $e$, $f$ and $h$ are denoted by the same names $e$, $f$ and $h$ in $\bF_n^b$. Thus, the intersection numbers between $e$, $f$ and $h$ are those mentioned above, and their intersections with $x_i$ are
\begin{align}
x_i\cdot x_j&=-\delta_{ij}\\
e\cdot x_i&=0\\
f\cdot x_i&=0\\
h\cdot x_i&=0
\end{align}
The possible holomorphic curves after the blowup can be decomposed as
\be
C=\alpha e+\beta f+\sum_i\gamma_i x_i
\ee
where $\alpha,\beta\ge0$ and $\gamma_i\in\Z$. Using adjunction formula (\ref{adj}), we see that we can write the canonical class of $\bF_n^b$ as
\be
K=-(e+h+2f)+\sum x_i
\ee
from which we can compute
\be
K^2=8-b
\ee

There is an isomorphism $\bF_n^b\to\bF_{n+1}^b$ under special circumstances. The isomorphism is given by
\begin{align}
e-x_i&\to e \label{Ins}\\
f-x_i&\to x_i\\
x_i&\to f-x_i\\
x_j&\to x_j~~~~~~~~~~\text{for $j\neq i$}\label{Ine}
\end{align}
where a special blowup $x_i$ is singled out. We will denote this isomorphism by $\cI_n$ and its inverse by $\cI_n^{-1}$. Notice that $\cI_n$ for $n\ge1$ can only be performed when $x_i$ is a non-generic blowup hitting the locus of the $-n$ curve $e$ to produce $e-x_i$.

For ease of notation, we define a surface $\bF_{-n}^b$ with $n>0$ as a surface isomorphic to $\bF_n^b$ with the isomorphism given by
\begin{align}
e&\lra h\label{ni}\\
f&\lra f\\
h&\lra e\\
x_i&\lra x_i\label{nf}
\end{align}

One can manufacture a transverse intersection of two surfaces $S_1$ and $S_2$ by gluing a curve $C_1$ in $S_1$ to a curve $C_2$ in $S_2$. Such a gluing is allowed only if certain conditions are satisfied. We can compute the genus $g(C_1)$ of $C_1$ and the genus $g(C_2)$ of $C_2$ using the adjunction formula (\ref{adj}). The gluing is sensible only if
\be
g(C_1)=g(C_2)=g\ge0
\ee
Moreover, for a local neighborhood of the resulting intersecting configuration of $S_1$ and $S_2$ to be Calabi-Yau, the following condition should be satisfied
\be\label{CY}
C_1^2+C_2^2=2g-2
\ee
where $C_1^2$ and $C_2^2$ are calculated in $S_1$ and $S_2$ respectively. In principle, one can glue multiple gluing curves between two surfaces, but only a single gluing (and no gluing) will be relevant for geometries considered in this paper. 

In a similar fashion, one can construct an intersecting configuration of surfaces $S_i$ by gluing curves in different surfaces. Let the curve inside $S_i$ participating in the gluing between $S_i$ and $S_j$ be denoted by $C_{ij}$. We define $C_{ij}=0$ if there is no intersection between $S_i$ and $S_j$. For such a collection of gluings to be consistent, we need to make sure that it leads to consistent triple intersections of the surfaces. The triple intersection numbers can be computed as
\begin{align}
S_{i}^3&=K^2_{i}\\
S^2_{i}\cdot S_{j}&=K_{i}\cdot C_{ij}=C_{ji}^2\label{CC2}\\
S_{i}\cdot S_{j}\cdot S_{k}&=C_{ij}\cdot C_{ik}=C_{jk}\cdot C_{ji}=C_{ki}\cdot C_{kj}\label{C3}
\end{align}
where $K_i$ is the canonical class of $S_i$ and $i\neq j\neq k$. The conditions for a collection of gluings $C_{ij}$ to be consistent arise from the equality of different ways of computing the left hand side of (\ref{C3}). That is, we must have
\be
C_{ij}\cdot C_{ik}=C_{jk}\cdot C_{ji}=C_{ki}\cdot C_{kj}
\ee
for $i\neq j\neq k$. The equality of two different ways to compute the left hand side of (\ref{CC2}) do not lead to any new consistency conditions since it is equivalent to the Calabi-Yau condition (\ref{CY}).

Finally, we denote an intersecting configuration of surfaces in a graphical form. For example, the graph
\be
\begin{tikzpicture} [scale=1.9]
\node (v1) at (-4.9,-0.5) {$S_1$};
\node (v2) at (-2.9,-0.5) {$S_2$};
\node at (-4.5,-0.4) {\scriptsize{$C_{12}$}};
\node at (-3.3,-0.4) {\scriptsize{$C_{21}$}};
\draw  (v1) edge (v2);
\begin{scope}[shift={(2,0)}]
\node (v3) at (-2.9,-0.5) {$S_3$};
\node at (-3.3,-0.4) {\scriptsize{$C_{32}$}};
\end{scope}
\draw  (v2) edge (v3);
\node at (-2.5,-0.4) {\scriptsize{$C_{23}$}};
\begin{scope}[shift={(4,0)}]
\node (v4) at (-2.9,-0.5) {$S_4$};
\node at (-3.3,-0.4) {\scriptsize{$C_{43}$}};
\end{scope}
\draw  (v4) edge (v3);
\node at (-0.5,-0.4) {\scriptsize{$C_{34}$}};
\draw (v2) .. controls (-2.7,-1.1) and (0.8,-1.1) .. (v4);
\node at (-2.8,-0.9) {\scriptsize{$C_{24}$}};
\node at (1,-0.9) {\scriptsize{$C_{42}$}};
\end{tikzpicture}
\ee
denotes a configuration of four surfaces such that there are no intersections between $S_1$ and $S_3$, and $S_1$ and $S_4$. An edge between two surfaces denotes that there is an intersection between the two surfaces. The gluing curve $C_{ij}$ in $S_{i}$ for the gluing to $S_j$ is denoted near $S_i$ at the end of the edge between $S_i$ and $S_j$.

Another intersection number inside a threefold is that of a curve $C$ and a surface $S_i$. If $C$ lies in $S_{i}$, then the intersection is computed via
\be\label{SC1}
C\cdot S_i=K_{i}\cdot C
\ee
If $C$ lies in some other surface $S_{j}$, then the intersection is computed via
\be\label{SC2}
C\cdot S_i= C_{ji}\cdot C
\ee

\subsection{Intersection matrix associated to a geometry}
Let us now associate a matrix $M_{ij}$ to every geometry composed out of Hirzebruch surfaces, which we dub as the \emph{intersection matrix} of that geometry. It is defined as
\be\label{IM}
M_{ij}=-f_i\cdot S_j
\ee
where $f_i$ is the fiber of the Hirzebruch surface $S_i$.

A diagonal entry of this matrix is
\be
M_{ii}=-K_i\cdot f_i =2
\ee
just like the diagonal entry of a Cartan matrix. We can write the gluing curve $C_{ij}$ as 
\be
C_{ij}=\alpha_{ij} e_i+\beta_{ij} f_i + \sum_a \gamma_{ij,a}x_a
\ee
where $e_i,f_i$ are the $e,f$ curves of the Hirzebruch surface $S_i$. Then the off-diagonal entry of the intersection matrix is
\be
M_{ij}=-\alpha_{ij}
\ee
which is non-positive just like an off-diagonal entry of a Cartan matrix. 

If a geometry is such that the off-diagonal entries of its intersection matrix are such that $M_{ij}$ can be identified with the Cartan matrix of a semi-simple Lie algebra $\fg$, then M-theory compactified on the geometry produces a supersymmetric $5d$ gauge theory with gauge algebra $\fg$ on its Coulomb branch. M2 branes compactified on the fibers $f_i$ of $S_i$ lead to W-bosons and thus $f_i$ correspond to roots of $\fg$. M5 branes wrapping $S_i$ give rise to monopole strings and hence $S_i$ are associated to co-roots of $\fg$. The pairing of (\ref{IM}) descends to the pairing of roots and co-roots and hence is indeed the Cartan matrix. The matter content of the gauge theory is encoded in the blowups on Hirzebruch surfaces and will be discussed later.

We can see that for a geometry to give rise to a gauge theory, if $\alpha_{ij}=0$ then $\alpha_{ji}$ must be zero. On the other hand, if $\alpha_{ij}\neq0$ then $\alpha_{ji}$ must be non-zero as well. Moreover, $\alpha_{ij}$ must be bounded above by three.

\subsection{Prepotential associated to a geometry}
The 3-form gauge field of M-theory reduces on each surface $S_i$ to give rise to a $\u(1)$ gauge field in $5d$. Thus, every geometry leads to an abelian gauge theory description in the far infrared irrespective of whether or not it can be described as the Coulomb branch of a non-abelian gauge theory. The prepotential $\cF$ of this abelian gauge theory is encoded in the triple intersection numbers of the surfaces $S_i$ as we now describe. Let $\phi_i$ be the scalar in the $\u(1)$ gauge multiplet arising from $S_i$. The prepotential is a cubic polynomial in $\phi_i$ if we ignore\footnote{We note that ``ignoring'' is not the same as setting the mass parameters to zero. It simply means that we truncate the prepotential to terms involving only $\phi_i$. It is possible that the phase under discussion does not exist at zero mass parameters.} the terms involving mass parameters. Let $c_{ijk}$ be the coefficient of term $\phi_i\phi_j\phi_k$ in $6\cF$. Then
\begin{align}
c_{iii}&=S_i^3\\
c_{iij}&=3S_i^2\cdot S_j\\
c_{ijk}&=6S_i\cdot S_j\cdot S_k
\end{align}
where $i,j,k$ are three distinct indices.

If the geometry admits a gauge theory interpretation, we can also compute $6\cF$ using a one-loop calculation
\be \label{PP}
6\cF=\sum_a k_ad_a^{ijk}\phi_{i,a}\phi_{j,a}\phi_{k,a}+\frac{1}{2}\left(\sum_{r}|r \cdot \phi|^3- \sum_f h_f \sum_{w(\cR_f)}|w(\cR_f) \cdot \phi + (1-2h_f) m_f|^3\right)
\ee
where $r$ are the roots of the gauge algebra $\fg$, $w(\cR_f)$ parametrize weights of an irrep $\cR_f$ of $\fg$ formed by the charged hypermultiplets, $m_f\in\R$ is a mass term for each full\footnote{Half-hypermultiplets do not admit mass paramters unless completed into a full hypermultiplet.} hypermultiplet $f$, and $h_f=1$ for a full hyper $f$ and $h_f=\half$ for a half hyper $f$. The notation $w\cdot\phi$ denotes the scalar product of the Dynkin coefficients of the weight $w$ with Coulomb branch parameters. The indices $a$ denote various $\su(n\ge3)$ subfactors of the gauge algebra $\fg$, $\phi_{i,a}$ denote the Coulomb branch parameters for the $a$-th $\su(n)$ subfactor, $d_a^{ijk}$ denotes the corresponding rank three invariant tensor, and $k_a$ denotes the corresponding CS level.

\subsection{Pure gauge theories}
Using the above relationship between prepotential and triple intersection numbers, along with the relationship of intersection matrix to the Cartan matrix, we can determine the geometries corresponding to pure gauge theories. They are composed of Hirzebruch surfaces without any blowups. We collect them below.

\medskip

\begin{center}
\ubf{$\su(n)_k$, $n\ge3$, $2-n<k<n-2$}:
\end{center}
\be\label{sun}
\begin{tikzpicture} [scale=1.9]
\node (v1) at (-4.9,-0.5) {$\bF_{n-2-k}$};
\node (v2) at (-3.1,-0.5) {$\bF_{n-4-k}$};
\node (v3) at (-1.3,-0.5) {$\bF_{n-6-k}$};
\node at (-4.4,-0.4) {\scriptsize{$e$}};
\node at (-3.6,-0.4) {\scriptsize{$h$}};
\draw  (v1) edge (v2);
\draw  (v2) edge (v3);
\begin{scope}[shift={(1.8,0)}]
\node at (-4.4,-0.4) {\scriptsize{$e$}};
\node at (-3.6,-0.4) {\scriptsize{$h$}};
\end{scope}
\node (v4) at (-0.1,-0.5) {$\cdots$};
\draw  (v3) edge (v4);
\node (v5) at (1.1,-0.5) {$\bF_{2-n-k}$};
\draw  (v4) edge (v5);
\begin{scope}[shift={(3.6,0)}]
\node at (-4.4,-0.4) {\scriptsize{$e$}};
\node at (-3,-0.4) {\scriptsize{$h$}};
\end{scope}
\end{tikzpicture}
\ee
We remind the reader that we are also using Hirzebruch surfaces of negative degree\footnote{Hirzebruch surface $\bF^b_n$ is said to have degree $n$.} which are defined in terms of Hirzebruch surfaces of positive degree via (\ref{ni}--\ref{nf}). Each surface corresponds to a node of the Dynkin diagram of $\su(n)$. One can see that interchanging the left and right sides of the above diagram sends $k$ to $-k$. Thus, $\su(n)_k$ is the same as $\su(n)_{-k}$ upto outer automorphism which is a symmetry of the pure gauge theory.

\medskip

\begin{center}
\ubf{$\su(n)_k$, $n\ge3$, $k=2-n-2m$, $m\ge0$}:
\end{center}
\be\label{sun0}
\begin{tikzpicture} [scale=1.9]
\node (v1) at (-4.9,-0.5) {$\bF_{n-2-k}$};
\node (v2) at (-3.1,-0.5) {$\bF_{n-4-k}$};
\node (v3) at (-1.8,-0.5) {$\cdots$};
\node at (-4.4,-0.4) {\scriptsize{$e$}};
\node at (-3.6,-0.4) {\scriptsize{$h$}};
\draw  (v1) edge (v2);
\draw  (v2) edge (v3);
\begin{scope}[shift={(1.8,0)}]
\node at (-4.4,-0.4) {\scriptsize{$e$}};
\node at (-3,-0.4) {\scriptsize{$h$}};
\end{scope}
\node (v4) at (-0.7,-0.5) {$\bF_{4-n-k}$};
\draw  (v3) edge (v4);
\node (v5) at (1.1,-0.5) {$\bF_{0}$};
\draw  (v4) edge (v5);
\begin{scope}[shift={(3.6,0)}]
\node at (-3.8,-0.4) {\scriptsize{$e$}};
\node at (-2.9,-0.4) {\scriptsize{$e$+$mf$}};
\end{scope}
\end{tikzpicture}
\ee

\medskip

\begin{center}
\ubf{$\su(n)_k$, $n\ge3$, $k=1-n-2m$, $m\ge0$}:
\end{center}
\be\label{sun1}
\begin{tikzpicture} [scale=1.9]
\node (v1) at (-4.9,-0.5) {$\bF_{n-2-k}$};
\node (v2) at (-3.1,-0.5) {$\bF_{n-4-k}$};
\node (v3) at (-1.8,-0.5) {$\cdots$};
\node at (-4.4,-0.4) {\scriptsize{$e$}};
\node at (-3.6,-0.4) {\scriptsize{$h$}};
\draw  (v1) edge (v2);
\draw  (v2) edge (v3);
\begin{scope}[shift={(1.8,0)}]
\node at (-4.4,-0.4) {\scriptsize{$e$}};
\node at (-3,-0.4) {\scriptsize{$h$}};
\end{scope}
\node (v4) at (-0.7,-0.5) {$\bF_{4-n-k}$};
\draw  (v3) edge (v4);
\node (v5) at (1.1,-0.5) {$\bF_{1}$};
\draw  (v4) edge (v5);
\begin{scope}[shift={(3.6,0)}]
\node at (-3.8,-0.4) {\scriptsize{$e$}};
\node at (-2.9,-0.4) {\scriptsize{$h$+$mf$}};
\end{scope}
\end{tikzpicture}
\ee
The geometries for pure $\su(n)_k$ with $k\ge n-2$ can be obtained by exchanging the left and right ends of the above two geometries for $k\le 2-n$.

\medskip

\begin{center}
\ubf{$\so(2n+1)$, $n\ge3$}:
\end{center}
\be
\begin{tikzpicture} [scale=1.9]
\node (v1) at (-2.9,-0.5) {$\bF_{2n-5}$};
\node (v3) at (-1.3,-0.5) {$\bF_{2n-7}$};
\node at (-2.5,-0.4) {\scriptsize{$e$}};
\node at (-1.7,-0.4) {\scriptsize{$h$}};
\node (v4) at (-0.1,-0.5) {$\cdots$};
\draw  (v3) edge (v4);
\node (v5) at (0.9,-0.5) {$\bF_{1}$};
\draw  (v4) edge (v5);
\begin{scope}[shift={(3.6,0)}]
\node at (-4.5,-0.4) {\scriptsize{$e$}};
\node at (-3,-0.4) {\scriptsize{$h$}};
\end{scope}
\draw  (v1) edge (v3);
\node (v2) at (2.2,-0.5) {$\bF_1$};
\draw  (v5) edge (v2);
\node at (1.2,-0.4) {\scriptsize{$e$}};
\node at (1.9,-0.4) {\scriptsize{$e$}};
\node (v6) at (3.5,-0.5) {$\bF_6$};
\draw  (v2) edge (v6);
\node at (2.5,-0.4) {\scriptsize{$2h$}};
\node at (3.2,-0.4) {\scriptsize{$e$}};
\end{tikzpicture}
\ee
The surfaces are in one-to-one correspondence with the nodes of the Dynkin diagram of $\so(2n+1)$ in such a way that the fundamental representation of $\so(2n+1)$ corresponds to the left-most node and the spinor representation of $\so(2n+1)$ corresponds to the rightmost node. 

In all the other cases that follow, the graph will always be oriented such that the left-most node would correspond to the fundamental representation of the corresponding Lie algebra.

\medskip

\begin{center}
\ubf{$\sp(n)_{\theta}$, $n\ge2$, $\theta=n\pi~(\text{\normalfont{mod}}~2\pi)$}:
\end{center}
\be
\begin{tikzpicture} [scale=1.9]
\node (v1) at (-2.9,-0.5) {$\bF_{2n+2}$};
\node (v3) at (-1.3,-0.5) {$\bF_{2n}$};
\node at (-2.5,-0.4) {\scriptsize{$e$}};
\node at (-1.6,-0.4) {\scriptsize{$h$}};
\node (v4) at (-0.2,-0.5) {$\cdots$};
\draw  (v3) edge (v4);
\node (v5) at (0.9,-0.5) {$\bF_{8}$};
\draw  (v4) edge (v5);
\begin{scope}[shift={(3.6,0)}]
\node at (-4.6,-0.4) {\scriptsize{$e$}};
\node at (-3,-0.4) {\scriptsize{$h$}};
\end{scope}
\draw  (v1) edge (v3);
\node (v2) at (2.2,-0.5) {$\bF_6$};
\draw  (v5) edge (v2);
\node at (1.2,-0.4) {\scriptsize{$e$}};
\node at (1.9,-0.4) {\scriptsize{$h$}};
\node (v6) at (3.5,-0.5) {$\bF_1$};
\draw  (v2) edge (v6);
\node at (2.5,-0.4) {\scriptsize{$e$}};
\node at (3.2,-0.4) {\scriptsize{$2h$}};
\end{tikzpicture}
\ee

\medskip

\begin{center}
\ubf{$\sp(n)_{\theta}$, $n\ge2$, $\theta=(n+1)\pi~(\text{\normalfont{mod}}~2\pi)$}:
\end{center}
\be\label{spd}
\begin{tikzpicture} [scale=1.9]
\node (v1) at (-2.9,-0.5) {$\bF_{2n+2}$};
\node (v3) at (-1.3,-0.5) {$\bF_{2n}$};
\node at (-2.5,-0.4) {\scriptsize{$e$}};
\node at (-1.6,-0.4) {\scriptsize{$h$}};
\node (v4) at (-0.2,-0.5) {$\cdots$};
\draw  (v3) edge (v4);
\node (v5) at (0.9,-0.5) {$\bF_{8}$};
\draw  (v4) edge (v5);
\begin{scope}[shift={(3.6,0)}]
\node at (-4.6,-0.4) {\scriptsize{$e$}};
\node at (-3,-0.4) {\scriptsize{$h$}};
\end{scope}
\draw  (v1) edge (v3);
\node (v2) at (2.2,-0.5) {$\bF_6$};
\draw  (v5) edge (v2);
\node at (1.2,-0.4) {\scriptsize{$e$}};
\node at (1.9,-0.4) {\scriptsize{$h$}};
\node (v6) at (3.9,-0.5) {$\bF_0$};
\draw  (v2) edge (v6);
\node at (2.5,-0.4) {\scriptsize{$e$}};
\node at (3.5,-0.4) {\scriptsize{$2e$+$f$}};
\end{tikzpicture}
\ee
The relationship between the theta angle of $\sp(n)$ and the geometry cannot be determined based on whatever we have discussed so far. The relationship is derived in Appendix B.3 of \cite{Bhardwaj:2019fzv}.

\medskip

\begin{center}
\ubf{$\su(2)_0$}:
\end{center}
\be
\begin{tikzpicture} [scale=1.9]
\node (v6) at (3.8,-0.5) {$\bF_0$};
\end{tikzpicture}
\ee

\medskip

\begin{center}
\ubf{$\su(2)_\pi$}:
\end{center}
\be
\begin{tikzpicture} [scale=1.9]
\node (v6) at (3.8,-0.5) {$\bF_1$};
\end{tikzpicture}
\ee
For pure $\su(2)$ one finds $6\cF=8\phi^3$ which implies that there is a single Hirzebruch surface with $K^2=8$ implying that the Hirzebruch surface has no blowups. But, it does not fix the degree of Hirzebruch surface. The degree can be fixed by taking the limit $n=1$ of the results for $\sp(n)$.

\medskip

\begin{center}
\ubf{$\so(2n)$, $n\ge4$}:
\end{center}
\be
\begin{tikzpicture} [scale=1.9]
\node (v1) at (-2.9,-0.5) {$\bF_{2n-6}$};
\node (v3) at (-1.3,-0.5) {$\bF_{2n-8}$};
\node at (-2.5,-0.4) {\scriptsize{$e$}};
\node at (-1.7,-0.4) {\scriptsize{$h$}};
\node (v4) at (-0.2,-0.5) {$\cdots$};
\draw  (v3) edge (v4);
\node (v5) at (0.9,-0.5) {$\bF_{2}$};
\draw  (v4) edge (v5);
\begin{scope}[shift={(3.6,0)}]
\node at (-4.5,-0.4) {\scriptsize{$e$}};
\node at (-3,-0.4) {\scriptsize{$h$}};
\end{scope}
\draw  (v1) edge (v3);
\node (v2) at (2.2,-0.5) {$\bF_0$};
\draw  (v5) edge (v2);
\node at (1.2,-0.4) {\scriptsize{$e$}};
\node at (1.9,-0.4) {\scriptsize{$e$}};
\node (v6) at (3.5,-0.5) {$\bF_2$};
\draw  (v2) edge (v6);
\node at (2.5,-0.4) {\scriptsize{$e$}};
\node at (3.2,-0.4) {\scriptsize{$e$}};
\node (v7) at (2.2,0.5) {$\bF_2$};
\draw  (v7) edge (v2);
\node at (2.1,-0.2) {\scriptsize{$e$}};
\node at (2.1,0.2) {\scriptsize{$e$}};
\end{tikzpicture}
\ee

\medskip

\begin{center}
\ubf{$\fe_6$}:
\end{center}
\be
\begin{tikzpicture} [scale=1.9]
\node at (0.6,-0.4) {\scriptsize{$h$}};
\node at (-0.1,-0.4) {\scriptsize{$e$}};
\node (v4) at (-0.4,-0.5) {$\bF_4$};
\node (v5) at (0.9,-0.5) {$\bF_{2}$};
\draw  (v4) edge (v5);
\node (v2) at (2.2,-0.5) {$\bF_0$};
\draw  (v5) edge (v2);
\node at (1.2,-0.4) {\scriptsize{$e$}};
\node at (1.9,-0.4) {\scriptsize{$e$}};
\node (v6) at (3.5,-0.5) {$\bF_2$};
\draw  (v2) edge (v6);
\node at (2.5,-0.4) {\scriptsize{$e$}};
\node at (3.2,-0.4) {\scriptsize{$e$}};
\node (v7) at (2.2,0.5) {$\bF_2$};
\draw  (v7) edge (v2);
\node at (2.1,-0.2) {\scriptsize{$e$}};
\node at (2.1,0.2) {\scriptsize{$e$}};
\node (v8) at (4.8,-0.5) {$\bF_4$};
\draw  (v6) edge (v8);
\node at (3.8,-0.4) {\scriptsize{$h$}};
\node at (4.5,-0.4) {\scriptsize{$e$}};
\end{tikzpicture}
\ee

\medskip

\begin{center}
\ubf{$\fe_7$}:
\end{center}
\be
\begin{tikzpicture} [scale=1.9]
\node at (0.6,-0.4) {\scriptsize{$h$}};
\node at (-0.1,-0.4) {\scriptsize{$e$}};
\node (v4) at (-0.4,-0.5) {$\bF_4$};
\node (v5) at (0.9,-0.5) {$\bF_{2}$};
\draw  (v4) edge (v5);
\node (v2) at (2.2,-0.5) {$\bF_0$};
\draw  (v5) edge (v2);
\node at (1.2,-0.4) {\scriptsize{$e$}};
\node at (1.9,-0.4) {\scriptsize{$e$}};
\node (v6) at (3.5,-0.5) {$\bF_2$};
\draw  (v2) edge (v6);
\node at (2.5,-0.4) {\scriptsize{$e$}};
\node at (3.2,-0.4) {\scriptsize{$e$}};
\node (v7) at (2.2,0.5) {$\bF_2$};
\draw  (v7) edge (v2);
\node at (2.1,-0.2) {\scriptsize{$e$}};
\node at (2.1,0.2) {\scriptsize{$e$}};
\node (v8) at (4.8,-0.5) {$\bF_4$};
\draw  (v6) edge (v8);
\node at (3.8,-0.4) {\scriptsize{$h$}};
\node at (4.5,-0.4) {\scriptsize{$e$}};
\node (v1) at (-1.7,-0.5) {$\bF_6$};
\draw  (v1) edge (v4);
\node at (-0.7,-0.4) {\scriptsize{$h$}};
\node at (-1.4,-0.4) {\scriptsize{$e$}};
\end{tikzpicture}
\ee

\medskip

\begin{center}
\ubf{$\fe_8$}:
\end{center}
\be
\begin{tikzpicture} [scale=1.9]
\node at (0.6,-0.4) {\scriptsize{$h$}};
\node at (-0.1,-0.4) {\scriptsize{$e$}};
\node (v4) at (-0.4,-0.5) {$\bF_4$};
\node (v5) at (0.9,-0.5) {$\bF_{2}$};
\draw  (v4) edge (v5);
\node (v2) at (2.2,-0.5) {$\bF_0$};
\draw  (v5) edge (v2);
\node at (1.2,-0.4) {\scriptsize{$e$}};
\node at (1.9,-0.4) {\scriptsize{$e$}};
\node (v6) at (3.5,-0.5) {$\bF_2$};
\draw  (v2) edge (v6);
\node at (2.5,-0.4) {\scriptsize{$e$}};
\node at (3.2,-0.4) {\scriptsize{$e$}};
\node (v7) at (2.2,0.5) {$\bF_2$};
\draw  (v7) edge (v2);
\node at (2.1,-0.2) {\scriptsize{$e$}};
\node at (2.1,0.2) {\scriptsize{$e$}};
\node (v8) at (4.8,-0.5) {$\bF_4$};
\draw  (v6) edge (v8);
\node at (3.8,-0.4) {\scriptsize{$h$}};
\node at (4.5,-0.4) {\scriptsize{$e$}};
\node (v1) at (-1.7,-0.5) {$\bF_6$};
\draw  (v1) edge (v4);
\node at (-0.7,-0.4) {\scriptsize{$h$}};
\node at (-1.4,-0.4) {\scriptsize{$e$}};
\node (v3) at (-3.1,-0.5) {$\bF_8$};
\draw  (v3) edge (v1);
\node at (-2,-0.4) {\scriptsize{$h$}};
\node at (-2.8,-0.4) {\scriptsize{$e$}};
\end{tikzpicture}
\ee

\medskip

\begin{center}
\ubf{$\ff_4$}:
\end{center}
\be
\begin{tikzpicture} [scale=1.9]
\node (v5) at (0.9,-0.5) {$\bF_{8}$};
\node (v2) at (2.2,-0.5) {$\bF_6$};
\draw  (v5) edge (v2);
\node at (1.2,-0.4) {\scriptsize{$e$}};
\node at (1.9,-0.4) {\scriptsize{$h$}};
\node (v6) at (3.5,-0.5) {$\bF_1$};
\draw  (v2) edge (v6);
\node at (2.5,-0.4) {\scriptsize{$e$}};
\node at (3.2,-0.4) {\scriptsize{$2h$}};
\node (v1) at (4.7,-0.5) {$\bF_1$};
\draw  (v6) edge (v1);
\node at (3.8,-0.4) {\scriptsize{$e$}};
\node at (4.4,-0.4) {\scriptsize{$e$}};
\end{tikzpicture}
\ee

\medskip

\begin{center}
\ubf{$\fg_2$}:
\end{center}
\be
\begin{tikzpicture} [scale=1.9]
\node (v2) at (2.2,-0.5) {$\bF_8$};
\node (v6) at (3.8,-0.5) {$\bF_0$};
\draw  (v2) edge (v6);
\node at (2.5,-0.4) {\scriptsize{$e$}};
\node at (3.4,-0.4) {\scriptsize{$3e$+$f$}};
\end{tikzpicture}
\ee

\subsection{Adding matter}\label{M}
Once we add matter to a gauge theory, the Coulomb branch is divided into different phases described by different prepotentials. These phases correspond to different possible signs generated by absolute values in (\ref{PP}). One can cycle through various signs by changing the values of mass parameters and Coulomb moduli with respect to each other. Suppose we are in a phase in which the quantity
\be\label{vv}
w(\cR_f) \cdot \phi + (1-2h_f) m_f
\ee
is positive for the weight $w(\cR_f)$. Then we can find a holomorphic curve $C_{w(\cR_f)}$ in the geometry that is composed out of the fibers and blowups such that the $i$-th Dynkin coefficient of $w(\cR_f)$ is given by
\be
-C_{w(\cR_f)}\cdot S_i
\ee
If the quantity (\ref{vv}) is negative for $w(\cR_f)$, then we can find a holomorphic curve $C_{w(\cR_f)}$ in the geometry that is composed out of the fibers and blowups such that the $i$-th Dynkin coefficient of $w(\cR_f)$ is given by
\be
C_{w(\cR_f)}\cdot S_i
\ee

A phase transition occurs whenever the sign of (\ref{vv}) for some particular weight $w(\cR_f)$ is flipped. Geometrically this corresponds to a flop of the curve $C_{w(\cR_f)}$. Said in the reverse direction, all the phases of the gauge theory can be generated by starting from the geometry corresponding to a particular phase and performing flop transitions of curves composed out of fibers and blowups\footnote{Flop transitions of curves involving the $e$ curve of some Hirzebruch surface lead to phases which are not described by the same gauge theory. Sometimes such phases cannot be described by any gauge theory and are referred to as non-gauge-theoretic phases. One such non-gauge-theoretic phase is described by the geometry which is a local neighborhood of $\P^2$.}.

We will focus our attention only on full hypermultiplets from this point on. By turning on a large negative mass for a full hypermultiplet $f$, we can reach a phase in which (\ref{vv}) is positive only for the highest weight and negative for all the other weights. We refer to this phase by saying that the hyper $f$ is \emph{marginally integrated in}. This is because making $m_f$ more negative leads to a final phase transition to the phase connected to $m_f=-\infty$ at which point the hyper $f$ is completely integrated out. The marginally integrated in phase has a very simple geometric construction. We start with the geometry for pure gauge theory. Now if the $i$-th Dynkin coefficient of the highest weight is $d_i$, then we perform $d_i$ number of blowups on the surface $S_i$. Finally, we glue together all the blowups pairwise with each other, so that all of them lead to a single curve $C_f$ in the full geometry. We claim that the curve associated to the highest weight $w^h_f$ is given by $C_f$. The lower weights can be written as
\be
w_f=w^h_f-n_i r_i
\ee
where $r_i$ are the roots of gauge algebra $\fg$. The curve corresponding to such a weight $w_f$ is
\be
-\left(C_f-n_i f_i\right)
\ee

As an example, consider marginally integrating in a fundamental to the pure $\su(3)_0$ theory. This gives rise to the theory $\su(3)_{\half}+\F$ since the CS level is changed by half-unit upon integrating in a fundamental. The geometry for the marginally integrated in phase of the gauge theory $\su(3)_{\half}+\F$ is obtained by simply adding a blowup onto the left-most node of the geometry (\ref{sun}) for $n=3$ and $k=0$. That is, the desired geometry is simply
\be\label{g1}
\begin{tikzpicture} [scale=1.9]
\node (v2) at (2.2,-0.5) {$\bF_1^1$};
\node (v6) at (3.8,-0.5) {$\bF_1$};
\draw  (v2) edge (v6);
\node at (2.5,-0.4) {\scriptsize{$e$}};
\node at (3.5,-0.4) {\scriptsize{$e$}};
\end{tikzpicture}
\ee
According to the above claim, the blowup $x$ is the curve corresponding to the highest weight given by Dynkin coefficients $(1,0)$. Indeed we can compute ($S_1$ is the left surface and $S_2$ is the right surface)
\be
-x\cdot S_1=-K_1\cdot x=1
\ee
and
\be
-x\cdot S_2=(x\cdot e)_{S_1}=0
\ee
where the notation $(x\cdot e)_{S_2}$ indicates that the intersection product of $x$ and $e$ is being taken in the surface $S_1$ and in particular the $e$ curve appearing in the intersection product is the $e$ curve of the Hirzebruch surface $S_1$. According to the above claim, the other weights $(-1,1)$ and $(0,-1)$ are associated to the curves $f_1-x$ and $f_1+f_2-x$ respectively. Indeed, we can check that
\begin{align}
(f_1-x)\cdot S_1&=K_1\cdot (f_1-x)=-1\\
(f_1-x)\cdot S_2&=(e\cdot (f-x))_{S_1}=1\\
(f_2+f_1-x)\cdot S_1&=(e\cdot f)_{S_2}+(K\cdot (f-x))_{S_1}=0\\
(f_2+f_1-x)\cdot S_2&=(K\cdot f)_{S_2}+(e\cdot (f-x))_{S_1}=-1
\end{align}

Other phases can now be accessed by doing flops. Suppose we want to now change the sign of (\ref{vv}) for the weight $(-1,1)$ from negative to positive. This corresponds to flop of the curve $f_1-x$ since that is the curve associated to the weight $(1,-1)$ before the phase transition. Notice that the self-intersection of $f_1-x$ in $S_1$ is $-1$ and the genus is zero, so indeed this curve can be flopped. To perform the flop, we first rewrite the geometry (\ref{g1}) by performing the isomorphism $\cI_0^{-1}$ (see equations (\ref{Ins}--\ref{Ine})) on the left surface $S_1$
\be
\begin{tikzpicture} [scale=1.9]
\node (v2) at (2.2,-0.5) {$\bF_0^1$};
\node (v6) at (3.8,-0.5) {$\bF_1$};
\draw  (v2) edge (v6);
\node at (2.5,-0.4) {\scriptsize{$e$-$x$}};
\node at (3.5,-0.4) {\scriptsize{$e$}};
\end{tikzpicture}
\ee
The curves corresponding to $(1,0),(-1,1),(0,-1)$ in this isomorphism frame are $f_1-x,x,f_2+x$ respectively. And the curve to be flopped has become the blowup $x$. In the first step of the flop, we blowdown $x$ changing $S_1$ from $\bF_0^1$ to $\bF_0$ and the gluing curve $C_{12}$ from $e-x$ to the curve $e$ of $\bF_0$. In the second step of the flop, we perform a blowup on $S_2$ thus changing it from $\bF_1$ to $\bF_1^1$. The blowup has to hit the gluing curve $C_{21}$ since the gluing curve $C_{12}$ was affected by the blowdown. Thus $C_{21}$ is changed from $e$ to $e-x$ and the geometry after flop is
\be
\begin{tikzpicture} [scale=1.9]
\node (v2) at (2.2,-0.5) {$\bF_0$};
\node (v6) at (3.8,-0.5) {$\bF^1_1$};
\draw  (v2) edge (v6);
\node at (2.5,-0.4) {\scriptsize{$e$}};
\node at (3.5,-0.4) {\scriptsize{$e$-$x$}};
\end{tikzpicture}
\ee
which can be rewritten as
\be
\begin{tikzpicture} [scale=1.9]
\node (v2) at (2.2,-0.5) {$\bF_0$};
\node (v6) at (3.8,-0.5) {$\bF^1_2$};
\draw  (v2) edge (v6);
\node at (2.5,-0.4) {\scriptsize{$e$}};
\node at (3.5,-0.4) {\scriptsize{$e$}};
\end{tikzpicture}
\ee
after performing the isomorphism $\cI_1$ on $S_2$. This geometry corresponds to the phase in which the quantity (\ref{vv}) is positive for the weights $(1,0)$ and $(-1,1)$, and negative for the weight $(0,-1)$. The curves corresponding to the three weights are $f_1+f_2-x,f_2-x,x$ respectively, as the reader can easily check.

As another example, consider marginally integrating in two fundamentals and an antisymmetric to the pure $\su(4)_1$ theory. The resulting gauge theory is $\su(4)_2+2\F+\Asym$ and the corresponding geometry is
\be
\begin{tikzpicture} [scale=1.9]
\node (v2) at (2.2,-0.5) {$\bF^2_1$};
\node (v6) at (3.8,-0.5) {$\bF^1_1$};
\draw  (v2) edge (v6);
\node at (2.5,-0.4) {\scriptsize{$e$}};
\node at (3.5,-0.4) {\scriptsize{$e$}};
\node (v1) at (5.4,-0.5) {$\bF_3$};
\draw  (v6) edge (v1);
\node at (4.1,-0.4) {\scriptsize{$h$}};
\node at (5.1,-0.4) {\scriptsize{$e$}};
\end{tikzpicture}
\ee

As a third example consider adding adjoint to $\su(4)_0$ to yield the gauge theory $\su(4)_0+\A$ for which the associated geometry is
\be\label{ex2}
\begin{tikzpicture} [scale=1.9]
\node (v2) at (2.2,-0.5) {$\bF^1_2$};
\node (v6) at (3.8,-0.5) {$\bF_0$};
\draw  (v2) edge (v6);
\node at (2.5,-0.4) {\scriptsize{$e$}};
\node at (3.5,-0.4) {\scriptsize{$e$}};
\node (v1) at (5.4,-0.5) {$\bF^1_2$};
\draw  (v6) edge (v1);
\node at (4.1,-0.4) {\scriptsize{$e$}};
\node at (5.1,-0.4) {\scriptsize{$e$}};
\draw (v2) .. controls (2.3,-1.1) and (5.3,-1.1) .. (v1);
\node at (2.1,-0.8) {\scriptsize{$x$}};
\node at (5.5,-0.8) {\scriptsize{$x$}};
\end{tikzpicture}
\ee
where $x$ near the left surface denotes the blowup in the left surface, and $x$ near the right surface denotes the blowup in the right surface.

As the last example, let us see how the theta angle becomes irrelevant when a fundamental is integrated into pure $\sp(2)$. Let us start from $\sp(2)_\pi$ and marginally integrate in a fundamental. The corresponding geometry is
\be
\begin{tikzpicture} [scale=1.9]
\node (v2) at (2.2,-0.5) {$\bF^1_6$};
\node (v6) at (3.9,-0.5) {$\bF_0$};
\draw  (v2) edge (v6);
\node at (2.5,-0.4) {\scriptsize{$e$}};
\node at (3.5,-0.4) {\scriptsize{$2e$+$f$}};
\end{tikzpicture}
\ee
Flopping the curve $f-x$ living in the left surface $S_1$ leads to
\be
\begin{tikzpicture} [scale=1.9]
\node (v2) at (2.2,-0.5) {$\bF_5$};
\node (v6) at (3.9,-0.5) {$\bF^1_0$};
\draw  (v2) edge (v6);
\node at (2.5,-0.4) {\scriptsize{$e$}};
\node at (3.4,-0.4) {\scriptsize{$2e$+$f$-$x$}};
\end{tikzpicture}
\ee
Now we perform $\cI_0$ on the right surface $S_2$ to rewrite the above geometry as
\be
\begin{tikzpicture} [scale=1.9]
\node (v2) at (2.2,-0.5) {$\bF_5$};
\node (v6) at (3.9,-0.5) {$\bF^1_1$};
\draw  (v2) edge (v6);
\node at (2.5,-0.4) {\scriptsize{$e$}};
\node at (3.5,-0.4) {\scriptsize{$2h$-$x$}};
\end{tikzpicture}
\ee
Flopping $x$ in $S_2$ leads to
\be
\begin{tikzpicture} [scale=1.9]
\node (v2) at (2.2,-0.5) {$\bF_6^1$};
\node (v6) at (3.9,-0.5) {$\bF^1_1$};
\draw  (v2) edge (v6);
\node at (2.5,-0.4) {\scriptsize{$e$}};
\node at (3.6,-0.4) {\scriptsize{$2h$}};
\end{tikzpicture}
\ee
which is precisely the geometry obtained when one marginally integrates in a fundamental into pure $\sp(2)_0$. Similarly, one can show that the marginally integrated in phases for $\sp(n)_0+\F$ and $\sp(n)_\pi+\F$ are flop equivalent to each other, and hence the theta angle is physically irrelevant.

Let us now discuss some subtleties about CS levels and theta angles when matter charged under a mixed representation of multiple simple gauge algebras is integrated in. For example, the geometry
\be\label{susu}
\begin{tikzpicture} [scale=1.9]
\node (v2) at (2.2,-0.5) {$\bF^1_1$};
\node (v6) at (3.8,-0.5) {$\bF_1$};
\draw  (v2) edge (v6);
\node at (2.5,-0.4) {\scriptsize{$e$}};
\node at (3.5,-0.4) {\scriptsize{$e$}};
\begin{scope}[shift={(0,-0.9)}]
\node (v2_1) at (2.2,-0.5) {$\bF^1_0$};
\node (v6_1) at (3.8,-0.5) {$\bF_2$};
\draw  (v2_1) edge (v6_1);
\node at (2.5,-0.4) {\scriptsize{$e$}};
\node at (3.5,-0.4) {\scriptsize{$e$}};
\end{scope}
\draw  (v2) edge (v2_1);
\node at (2.1,-0.8) {\scriptsize{$x$}};
\node at (2.1,-1.1) {\scriptsize{$x$}};
\end{tikzpicture}
\ee
describes the theory $\su(3)_\frac32\oplus\su(3)_{-\frac52}$ with a bifundamental marginally integrated in. To see this, we have to remember that a bifundamental is defined as $\F\otimes\bar\F$, so we should flip the lower $\su(3)$ and rewrite the above geometry as
\be
\begin{tikzpicture} [scale=1.9]
\node (v2) at (2.2,-0.5) {$\bF^1_1$};
\node (v6) at (3.8,-0.5) {$\bF_1$};
\draw  (v2) edge (v6);
\node at (2.5,-0.4) {\scriptsize{$e$}};
\node at (3.5,-0.4) {\scriptsize{$e$}};
\begin{scope}[shift={(0,-0.9)}]
\node (v2_1) at (2.2,-0.5) {$\bF_2$};
\node (v6_1) at (3.8,-0.5) {$\bF_0^1$};
\draw  (v2_1) edge (v6_1);
\node at (2.5,-0.6) {\scriptsize{$e$}};
\node at (3.5,-0.6) {\scriptsize{$e$}};
\end{scope}
\draw  (v2) edge (v6_1);
\node at (2.65,-0.65) {\scriptsize{$x$}};
\node at (3.55,-1.15) {\scriptsize{$x$}};
\end{tikzpicture}
\ee
Now we see that we have integrated in $\F\otimes\bar\F$ into pure $\su(3)_0\oplus\su(3)_{-1}$ theory. That is, we have integrated in $3\F$ into the $\su(3)_0$ subfactor and $3\bar\F$ into the $\su(3)_{-1}$ subfactor, thus yielding the theory $\su(3)_\frac32\oplus\su(3)_{-\frac52}$ with a bifundamental.

We could have also thought of the bifundamental as $\bar\F\otimes\F$, and then we would have flipped the upper $\su(3)$ in (\ref{susu}) to rewrite it as
\be
\begin{tikzpicture} [scale=1.9]
\node (v2) at (2.2,-0.5) {$\bF_1$};
\node (v6) at (3.8,-0.5) {$\bF^1_1$};
\draw  (v2) edge (v6);
\node at (2.5,-0.4) {\scriptsize{$e$}};
\node at (3.5,-0.4) {\scriptsize{$e$}};
\begin{scope}[shift={(0,-0.9)}]
\node (v2_1) at (2.2,-0.5) {$\bF^1_0$};
\node (v6_1) at (3.8,-0.5) {$\bF_2$};
\draw  (v2_1) edge (v6_1);
\node at (2.5,-0.6) {\scriptsize{$e$}};
\node at (3.5,-0.6) {\scriptsize{$e$}};
\end{scope}
\draw  (v6) edge (v2_1);
\node at (3.35,-0.65) {\scriptsize{$x$}};
\node at (2.4,-1.15) {\scriptsize{$x$}};
\end{tikzpicture}
\ee
which describes marginal integration of $\bar\F\otimes\F$ into pure $\su(3)_0\oplus\su(3)_1$ theory. In this way, we find that we can also describe the theory associated to the geometry (\ref{susu}) as $\su(3)_{-\frac32}\oplus\su(3)_{\frac52}$ with a bifundamental. That is, the overall sign of the CS levels is physically irrelevant. This is true for any general $\su(m)\oplus\su(n)$.

Now, let us consider the geometry
\be
\begin{tikzpicture} [scale=1.9]
\node (v2) at (2.2,-0.5) {$\bF^1_1$};
\node (v6) at (3.8,-0.5) {$\bF_1$};
\draw  (v2) edge (v6);
\node at (2.5,-0.4) {\scriptsize{$e$}};
\node at (3.5,-0.4) {\scriptsize{$e$}};
\begin{scope}[shift={(0,-0.9)}]
\node (v2_1) at (2.2,-0.5) {$\bF^1_0$};
\end{scope}
\draw  (v2) edge (v2_1);
\node at (2.1,-0.8) {\scriptsize{$x$}};
\node at (2.1,-1.1) {\scriptsize{$x$}};
\end{tikzpicture}
\ee
The way the geometry is written, we are treating the bifundamental as\footnote{Even though $\F\simeq\bar\F$ for $\su(2)$, it is conventional to distinguish between $\F$ and $\bar\F$ as described in what follows.} $\F\otimes\bar\F$. In other words, we are integrating $2\F$ into $\su(3)_{0}$ and $3\bar\F$ into $\su(2)_0$. It is conventional to assume that integrating in $\bar\F$ into $\su(2)$ does not change theta angle, while integrating in $\F$ changes the theta angle. In other words, bringing in a fundamental from negative infinity mass to zero mass changes theta angle but taking mass to positive infinity from zero does not. So, the gauge theory associated to the above geometry can be identified as $\su(3)_1\oplus\su(2)_0$ with a bifundamental.

If we treat the bifundamental as $\bar\F\otimes\F$, then we write the geometry as
\be
\begin{tikzpicture} [scale=1.9]
\node (v2) at (2.2,-0.5) {$\bF_1$};
\node (v6) at (3.8,-0.5) {$\bF^1_1$};
\draw  (v2) edge (v6);
\node at (2.5,-0.4) {\scriptsize{$e$}};
\node at (3.5,-0.4) {\scriptsize{$e$}};
\begin{scope}[shift={(0,-0.9)}]
\node (v2_1) at (2.2,-0.5) {$\bF^1_0$};
\end{scope}
\draw  (v6) edge (v2_1);
\node at (3.35,-0.65) {\scriptsize{$x$}};
\node at (2.45,-1.1) {\scriptsize{$x$}};
\end{tikzpicture}
\ee
Now we are integrating in $2\bar\F$ into $\su(3)_{0}$ and $3\F$ into $\su(2)_0$, and hence can identify the gauge theory as $\su(3)_{-1}\oplus\su(2)_\pi$ with a bifundamental. That is, simultaneously flipping the sign of CS level and the theta angle for $\su(3)\oplus\su(2)$ leaves the theory invariant. This is true for any general $\su(2m+1)\oplus\sp(n)$ with a bifundamental. For $\su(2m)\oplus\sp(n)$ with a bifundamental, changing the CS level of $\su(2m)$ leaves the theory invariant if the theta angle of $\sp(n)$ is left unchanged.

\subsection{S-duality}\label{S}
Consider the surface $\bF^b_0$. This surface has an automorphism $\bF^b_0\to\bF^b_0$ described by
\begin{align}
e&\to f\label{Ss}\\
f&\to e\\
x_{i}&\to x_i\label{Se}
\end{align}
which we claim implements S-duality of Type IIB in M-theory language, and hence we denote this automorphism by $\cS$. 

An M2 brane wrapping $f$ in $\bF_0^b$ give rise to a W-boson and an M2 brane wrapping the $e$ curve in $\bF_0^b$ gives rise to an instanton. Thus the automorphism $\cS$ interchanges a W-boson with an instanton, which means that it naturally implements a strong-weak coupling duality, suggesting that it should be related to S-duality of Type IIB superstring theory.

Indeed, we can see this relationship by considering the geometry described by a single surface $\bF_0$ which manufactures the pure $\su(2)_0$ theory. The dual brane web in Type IIB is
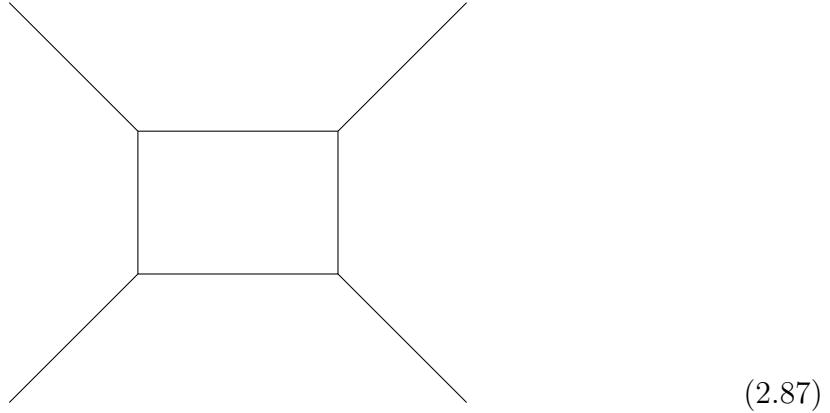
\be
\begin{tikzpicture} [scale=1.9]
\draw (-1.4,1.1) -- (-2.3,2) (0,1.1) -- (0.9,2) (-1.4,0.1) -- (-2.3,-0.8) (0,0.1) -- (0.9,-0.8);
\draw (-1.4,1.1) -- (-1.4,0.1) -- (0,0.1) -- (0,1.1);
\draw (0,1.1) -- (-1.4,1.1);
\end{tikzpicture}
\ee
where two horizontal compact branes are two different representatives of the $e$ curve and two vertical compact branes are two different representatives of the $f$ curve. Application of the automorphism $\cS$ on this geometry interchanges the $e$ and $f$ curves, and thus rotates the brane diagram by ninety degrees, which is indeed the same effect as the application of S-duality on the brane web.

Notice that the application of $\cS$ changes the intersection matrix associated to the geometry. Thus, by applying $\cS$ on some $S_i$ in the geometry, the intersection matrix can go from Cartan matrix for a gauge algebra $\fg$ to the Cartan matrix of a gauge algebra $\fh$. Under this process, the blowups on the geometry go from describing the matter content for $\fg$ to describing the matter content for $\fh$. Thus, we obtain a duality between a gauge theory with algebra $\fg$ and a gauge theory with gauge algebra $\fh$. 

For example, consider the geometry for $\sp(2)+\F+\Asym$
\be
\begin{tikzpicture} [scale=1.9]
\node (v2) at (2.2,-0.5) {$\bF^1_6$};
\node (v6) at (3.9,-0.5) {$\bF^1_0$};
\draw  (v2) edge (v6);
\node at (2.5,-0.4) {\scriptsize{$e$}};
\node at (3.5,-0.4) {\scriptsize{$2e$+$f$}};
\end{tikzpicture}
\ee
Applying $\cS$ on the right surface $S_2$ leads to the geometry
\be
\begin{tikzpicture} [scale=1.9]
\node (v2) at (2.2,-0.5) {$\bF^1_6$};
\node (v6) at (3.9,-0.5) {$\bF^1_0$};
\draw  (v2) edge (v6);
\node at (2.5,-0.4) {\scriptsize{$e$}};
\node at (3.5,-0.4) {\scriptsize{$e$+$2f$}};
\end{tikzpicture}
\ee
which can be recognized as marginally integrating in a fundamental and an antifundamental\footnote{By integrating in an antifundamental, we mean that we integrate in a fundamental by decreasing its associated mass parameter from positive infinity to a finite value. This has opposite effect on the CS level as integrating in a fundamental does, that is the CS level decreases by half when an antifundamental is integrated. Similar statements are true for other representations of $\su(n)$: Integrating in the complex conjugate $\bar\cR$ has opposite effect on the CS level when compared to the effect on CS level caused by integrating in $\cR$.} into pure $\su(3)_{-5}$ and thus can be recognized as the gauge theory $\su(3)_{-5}+2\F$. The CS levels are unchanged since the change caused by integrating in a fundamental is canceled by the change caused by integrating in an antifundamental. So, the application of $\cS$ has led to the following duality between two $5d$ gauge theories
\be\label{ex1}
\sp(2)+\F+\Asym=\su(3)_{-5}+2\F
\ee
where, throughout this paper, equality sign between two gauge theories will represent duality between the two gauge theories.

Another feature one can notice is that as soon as one finds a duality, one can add matter to both sides of the duality. For example, let us add $m\F$ and $n\Asym$ to the left hand side of the above duality. The corresponding geometry is
\be
\begin{tikzpicture} [scale=1.9]
\node (v2) at (2.2,-0.5) {$\bF^{1+m}_6$};
\node (v6) at (3.9,-0.5) {$\bF^{1+n}_0$};
\draw  (v2) edge (v6);
\node at (2.6,-0.4) {\scriptsize{$e$}};
\node at (3.4,-0.4) {\scriptsize{$2e$+$f$}};
\end{tikzpicture}
\ee
which is isomorphic to
\be
\begin{tikzpicture} [scale=1.9]
\node (v2) at (2.2,-0.5) {$\bF^{1+m}_6$};
\node (v6) at (3.9,-0.5) {$\bF^{1+n}_0$};
\draw  (v2) edge (v6);
\node at (2.6,-0.4) {\scriptsize{$e$}};
\node at (3.4,-0.4) {\scriptsize{$e$+$2f$}};
\end{tikzpicture}
\ee
that is we obtain the duality
\be
\sp(2)+(1+m)\F+(1+n)\Asym=\su(3)_{-5+\half(m-n)}+(2+m+n)\F
\ee
This is possible because the extra matter that we add simply sits there unaffected by the operations performed to obtain the duality. The mapping of matter from $\sp(2)$ to $\su(3)$ is
\begin{align}
\F&\to\F\\
\Asym&\to\bar\F
\end{align}
That is integrating in $\F$ of $\sp(2)$ is dual to integrating in $\F$ of $\su(3)$ and integrating in $\Asym$ of $\sp(2)$ is dual to integrating in $\bar\F$ of $\su(3)$.

The possibility of obtaining new dualities from old dualities by adding matter allows us to define the notion of an \emph{irreducible duality} which cannot be obtained from another duality by adding matter to it. For example, the irreducible duality responsible for the duality (\ref{ex1}) is
\be\label{exS}
\sp(2)_\pi=\su(3)_{-5}
\ee

\section{A special class of irreducible dualities}\label{ID}
The irreducible dualities can be explored systematically in a bottom-up fashion by using the geometric approach reviewed in the last section. We demonstrate this by systematically deducing a special class of irreducible dualities defined by the following requirements:
\ben
\item We require that the gauge theory on one side of the duality must carry a simple gauge algebra $\fg$. The gauge theory on the other side of the duality is allowed to carry any general semi-simple gauge algebra. 
\item Moreover, at the point when $\cS$ transformations are applied on the geometry, the geometry should take the form of the Dynkin diagram of the simple gauge algebra $\fg$ without any extra edges. For example, the geometry (\ref{ex2}) does not satisfy this requirement. Even though the geometry gives rise to $\su(3)$ gauge algebra in $5d$, the geometry has an extra edge between the leftmost and the rightmost nodes which does not appear in the Dynkin diagram of $\su(3)$. 
\item We also assume that the geometries on two sides of the irreducible duality are related only by a sequence of $\cS$ transformations on the surfaces, without any mixing with the $\cI_n$ transformations.
\item The matter appearing on both sides of an irreducible duality is assumed to be such that all the representations have highest weights with Dynkin indices bounded above by one. The reducible dualities built on top of an irreducible duality have no restrictions on the allowed matter content.
\een

Let us consider a geometry satisfying the above requirements and giving rise to a simple gauge algebra $\fg$. Suppose a local portion of the geometry looks as follows
\be\label{C1}
\begin{tikzpicture} [scale=1.9]
\node (v2) at (2.1,-0.5) {$\bF^{b_1}_0$};
\node (v6) at (4.3,-0.5) {$\bF^{b_2}_n$};
\draw  (v2) edge (v6);
\node at (2.6,-0.4) {\scriptsize{$e$+$\gamma_i x_i$}};
\node at (3.6,-0.4) {\scriptsize{$\alpha e$+$\beta f$+$\delta_a x_a$}};
\end{tikzpicture}
\ee
where $x_i$ are the blowups on the left surface and $x_a$ are the blowups on the right surface. Since the above geometry is a piece of geometry admitting a gauge theory description, $\alpha$ must be strictly positive. Now suppose that we perform $\cS$ on the left surface to obtain
\be
\begin{tikzpicture} [scale=1.9]
\node (v2) at (2.1,-0.5) {$\bF^{b_1}_0$};
\node (v6) at (4.3,-0.5) {$\bF^{b_2}_n$};
\draw  (v2) edge (v6);
\node at (2.6,-0.4) {\scriptsize{$f$+$\gamma_i x_i$}};
\node at (3.6,-0.4) {\scriptsize{$\alpha e$+$\beta f$+$\delta_a x_a$}};
\end{tikzpicture}
\ee
Since $\alpha>0$, we have lost the gauge theory interpretation. To restore it, we must perform $\cS$ on the right surface as well. This means that we must have $n=0$. After performing $\cS$ on the right surface we obtain
\be\label{C2}
\begin{tikzpicture} [scale=1.9]
\node (v2) at (2.1,-0.5) {$\bF^{b_1}_0$};
\node (v6) at (4.3,-0.5) {$\bF^{b_2}_n$};
\draw  (v2) edge (v6);
\node at (2.6,-0.4) {\scriptsize{$f$+$\gamma_i x_i$}};
\node at (3.6,-0.4) {\scriptsize{$\beta e$+$\alpha f$+$\delta_a x_a$}};
\end{tikzpicture}
\ee
Since we are looking for a duality between gauge theories, at this point we must be able to restore a gauge theory interpretation, thus forcing $\beta=0$. Moreover, $\alpha$ must be equal to one, since no curve of the form
\be
\alpha f+\delta_a x_a
\ee
has a non-negative genus\footnote{This is also the reason why we chose the gluing curve in the left surface of (\ref{C1}) to be $e+\gamma_i x_i$ rather than the more general $\mu e +\gamma_i x_i$ for $\mu\ge1$.} if $\alpha>1$. We can in fact completely fix the form of (\ref{C2}) according to the conditions that must be satisfied by gluing curves discussed in last section. Only the following three possibilities are allowed:
\bit
\item \be\label{ff1}
\begin{tikzpicture} [scale=1.9]
\node (v2) at (2.1,-0.5) {$\bF^{1+1+b_1}_0$};
\node (v6) at (4.3,-0.5) {$\bF^{b_2}_0$};
\draw  (v2) edge (v6);
\node at (2.7,-0.4) {\scriptsize{$f$-$x$-$y$}};
\node at (3.9,-0.4) {\scriptsize{$f$}};
\end{tikzpicture}
\ee
where $x$ and $y$ denote the first two blowups on the left surface. The rest of the $b_1$ blowups on the left surface do not participate in the gluing. Similarly, all the $b_2$ blowups on the right surface do not participate in the gluing.
\item \be\label{ff2}
\begin{tikzpicture} [scale=1.9]
\node (v2) at (2.1,-0.5) {$\bF^{b_1}_0$};
\node (v6) at (4.3,-0.5) {$\bF^{1+1+b_2}_0$};
\draw  (v2) edge (v6);
\node at (2.4,-0.4) {\scriptsize{$f$}};
\node at (3.7,-0.4) {\scriptsize{$f$-$x$-$y$}};
\end{tikzpicture}
\ee
which is just obtained by exchanging the left and right surfaces in the configuration displayed above.
\item The final allowed configuration is
\be\label{ff3}
\begin{tikzpicture} [scale=1.9]
\node (v2) at (2.1,-0.5) {$\bF^{1+b_1}_0$};
\node (v6) at (4.3,-0.5) {$\bF^{1+b_2}_0$};
\draw  (v2) edge (v6);
\node at (2.6,-0.4) {\scriptsize{$f$-$x$}};
\node at (3.8,-0.4) {\scriptsize{$f$-$x$}};
\end{tikzpicture}
\ee
where a single blowup $x$ on each surface participates in the gluing. The rest of the blowups do not participate in the gluing.
\eit

The moral of the above analysis is that if after performing an $\cS$ transformation on a surface $S_i$, a gluing curve $C_{ij}$ to another surface $S_j$ contains no $e$ curve, then we must perform an $\cS$ transformation on $S_j$ as well. After performing $\cS$ transformation on both $S_i$ and $S_j$ the geometry necessarily describes a gauge theory with semi-simple gauge algebra. In other words, such a combined $\cS$ transformation on $S_i$ and $S_j$ breaks the edge between the $i$-th and $j$-th node in the Dynkin diagram of $\fg$.

Thus each irreducible duality in the special class being studied in this paper is characterized by first choosing a connected set $\cP$ of nodes in the Dynkin diagram of a simple Lie algebra $\fg$. The edges between these nodes are deleted while going to the other side of the duality. For an edge connecting a node $i$ lying in $\cP$ to a node $j$ lying outside $\cP$, we have the following options:
\bit
\item If there is a single edge between $i$ and $j$, on the other side of the duality also we can have a single edge between $i$ and $j$. This is the case when $C_{ij}$ is of the form
\be
C_{ij}=e+f+\gamma_a x_a
\ee
before the duality. After the $\cS$ transformation on $S_i$, $e$ and $f$ are exchanged and $C_{ij}$ is left invariant. In terms of Dynkin diagrams, the transition can be represented as
\be
\begin{tikzpicture} [scale=1.9]
\draw  (-0.4,-0.2) ellipse (0.1 and 0.1);
\draw  (0.5,-0.2) ellipse (0.1 and 0.1);
\draw (-0.3,-0.2) -- (0.4,-0.2);
\node at (-0.4,-0.2) {\scriptsize{$i$}};
\node at (0.5,-0.2) {\scriptsize{$j$}};
\draw[red] (1,-0.2) -- (1.8,-0.2);
\draw[red] (1.75,-0.15) -- (1.8,-0.2) -- (1.75,-0.25);
\begin{scope}[shift={(2.7,0)}]
\draw  (-0.4,-0.2) ellipse (0.1 and 0.1);
\draw  (0.5,-0.2) ellipse (0.1 and 0.1);
\draw (-0.3,-0.2) -- (0.4,-0.2);
\node at (-0.4,-0.2) {\scriptsize{$i$}};
\node at (0.5,-0.2) {\scriptsize{$j$}};
\end{scope}
\end{tikzpicture}
\ee
\item If there is a single edge between $i$ and $j$, on the other side we can have two edges pointing from $i$ to $j$. That is, we can have the transition
\be
\begin{tikzpicture} [scale=1.9]
\draw  (-0.4,-0.2) ellipse (0.1 and 0.1);
\draw  (0.5,-0.2) ellipse (0.1 and 0.1);
\draw (-0.3,-0.2) -- (0.4,-0.2);
\node at (-0.4,-0.2) {\scriptsize{$i$}};
\node at (0.5,-0.2) {\scriptsize{$j$}};
\draw[red] (1,-0.2) -- (1.8,-0.2);
\draw[red] (1.75,-0.15) -- (1.8,-0.2) -- (1.75,-0.25);
\begin{scope}[shift={(2.7,0)}]
\draw  (-0.4,-0.2) ellipse (0.1 and 0.1);
\draw  (0.5,-0.2) ellipse (0.1 and 0.1);
\draw (-0.3,-0.175) -- (0.375,-0.175);
\begin{scope}[shift={(0,-0.05)}]
\draw (-0.3,-0.175) -- (0.375,-0.175);
\end{scope}
\node at (-0.4,-0.2) {\scriptsize{$i$}};
\node at (0.5,-0.2) {\scriptsize{$j$}};
\end{scope}
\draw (3.05,-0.15) -- (3.1,-0.2) -- (3.05,-0.25);
\end{tikzpicture}
\ee
indicating that the root corresponding to the $j$-th node becomes shorter after the duality. This is the case when $C_{ij}$ is of the form
\be
C_{ij}=e+2f+\gamma_a x_a
\ee
before the duality and becomes
\be
C_{ij}=2e+f+\gamma_a x_a
\ee
after the duality.
\item The last case involving a single node between $i$ and $j$ implements the following transition under the duality transformation
\be
\begin{tikzpicture} [scale=1.9]
\draw  (-0.4,-0.2) ellipse (0.1 and 0.1);
\draw  (0.5,-0.2) ellipse (0.1 and 0.1);
\draw (-0.3,-0.2) -- (0.4,-0.2);
\node at (-0.4,-0.2) {\scriptsize{$i$}};
\node at (0.5,-0.2) {\scriptsize{$j$}};
\draw[red] (1,-0.2) -- (1.8,-0.2);
\draw[red] (1.75,-0.15) -- (1.8,-0.2) -- (1.75,-0.25);
\begin{scope}[shift={(2.7,0)}]
\draw  (-0.4,-0.2) ellipse (0.1 and 0.1);
\draw  (0.5,-0.2) ellipse (0.1 and 0.1);
\draw (-0.3,-0.175) -- (0.375,-0.175);
\begin{scope}[shift={(0,-0.05)}]
\draw (-0.3,-0.175) -- (0.375,-0.175);
\end{scope}
\node at (-0.4,-0.2) {\scriptsize{$i$}};
\node at (0.5,-0.2) {\scriptsize{$j$}};
\end{scope}
\draw (3.05,-0.15) -- (3.1,-0.2) -- (3.05,-0.25);
\draw (2.4,-0.2) -- (3.1,-0.2);
\end{tikzpicture}
\ee
This is the case when $C_{ij}$ is of the form
\be
C_{ij}=e+3f+\gamma_a x_a
\ee
before the duality.

However, this transition is possible only if $\cP$ contains the single node $i$. Otherwise, there is another node $k$ in $\cP$ joined to $i$, and after the duality the gluing curve $C_{ik}$ must take one of the three possibilities $f-x-y$, $f$, $f-x$ as displayed in (\ref{ff1}), (\ref{ff2}), (\ref{ff3}) respectively. $C_ij$ after duality takes the form
\be
C_{ij}=3e+f+\gamma_a x_a
\ee
One can see that it is not possible to ensure $C_{ij}\cdot C_{ik}=0$ while keeping the genus of $C_{ij}$ non-negative. A non-zero $C_{ij}\cdot C_{ik}$ would imply that there must be an edge between the $j$-th and $k$-th node in the Dynkin diagram of $\fg$, but this is a contradiction since Dynkin diagrams for finite Lie algebras do not contain any loops.
\item The possibilities with two edges are
\be
\begin{tikzpicture} [scale=1.9]
\draw[red] (1,-0.2) -- (1.8,-0.2);
\draw[red] (1.75,-0.15) -- (1.8,-0.2) -- (1.75,-0.25);
\begin{scope}[shift={(2.7,0)}]
\draw  (-0.4,-0.2) ellipse (0.1 and 0.1);
\draw  (0.5,-0.2) ellipse (0.1 and 0.1);
\draw (-0.3,-0.2) -- (0.4,-0.2);
\node at (-0.4,-0.2) {\scriptsize{$i$}};
\node at (0.5,-0.2) {\scriptsize{$j$}};
\end{scope}
\begin{scope}[shift={(-2.7,0)}]
\begin{scope}[shift={(2.7,0)}]
\draw  (-0.4,-0.2) ellipse (0.1 and 0.1);
\draw  (0.5,-0.2) ellipse (0.1 and 0.1);
\draw (-0.3,-0.175) -- (0.375,-0.175);
\begin{scope}[shift={(0,-0.05)}]
\draw (-0.3,-0.175) -- (0.375,-0.175);
\end{scope}
\node at (-0.4,-0.2) {\scriptsize{$i$}};
\node at (0.5,-0.2) {\scriptsize{$j$}};
\end{scope}
\draw (3.05,-0.15) -- (3.1,-0.2) -- (3.05,-0.25);
\end{scope}
\end{tikzpicture}
\ee
corresponding to
\be
C_{ij}=2e+f+\gamma_a x_a
\ee
before the duality;
\be
\begin{tikzpicture} [scale=1.9]
\draw[red] (1,-0.2) -- (1.8,-0.2);
\draw[red] (1.75,-0.15) -- (1.8,-0.2) -- (1.75,-0.25);
\begin{scope}[shift={(2.7,0)}]
\draw  (-0.4,-0.2) ellipse (0.1 and 0.1);
\draw  (0.5,-0.2) ellipse (0.1 and 0.1);
\draw (-0.3,-0.175) -- (0.375,-0.175);
\begin{scope}[shift={(0,-0.05)}]
\draw (-0.3,-0.175) -- (0.375,-0.175);
\end{scope}
\node at (-0.4,-0.2) {\scriptsize{$i$}};
\node at (0.5,-0.2) {\scriptsize{$j$}};
\end{scope}
\draw (3.05,-0.15) -- (3.1,-0.2) -- (3.05,-0.25);
\begin{scope}[shift={(-2.7,0)}]
\begin{scope}[shift={(2.7,0)}]
\draw  (-0.4,-0.2) ellipse (0.1 and 0.1);
\draw  (0.5,-0.2) ellipse (0.1 and 0.1);
\draw (-0.3,-0.175) -- (0.375,-0.175);
\begin{scope}[shift={(0,-0.05)}]
\draw (-0.3,-0.175) -- (0.375,-0.175);
\end{scope}
\node at (-0.4,-0.2) {\scriptsize{$i$}};
\node at (0.5,-0.2) {\scriptsize{$j$}};
\end{scope}
\draw (3.05,-0.15) -- (3.1,-0.2) -- (3.05,-0.25);
\end{scope}
\end{tikzpicture}
\ee
corresponding to
\be
C_{ij}=2e+2f+\gamma_a x_a
\ee
before the duality;
\be
\begin{tikzpicture} [scale=1.9]
\draw[red] (1,-0.2) -- (1.8,-0.2);
\draw[red] (1.75,-0.15) -- (1.8,-0.2) -- (1.75,-0.25);
\begin{scope}[shift={(2.7,0)}]
\draw  (-0.4,-0.2) ellipse (0.1 and 0.1);
\draw  (0.5,-0.2) ellipse (0.1 and 0.1);
\draw (-0.3,-0.175) -- (0.375,-0.175);
\begin{scope}[shift={(0,-0.05)}]
\draw (-0.3,-0.175) -- (0.375,-0.175);
\end{scope}
\node at (-0.4,-0.2) {\scriptsize{$i$}};
\node at (0.5,-0.2) {\scriptsize{$j$}};
\end{scope}
\draw (3.05,-0.15) -- (3.1,-0.2) -- (3.05,-0.25);
\draw (2.4,-0.2) -- (3.1,-0.2);
\begin{scope}[shift={(-2.7,0)}]
\begin{scope}[shift={(2.7,0)}]
\draw  (-0.4,-0.2) ellipse (0.1 and 0.1);
\draw  (0.5,-0.2) ellipse (0.1 and 0.1);
\draw (-0.3,-0.175) -- (0.375,-0.175);
\begin{scope}[shift={(0,-0.05)}]
\draw (-0.3,-0.175) -- (0.375,-0.175);
\end{scope}
\node at (-0.4,-0.2) {\scriptsize{$i$}};
\node at (0.5,-0.2) {\scriptsize{$j$}};
\end{scope}
\draw (3.05,-0.15) -- (3.1,-0.2) -- (3.05,-0.25);
\end{scope}
\end{tikzpicture}
\ee
corresponding to
\be
C_{ij}=2e+3f+\gamma_a x_a
\ee
before the duality; and
\be
\begin{tikzpicture} [scale=1.9]
\draw[red] (1,-0.2) -- (1.8,-0.2);
\draw[red] (1.75,-0.15) -- (1.8,-0.2) -- (1.75,-0.25);
\begin{scope}[shift={(-2.7,0)}]
\begin{scope}[shift={(2.7,0)}]
\draw  (-0.4,-0.2) ellipse (0.1 and 0.1);
\draw  (0.5,-0.2) ellipse (0.1 and 0.1);
\draw (-0.275,-0.175) -- (0.4,-0.175);
\begin{scope}[shift={(0,-0.05)}]
\draw (-0.275,-0.175) -- (0.4,-0.175);
\end{scope}
\node at (-0.4,-0.2) {\scriptsize{$i$}};
\node at (0.5,-0.2) {\scriptsize{$j$}};
\end{scope}
\draw (2.45,-0.15) -- (2.4,-0.2) -- (2.45,-0.25);
\end{scope}
\begin{scope}[shift={(-0.025,0)}]
\begin{scope}[shift={(2.7,0)}]
\draw  (-0.4,-0.2) ellipse (0.1 and 0.1);
\draw  (0.5,-0.2) ellipse (0.1 and 0.1);
\draw (-0.275,-0.175) -- (0.4,-0.175);
\begin{scope}[shift={(0,-0.05)}]
\draw (-0.275,-0.175) -- (0.4,-0.175);
\end{scope}
\node at (-0.4,-0.2) {\scriptsize{$i$}};
\node at (0.5,-0.2) {\scriptsize{$j$}};
\end{scope}
\draw (2.45,-0.15) -- (2.4,-0.2) -- (2.45,-0.25);
\end{scope}
\end{tikzpicture}
\ee
corresponding to
\be
C_{ij}=e+f+\gamma_a x_a
\ee
before the duality.
\item The possibilities with three edges similarly are
\be
\begin{tikzpicture} [scale=1.9]
\begin{scope}[shift={(2.9,0)}]
\draw  (-0.4,-0.2) ellipse (0.1 and 0.1);
\draw  (0.5,-0.2) ellipse (0.1 and 0.1);
\draw (-0.3,-0.2) -- (0.4,-0.2);
\node at (-0.4,-0.2) {\scriptsize{$i$}};
\node at (0.5,-0.2) {\scriptsize{$j$}};
\end{scope}
\draw[red] (1,-0.2) -- (1.8,-0.2);
\draw[red] (1.75,-0.15) -- (1.8,-0.2) -- (1.75,-0.25);
\begin{scope}[shift={(-0.1,0)}]
\draw  (-0.4,-0.2) ellipse (0.1 and 0.1);
\draw  (0.5,-0.2) ellipse (0.1 and 0.1);
\draw (-0.3,-0.175) -- (0.375,-0.175);
\begin{scope}[shift={(0,-0.05)}]
\draw (-0.3,-0.175) -- (0.375,-0.175);
\end{scope}
\node at (-0.4,-0.2) {\scriptsize{$i$}};
\node at (0.5,-0.2) {\scriptsize{$j$}};
\end{scope}
\draw (0.25,-0.15) -- (0.3,-0.2) -- (0.25,-0.25);
\draw (-0.4,-0.2) -- (0.3,-0.2);
\end{tikzpicture}
\ee
corresponding to
\be
C_{ij}=3e+f+\gamma_a x_a
\ee
before the duality;
\be
\begin{tikzpicture} [scale=1.9]
\draw[red] (1,-0.2) -- (1.8,-0.2);
\draw[red] (1.75,-0.15) -- (1.8,-0.2) -- (1.75,-0.25);
\begin{scope}[shift={(0,0)}]
\draw  (-0.4,-0.2) ellipse (0.1 and 0.1);
\draw  (0.5,-0.2) ellipse (0.1 and 0.1);
\draw (-0.3,-0.175) -- (0.375,-0.175);
\begin{scope}[shift={(0,-0.05)}]
\draw (-0.3,-0.175) -- (0.375,-0.175);
\end{scope}
\node at (-0.4,-0.2) {\scriptsize{$i$}};
\node at (0.5,-0.2) {\scriptsize{$j$}};
\end{scope}
\draw (0.35,-0.15) -- (0.4,-0.2) -- (0.35,-0.25);
\draw (-0.3,-0.2) -- (0.4,-0.2);
\begin{scope}[shift={(0.1,0)}]
\begin{scope}[shift={(2.7,0)}]
\draw  (-0.4,-0.2) ellipse (0.1 and 0.1);
\draw  (0.5,-0.2) ellipse (0.1 and 0.1);
\draw (-0.3,-0.175) -- (0.375,-0.175);
\begin{scope}[shift={(0,-0.05)}]
\draw (-0.3,-0.175) -- (0.375,-0.175);
\end{scope}
\node at (-0.4,-0.2) {\scriptsize{$i$}};
\node at (0.5,-0.2) {\scriptsize{$j$}};
\end{scope}
\draw (3.05,-0.15) -- (3.1,-0.2) -- (3.05,-0.25);
\end{scope}
\end{tikzpicture}
\ee
corresponding to
\be
C_{ij}=3e+2f+\gamma_a x_a
\ee
before the duality;
\be
\begin{tikzpicture} [scale=1.9]
\draw[red] (1,-0.2) -- (1.8,-0.2);
\draw[red] (1.75,-0.15) -- (1.8,-0.2) -- (1.75,-0.25);
\begin{scope}[shift={(0,0)}]
\draw  (-0.4,-0.2) ellipse (0.1 and 0.1);
\draw  (0.5,-0.2) ellipse (0.1 and 0.1);
\draw (-0.3,-0.175) -- (0.375,-0.175);
\begin{scope}[shift={(0,-0.05)}]
\draw (-0.3,-0.175) -- (0.375,-0.175);
\end{scope}
\node at (-0.4,-0.2) {\scriptsize{$i$}};
\node at (0.5,-0.2) {\scriptsize{$j$}};
\end{scope}
\draw (0.35,-0.15) -- (0.4,-0.2) -- (0.35,-0.25);
\draw (-0.3,-0.2) -- (0.4,-0.2);
\begin{scope}[shift={(2.7,0)}]
\draw  (-0.4,-0.2) ellipse (0.1 and 0.1);
\draw  (0.5,-0.2) ellipse (0.1 and 0.1);
\draw (-0.3,-0.175) -- (0.375,-0.175);
\begin{scope}[shift={(0,-0.05)}]
\draw (-0.3,-0.175) -- (0.375,-0.175);
\end{scope}
\node at (-0.4,-0.2) {\scriptsize{$i$}};
\node at (0.5,-0.2) {\scriptsize{$j$}};
\end{scope}
\draw (3.05,-0.15) -- (3.1,-0.2) -- (3.05,-0.25);
\draw (2.4,-0.2) -- (3.1,-0.2);
\end{tikzpicture}
\ee
corresponding to
\be
C_{ij}=3e+2f+\gamma_a x_a
\ee
before the duality; and
\be
\begin{tikzpicture} [scale=1.9]
\draw[red] (1,-0.2) -- (1.8,-0.2);
\draw[red] (1.75,-0.15) -- (1.8,-0.2) -- (1.75,-0.25);
\begin{scope}[shift={(-2.7,0)}]
\begin{scope}[shift={(2.7,0)}]
\draw  (-0.4,-0.2) ellipse (0.1 and 0.1);
\draw  (0.5,-0.2) ellipse (0.1 and 0.1);
\draw (-0.275,-0.175) -- (0.4,-0.175);
\begin{scope}[shift={(0,-0.05)}]
\draw (-0.275,-0.175) -- (0.4,-0.175);
\end{scope}
\node at (-0.4,-0.2) {\scriptsize{$i$}};
\node at (0.5,-0.2) {\scriptsize{$j$}};
\end{scope}
\draw (2.45,-0.15) -- (2.4,-0.2) -- (2.45,-0.25);
\end{scope}
\begin{scope}[shift={(-0.025,0)}]
\begin{scope}[shift={(2.7,0)}]
\draw  (-0.4,-0.2) ellipse (0.1 and 0.1);
\draw  (0.5,-0.2) ellipse (0.1 and 0.1);
\draw (-0.275,-0.175) -- (0.4,-0.175);
\begin{scope}[shift={(0,-0.05)}]
\draw (-0.275,-0.175) -- (0.4,-0.175);
\end{scope}
\node at (-0.4,-0.2) {\scriptsize{$i$}};
\node at (0.5,-0.2) {\scriptsize{$j$}};
\end{scope}
\draw (2.45,-0.15) -- (2.4,-0.2) -- (2.45,-0.25);
\end{scope}
\draw (0.4,-0.2) -- (-0.3,-0.2);
\draw (3.075,-0.2) -- (2.375,-0.2);
\end{tikzpicture}
\ee
corresponding to
\be
C_{ij}=e+f+\gamma_a x_a
\ee
before the duality.

In these cases, the gauge algebra before the duality must be $\fg=\fg_2$ and hence $\cP$ can only contain the single node $i$. Thus, these possibilities characterized by three edges between $i$ and $j$ before the duality actually completely classify all the irreducible dualities (of the special class being studied here\footnote{From this point on, whenever we say irreducible duality, we always mean an irreducible duality belonging to the special class of irreducible dualities being studied in this paper.}) originating from $\fg_2$.
\eit

In the rest of this section, we obtain all irreducible dualities in the special class discussed above and organize them in various subsections according to the identity of the simple gauge algebra $\fg$ appearing on one side of the duality. If both sides of the duality have a simple gauge algebra $\fg$ and $\fh$ such that the number of edges in the Dynkin diagram for $\fg$ are greater than or equal to the number of edges in the Dynkin diagram for $\fh$, then we place the duality in the subsection corresponding to $\fg$.

Before moving on to the discussion of irreducible dualities, we collect our namings for various irreducible representations. Let us label the nodes of Dynkin diagram of $\su(r+1)$ as
\be
\begin{tikzpicture} [scale=1.9]
\draw  (-2.3,-0.2) ellipse (0.1 and 0.1);
\draw (-2.2,-0.2) -- (-2,-0.2);
\node at (-1.8,-0.2) {$\cdots$};
\draw (-1.6,-0.2) -- (-1.4,-0.2);
\draw  (-1.3,-0.2) ellipse (0.1 and 0.1);
\begin{scope}[shift={(-5.2,-0.4)},rotate=180]
\draw  (-2.3,-0.2) ellipse (0.1 and 0.1);
\draw (-2.8,-0.2) -- (-2.4,-0.2);
\end{scope}
\begin{scope}[shift={(-5.8,-0.4)},rotate=180]
\draw  (-2.3,-0.2) ellipse (0.1 and 0.1);
\draw (-2.8,-0.2) -- (-2.4,-0.2);
\end{scope}
\node at (-3.5,-0.2) {\scriptsize{1}};
\node at (-2.9,-0.2) {\scriptsize{2}};
\node at (-2.3,-0.2) {\scriptsize{3}};
\node at (-1.3,-0.2) {\scriptsize{$r$}};
\end{tikzpicture}
\ee
Let $\cR_i$ be the representation such that its highest weight has $i$-th Dynkin coefficient equal to one and all other Dynkin coefficients equal to zero. We denote $\cR_i$ by $\L^i$ for all $i$. We also denote $\cR_1$ by $\F$ and $\cR_r$ by $\bar\F$.

Let us label the nodes of Dynkin diagram of $\so(2r+1)$ as
\be
\begin{tikzpicture} [scale=1.9]
\draw  (-2.3,-0.2) ellipse (0.1 and 0.1);
\draw (-2.2,-0.2) -- (-2,-0.2);
\node at (-1.8,-0.2) {$\cdots$};
\draw (-1.6,-0.2) -- (-1.4,-0.2);
\draw  (-1.1,-0.2) ellipse (0.3 and 0.1);
\begin{scope}[shift={(-5.2,-0.4)},rotate=180]
\draw  (-2.3,-0.2) ellipse (0.1 and 0.1);
\draw (-2.8,-0.2) -- (-2.4,-0.2);
\end{scope}
\begin{scope}[shift={(-5.8,-0.4)},rotate=180]
\draw  (-2.3,-0.2) ellipse (0.1 and 0.1);
\draw (-2.8,-0.2) -- (-2.4,-0.2);
\end{scope}
\node at (-3.5,-0.2) {\scriptsize{1}};
\node at (-2.9,-0.2) {\scriptsize{2}};
\node at (-2.3,-0.2) {\scriptsize{3}};
\node at (-1.1,-0.2) {\scriptsize{$r-1$}};
\draw  (-0.3,-0.2) ellipse (0.1 and 0.1);
\node at (-0.3,-0.2) {\scriptsize{$r$}};
\begin{scope}[shift={(-0.5,-0.825)},rotate=180]
\draw (-0.075,-0.6) -- (0.3,-0.6);
\draw (-0.075,-0.65) -- (0.3,-0.65);
\draw (-0.05,-0.575) -- (-0.1,-0.625) -- (-0.05,-0.675);
\end{scope}
\end{tikzpicture}
\ee
We denote $\cR_i$ by $\L^i$ for $1\le i\le r-1$. We also denote $\cR_1$ by $\F$. We denote $\cR_r$ by $\S$.

Let us label the nodes of Dynkin diagram of $\sp(r)$ as
\be
\begin{tikzpicture} [scale=1.9]
\draw  (-2.3,-0.2) ellipse (0.1 and 0.1);
\draw (-2.2,-0.2) -- (-2,-0.2);
\node at (-1.8,-0.2) {$\cdots$};
\draw (-1.6,-0.2) -- (-1.4,-0.2);
\draw  (-1.1,-0.2) ellipse (0.3 and 0.1);
\begin{scope}[shift={(-5.2,-0.4)},rotate=180]
\draw  (-2.3,-0.2) ellipse (0.1 and 0.1);
\draw (-2.8,-0.2) -- (-2.4,-0.2);
\end{scope}
\begin{scope}[shift={(-5.8,-0.4)},rotate=180]
\draw  (-2.3,-0.2) ellipse (0.1 and 0.1);
\draw (-2.8,-0.2) -- (-2.4,-0.2);
\end{scope}
\node at (-3.5,-0.2) {\scriptsize{1}};
\node at (-2.9,-0.2) {\scriptsize{2}};
\node at (-2.3,-0.2) {\scriptsize{3}};
\node at (-1.1,-0.2) {\scriptsize{$r-1$}};
\draw  (-0.3,-0.2) ellipse (0.1 and 0.1);
\node at (-0.3,-0.2) {\scriptsize{$r$}};
\begin{scope}[shift={(-0.7,0.425)},rotate=0]
\draw (-0.075,-0.6) -- (0.3,-0.6);
\draw (-0.075,-0.65) -- (0.3,-0.65);
\draw (-0.05,-0.575) -- (-0.1,-0.625) -- (-0.05,-0.675);
\end{scope}
\end{tikzpicture}
\ee
We denote $\cR_i$ by $\L^i$ for all $i$. We also denote $\cR_1$ by $\F$.

Let us label the nodes of Dynkin diagram of $\so(2r)$ as
\be
\begin{tikzpicture} [scale=1.9]
\draw  (-2.3,-0.2) ellipse (0.1 and 0.1);
\draw (-2.2,-0.2) -- (-2,-0.2);
\node at (-1.8,-0.2) {$\cdots$};
\draw (-1.6,-0.2) -- (-1.4,-0.2);
\draw  (-1.1,-0.2) ellipse (0.3 and 0.1);
\begin{scope}[shift={(-5.2,-0.4)},rotate=180]
\draw  (-2.3,-0.2) ellipse (0.1 and 0.1);
\draw (-2.8,-0.2) -- (-2.4,-0.2);
\end{scope}
\begin{scope}[shift={(-5.8,-0.4)},rotate=180]
\draw  (-2.3,-0.2) ellipse (0.1 and 0.1);
\draw (-2.8,-0.2) -- (-2.4,-0.2);
\end{scope}
\node at (-3.5,-0.2) {\scriptsize{1}};
\node at (-2.9,-0.2) {\scriptsize{2}};
\node at (-2.3,-0.2) {\scriptsize{3}};
\node at (-1.1,-0.2) {\scriptsize{$r-2$}};
\draw  (-1.1,-0.8) ellipse (0.1 and 0.1);
\node at (-1.1,-0.8) {\scriptsize{$r$}};
\draw (-0.8,-0.2) -- (-0.4,-0.2);
\draw  (-0.1,-0.2) ellipse (0.3 and 0.1);
\node at (-0.1,-0.2) {\scriptsize{$r-1$}};
\draw (-1.1,-0.3) -- (-1.1,-0.7);
\end{tikzpicture}
\ee
We denote $\cR_i$ by $\L^i$ for $1\le i\le r-2$. We also denote $\cR_1$ by $\F$. We denote $\cR_{r-1}$ and $\cR_r$ by $\S$ and $\C$.

Let us label the nodes of Dynkin diagram of $\fe_r$ as
\be
\begin{tikzpicture} [scale=1.9]
\draw  (-2.3,-0.2) ellipse (0.1 and 0.1);
\draw (-2.2,-0.2) -- (-2,-0.2);
\node at (-1.8,-0.2) {$\cdots$};
\draw (-1.6,-0.2) -- (-1.4,-0.2);
\draw  (-1.1,-0.2) ellipse (0.3 and 0.1);
\begin{scope}[shift={(-5.2,-0.4)},rotate=180]
\draw  (-2.3,-0.2) ellipse (0.1 and 0.1);
\draw (-2.8,-0.2) -- (-2.4,-0.2);
\end{scope}
\begin{scope}[shift={(-5.8,-0.4)},rotate=180]
\draw  (-2.3,-0.2) ellipse (0.1 and 0.1);
\draw (-2.8,-0.2) -- (-2.4,-0.2);
\end{scope}
\node at (-3.5,-0.2) {\scriptsize{1}};
\node at (-2.9,-0.2) {\scriptsize{2}};
\node at (-2.3,-0.2) {\scriptsize{3}};
\node at (-1.1,-0.2) {\scriptsize{$r-3$}};
\draw  (-1.1,-0.8) ellipse (0.1 and 0.1);
\node at (-1.1,-0.8) {\scriptsize{$r$}};
\draw (-0.8,-0.2) -- (-0.4,-0.2);
\draw  (-0.1,-0.2) ellipse (0.3 and 0.1);
\node at (-0.1,-0.2) {\scriptsize{$r-2$}};
\draw (-1.1,-0.3) -- (-1.1,-0.7);
\draw  (0.9,-0.2) ellipse (0.3 and 0.1);
\node at (0.9,-0.2) {\scriptsize{$r-1$}};
\draw (0.2,-0.2) -- (0.6,-0.2);
\end{tikzpicture}
\ee
We denote $\cR_i$ by $\L^i$ for $1\le i\le r-3$. We also denote $\cR_1$ by $\F$. We denote $\cR_{r-1}$ by $\C$, $\cR_{r-2}$ by $\C^2$ and$\cR_r$ by $\S$.

Let us label the nodes of Dynkin diagram of $\ff_4$ as
\be
\begin{tikzpicture} [scale=1.9]
\draw  (-2.3,-0.2) ellipse (0.1 and 0.1);
\begin{scope}[shift={(-5.2,-0.4)},rotate=180]
\draw  (-2.3,-0.2) ellipse (0.1 and 0.1);
\end{scope}
\begin{scope}[shift={(-5.8,-0.4)},rotate=180]
\draw  (-2.3,-0.2) ellipse (0.1 and 0.1);
\draw (-2.8,-0.2) -- (-2.4,-0.2);
\end{scope}
\node at (-3.5,-0.2) {\scriptsize{1}};
\node at (-2.9,-0.2) {\scriptsize{2}};
\node at (-2.3,-0.2) {\scriptsize{3}};
\draw  (-1.7,-0.2) ellipse (0.1 and 0.1);
\draw (-2.2,-0.2) -- (-1.8,-0.2);
\node at (-1.7,-0.2) {\scriptsize{4}};
\begin{scope}[shift={(-2.7,0.425)},rotate=0]
\draw (-0.075,-0.6) -- (0.3,-0.6);
\draw (-0.075,-0.65) -- (0.3,-0.65);
\draw (-0.05,-0.575) -- (-0.1,-0.625) -- (-0.05,-0.675);
\end{scope}
\end{tikzpicture}
\ee
We denote $\cR_1,\cR_2,\cR_3,\cR_4$ by $\F,\L^2,\L^3,\A$ respectively.

Let us label the nodes of Dynkin diagram of $\fg_2$ as
\be
\begin{tikzpicture} [scale=1.9]
\draw  (-2.3,-0.2) ellipse (0.1 and 0.1);
\begin{scope}[shift={(-5.2,-0.4)},rotate=180]
\draw  (-2.3,-0.2) ellipse (0.1 and 0.1);
\end{scope}
\node at (-2.9,-0.2) {\scriptsize{1}};
\node at (-2.3,-0.2) {\scriptsize{2}};
\begin{scope}[shift={(-2.7,0.425)},rotate=0]
\draw (-0.075,-0.6) -- (0.3,-0.6);
\draw (-0.075,-0.65) -- (0.3,-0.65);
\draw (-0.05,-0.575) -- (-0.1,-0.625) -- (-0.05,-0.675);
\end{scope}
\draw (-2.4,-0.2) -- (-2.8,-0.2);
\end{tikzpicture}
\ee
We denote $\cR_1,\cR_2$ by $\F,\A$ respectively.

\subsection{$\su(r+1)$ dualities}
Let us consider performing the transition
\be\label{suT}
\begin{tikzpicture} [scale=1.9]
\draw[red] (-0.2,-0.8) -- (-0.2,-1.4);
\draw[red] (-0.25,-1.35) -- (-0.2,-1.4) -- (-0.15,-1.35);
\draw  (-2.3,-0.2) ellipse (0.1 and 0.1);
\draw (-2.2,-0.2) -- (-2,-0.2);
\node at (-1.8,-0.2) {$\cdots$};
\draw (-1.6,-0.2) -- (-1.4,-0.2);
\draw  (-1.3,-0.2) ellipse (0.1 and 0.1);
\begin{scope}[shift={(1.6,0)}]
\draw  (-2.3,-0.2) ellipse (0.1 and 0.1);
\draw (-2.8,-0.2) -- (-2.4,-0.2);
\end{scope}
\begin{scope}[shift={(1.6,0)}]
\draw (-2.2,-0.2) -- (-2,-0.2);
\node (v1) at (-1.8,-0.2) {$\cdots$};
\draw (-1.6,-0.2) -- (-1.4,-0.2);
\draw  (-1.3,-0.2) ellipse (0.1 and 0.1);
\end{scope}
\begin{scope}[shift={(3.2,0)}]
\draw  (-2.3,-0.2) ellipse (0.1 and 0.1);
\draw (-2.8,-0.2) -- (-2.4,-0.2);
\end{scope}
\begin{scope}[shift={(3.2,0)}]
\draw (-2.2,-0.2) -- (-2,-0.2);
\node at (-1.8,-0.2) {$\cdots$};
\draw (-1.6,-0.2) -- (-1.4,-0.2);
\draw  (-1.3,-0.2) ellipse (0.1 and 0.1);
\end{scope}
\draw[dashed]  (v1) ellipse (0.8 and 0.3);
\node at (-1.8,0.2) {$m$};
\begin{scope}[shift={(0.1,-1.95)}]
\draw  (-2.5,-0.2) ellipse (0.1 and 0.1);
\draw  (-0.6,-0.2) ellipse (0.1 and 0.1);
\draw  (0,-0.2) ellipse (0.1 and 0.1);
\draw (-2.4,-0.2) -- (-2.2,-0.2);
\node at (-2,-0.2) {$\cdots$};
\draw (-1.8,-0.2) -- (-1.6,-0.2);
\draw  (-1.5,-0.2) ellipse (0.1 and 0.1);
\begin{scope}[shift={(1.4,0)}]
\draw  (-2.3,-0.2) ellipse (0.1 and 0.1);
\draw (-2.8,-0.2) -- (-2.4,-0.2);
\end{scope}
\begin{scope}[shift={(1.6,0)}]
\node (v1) at (-1.9,-0.2) {$\cdots$};
\draw  (-1.3,-0.2) ellipse (0.1 and 0.1);
\end{scope}
\begin{scope}[shift={(3.2,0)}]
\draw  (-2.3,-0.2) ellipse (0.1 and 0.1);
\draw (-2.8,-0.2) -- (-2.4,-0.2);
\end{scope}
\begin{scope}[shift={(3.2,0)}]
\draw (-2.2,-0.2) -- (-2,-0.2);
\node at (-1.8,-0.2) {$\cdots$};
\draw (-1.6,-0.2) -- (-1.4,-0.2);
\draw  (-1.3,-0.2) ellipse (0.1 and 0.1);
\end{scope}
\draw[dashed]  (v1) ellipse (0.9 and 0.3);
\node at (-0.3,0.2) {$n$};
\end{scope}
\begin{scope}[shift={(3.2,0)}]
\node at (-1.8,0.2) {$p$};
\draw (-2.4,-0.05) .. controls (-2.4,0) and (-2.4,0.05) .. (-2.3,0.05);
\draw (-2.3,0.05) -- (-1.9,0.05);
\draw (-1.8,0.1) .. controls (-1.8,0.05) and (-1.85,0.05) .. (-1.9,0.05);
\draw (-1.8,0.1) .. controls (-1.8,0.05) and (-1.75,0.05) .. (-1.7,0.05);
\draw (-1.7,0.05) -- (-1.3,0.05);
\draw (-1.2,-0.05) .. controls (-1.2,0) and (-1.2,0.05) .. (-1.3,0.05);
\end{scope}
\begin{scope}[shift={(-0.1,-2)}]
\node at (-1.8,0.2) {$m$};
\draw (-2.4,-0.05) .. controls (-2.4,0) and (-2.4,0.05) .. (-2.3,0.05);
\draw (-2.3,0.05) -- (-1.9,0.05);
\draw (-1.8,0.1) .. controls (-1.8,0.05) and (-1.85,0.05) .. (-1.9,0.05);
\draw (-1.8,0.1) .. controls (-1.8,0.05) and (-1.75,0.05) .. (-1.7,0.05);
\draw (-1.7,0.05) -- (-1.3,0.05);
\draw (-1.2,-0.05) .. controls (-1.2,0) and (-1.2,0.05) .. (-1.3,0.05);
\end{scope}
\begin{scope}[shift={(3.3,-2)}]
\node at (-1.8,0.2) {$p$};
\draw (-2.4,-0.05) .. controls (-2.4,0) and (-2.4,0.05) .. (-2.3,0.05);
\draw (-2.3,0.05) -- (-1.9,0.05);
\draw (-1.8,0.1) .. controls (-1.8,0.05) and (-1.85,0.05) .. (-1.9,0.05);
\draw (-1.8,0.1) .. controls (-1.8,0.05) and (-1.75,0.05) .. (-1.7,0.05);
\draw (-1.7,0.05) -- (-1.3,0.05);
\draw (-1.2,-0.05) .. controls (-1.2,0) and (-1.2,0.05) .. (-1.3,0.05);
\end{scope}
\draw (-2.4,-0.05) .. controls (-2.4,0) and (-2.4,0.05) .. (-2.3,0.05);
\draw (-2.3,0.05) -- (-1.9,0.05);
\draw (-1.8,0.1) .. controls (-1.8,0.05) and (-1.85,0.05) .. (-1.9,0.05);
\draw (-1.8,0.1) .. controls (-1.8,0.05) and (-1.75,0.05) .. (-1.7,0.05);
\draw (-1.7,0.05) -- (-1.3,0.05);
\draw (-1.2,-0.05) .. controls (-1.2,0) and (-1.2,0.05) .. (-1.3,0.05);
\node at (-0.2,0.2) {$n$};
\end{tikzpicture}
\ee
where the dashed ellipse encircles the nodes in the set $\cP$ (a total of $n$ in number), that is we will perform the transformation $\cS$ on the surfaces corresponding to these nodes. For $n\ge2$, the above transition acts non-trivially on the gauge algebra
\be
\su(m+n+p+1)\longrightarrow\su(m+2)\oplus \su(2)^{\oplus (n-2)}\oplus\su(p+2)
\ee

The geometry corresponding to the right hand side of this yet to be determined duality (for $n\ge2$) is uniquely determined to be
\be\label{suR}
\begin{tikzpicture} [scale=1.9]
\node (v1) at (-2.9,-0.5) {$\bF_{2m+1}$};
\node (v3) at (-2.1,-0.5) {$\cdots$};
\node at (-2.5,-0.4) {\scriptsize{$e$}};
\node at (-1.7,-0.4) {\scriptsize{$h$}};
\node (v4) at (-1.4,-0.5) {$\bF_{5}$};
\draw  (v3) edge (v4);
\node (v5) at (-0.4,-0.5) {$\bF_{3}$};
\draw  (v4) edge (v5);
\draw  (v1) edge (v3);
\node (v2) at (0.8,-0.5) {$\bF^1_0$};
\draw  (v5) edge (v2);
\node at (-0.1,-0.4) {\scriptsize{$e$}};
\node (v6) at (1.8,-0.5) {$\bF^{2}_0$};
\draw  (v2) edge (v6);
\node at (-1.1,-0.4) {\scriptsize{$e$}};
\node at (-0.7,-0.4) {\scriptsize{$h$}};
\node at (0.4,-0.4) {\scriptsize{$e$+$f$-$x$}};
\node at (1.1,-0.4) {\scriptsize{$f$-$x$}};
\node at (1.5,-0.4) {\scriptsize{$f$-$x$}};
\node (v7) at (2.8,-0.5) {$\bF^{2}_0$};
\draw  (v6) edge (v7);
\node at (2.1,-0.4) {\scriptsize{$f$-$y$}};
\node at (2.5,-0.4) {\scriptsize{$f$-$x$}};
\node (v8) at (3.5,-0.5) {$\cdots$};
\draw  (v7) edge (v8);
\node at (3.1,-0.4) {\scriptsize{$f$-$y$}};
\node (v9) at (4.2,-0.5) {$\bF^1_0$};
\draw  (v8) edge (v9);
\node at (3.9,-0.4) {\scriptsize{$f$-$x$}};
\node (v10) at (4.2,-1.4) {$\bF_3$};
\node (v11) at (3.2,-1.4) {$\bF_5$};
\node (v12) at (2.5,-1.4) {$\cdots$};
\node (v13) at (1.7,-1.4) {$\bF_{2p+1}$};
\draw  (v9) edge (v10);
\draw  (v10) edge (v11);
\draw  (v11) edge (v12);
\draw  (v12) edge (v13);
\node at (4.5,-0.8) {\scriptsize{$e$+$f$-$x$}};
\node at (4.4,-1.2) {\scriptsize{$e$}};
\node at (3.9,-1.3) {\scriptsize{$h$}};
\node at (3.5,-1.3) {\scriptsize{$e$}};
\node at (2.9,-1.3) {\scriptsize{$h$}};
\node at (2.1,-1.3) {\scriptsize{$e$}};
\begin{scope}[shift={(0,-0.2)}]
\node at (-1.8,0.2) {$m$};
\draw (-3.2,-0.05) .. controls (-3.2,0) and (-3.2,0.05) .. (-3.1,0.05);
\draw (-3.1,0.05) -- (-1.9,0.05);
\draw (-1.8,0.1) .. controls (-1.8,0.05) and (-1.85,0.05) .. (-1.9,0.05);
\draw (-1.8,0.1) .. controls (-1.8,0.05) and (-1.75,0.05) .. (-1.7,0.05);
\draw (-1.7,0.05) -- (-0.35,0.05);
\draw (-0.25,-0.05) .. controls (-0.25,0) and (-0.25,0.05) .. (-0.35,0.05);
\end{scope}
\begin{scope}[shift={(4.6,-1.75)}]
\node at (-1.8,-0.1) {$p$};
\draw (-3.2,0.15) .. controls (-3.2,0.1) and (-3.2,0.05) .. (-3.1,0.05);
\draw (-3.1,0.05) -- (-1.9,0.05);
\draw (-1.8,0) .. controls (-1.8,0.05) and (-1.85,0.05) .. (-1.9,0.05);
\draw (-1.8,0) .. controls (-1.8,0.05) and (-1.75,0.05) .. (-1.7,0.05);
\draw (-1.7,0.05) -- (-0.35,0.05);
\draw (-0.25,0.15) .. controls (-0.25,0.1) and (-0.25,0.05) .. (-0.35,0.05);
\end{scope}
\end{tikzpicture}
\ee
where, since we are searching for irreducible dualities, we have included only the minimum amount of blowups required to construct the geometry. After performing the $\cS$ transformations on surfaces contained in $\cP$, (\ref{suR}) can be written as
\be
\begin{tikzpicture} [scale=1.9]
\node (v1) at (-2.9,-0.5) {$\bF_{2m+1}$};
\node (v3) at (-2.1,-0.5) {$\cdots$};
\node at (-2.5,-0.4) {\scriptsize{$e$}};
\node at (-1.7,-0.4) {\scriptsize{$h$}};
\node (v4) at (-1.4,-0.5) {$\bF_{5}$};
\draw  (v3) edge (v4);
\node (v5) at (-0.4,-0.5) {$\bF_{3}$};
\draw  (v4) edge (v5);
\draw  (v1) edge (v3);
\node (v2) at (0.8,-0.5) {$\bF^1_0$};
\draw  (v5) edge (v2);
\node at (-0.1,-0.4) {\scriptsize{$e$}};
\node (v6) at (1.8,-0.5) {$\bF^{2}_0$};
\draw  (v2) edge (v6);
\node at (-1.1,-0.4) {\scriptsize{$e$}};
\node at (-0.7,-0.4) {\scriptsize{$h$}};
\node at (0.4,-0.4) {\scriptsize{$e$+$f$-$x$}};
\node at (1.1,-0.4) {\scriptsize{$e$-$x$}};
\node at (1.5,-0.4) {\scriptsize{$e$-$x$}};
\node (v7) at (2.8,-0.5) {$\bF^{2}_0$};
\draw  (v6) edge (v7);
\node at (2.1,-0.4) {\scriptsize{$e$-$y$}};
\node at (2.5,-0.4) {\scriptsize{$e$-$x$}};
\node (v8) at (3.5,-0.5) {$\cdots$};
\draw  (v7) edge (v8);
\node at (3.1,-0.4) {\scriptsize{$e$-$y$}};
\node (v9) at (4.2,-0.5) {$\bF^1_0$};
\draw  (v8) edge (v9);
\node at (3.9,-0.4) {\scriptsize{$e$-$x$}};
\node (v10) at (4.2,-1.4) {$\bF_3$};
\node (v11) at (3.2,-1.4) {$\bF_5$};
\node (v12) at (2.5,-1.4) {$\cdots$};
\node (v13) at (1.7,-1.4) {$\bF_{2p+1}$};
\draw  (v9) edge (v10);
\draw  (v10) edge (v11);
\draw  (v11) edge (v12);
\draw  (v12) edge (v13);
\node at (4.5,-0.8) {\scriptsize{$e$+$f$-$x$}};
\node at (4.4,-1.2) {\scriptsize{$e$}};
\node at (3.9,-1.3) {\scriptsize{$h$}};
\node at (3.5,-1.3) {\scriptsize{$e$}};
\node at (2.9,-1.3) {\scriptsize{$h$}};
\node at (2.1,-1.3) {\scriptsize{$e$}};
\begin{scope}[shift={(0,-0.2)}]
\node at (-1.8,0.2) {$m$};
\draw (-3.2,-0.05) .. controls (-3.2,0) and (-3.2,0.05) .. (-3.1,0.05);
\draw (-3.1,0.05) -- (-1.9,0.05);
\draw (-1.8,0.1) .. controls (-1.8,0.05) and (-1.85,0.05) .. (-1.9,0.05);
\draw (-1.8,0.1) .. controls (-1.8,0.05) and (-1.75,0.05) .. (-1.7,0.05);
\draw (-1.7,0.05) -- (-0.35,0.05);
\draw (-0.25,-0.05) .. controls (-0.25,0) and (-0.25,0.05) .. (-0.35,0.05);
\end{scope}
\begin{scope}[shift={(4.6,-1.75)}]
\node at (-1.8,-0.1) {$p$};
\draw (-3.2,0.15) .. controls (-3.2,0.1) and (-3.2,0.05) .. (-3.1,0.05);
\draw (-3.1,0.05) -- (-1.9,0.05);
\draw (-1.8,0) .. controls (-1.8,0.05) and (-1.85,0.05) .. (-1.9,0.05);
\draw (-1.8,0) .. controls (-1.8,0.05) and (-1.75,0.05) .. (-1.7,0.05);
\draw (-1.7,0.05) -- (-0.35,0.05);
\draw (-0.25,0.15) .. controls (-0.25,0.1) and (-0.25,0.05) .. (-0.35,0.05);
\end{scope}
\end{tikzpicture}
\ee
which is flop equivalent to
\be\label{suL}
\begin{tikzpicture} [scale=1.9]
\node (v1) at (-2.5,-0.5) {$\bF_{2m+2n-5}$};
\node (v3) at (-1.5,-0.5) {$\cdots$};
\node at (-3.5,-0.4) {\scriptsize{$e$}};
\node at (-3.1,-0.4) {\scriptsize{$h$}};
\node (v4) at (-0.7,-0.5) {$\bF_{2n-1}$};
\draw  (v3) edge (v4);
\node (v5) at (0.5,-0.5) {$\bF_{2n-3}^1$};
\draw  (v4) edge (v5);
\draw  (v1) edge (v3);
\node (v2) at (1.7,-0.5) {$\bF_{2n-5}$};
\draw  (v5) edge (v2);
\node at (-0.3,-0.4) {\scriptsize{$e$}};
\node (v6) at (2.5,-0.5) {$\cdots$};
\draw  (v2) edge (v6);
\node at (-1.9,-0.4) {\scriptsize{$e$}};
\node at (-1.1,-0.4) {\scriptsize{$h$}};
\node at (0.9,-0.4) {\scriptsize{$e$}};
\node at (1.3,-0.4) {\scriptsize{$h$}};
\node at (2.1,-0.4) {\scriptsize{$e$}};
\node (v7) at (3.2,-0.5) {$\bF_1$};
\draw  (v6) edge (v7);
\node at (2.9,-0.4) {\scriptsize{$h$}};
\node at (3.3,-1.2) {\scriptsize{$e$}};
\node (v8) at (3.2,-1.5) {$\bF_1^1$};
\draw  (v7) edge (v8);
\node at (3.3,-0.8) {\scriptsize{$e$}};
\node (v9) at (2.2,-1.5) {$\bF_3$};
\draw  (v8) edge (v9);
\node at (1.9,-1.4) {\scriptsize{$h$}};
\node (v10) at (1.5,-1.5) {$\cdots$};
\draw  (v9) edge (v10);
\node at (2.5,-1.4) {\scriptsize{$e$}};
\node at (1.1,-1.4) {\scriptsize{$e$}};
\node at (2.9,-1.4) {\scriptsize{$h$}};
\node (v14) at (-4.1,-0.5) {$\bF^{2n-4}_{2m+2n-3}$};
\draw  (v14) edge (v1);
\node at (0.1,-0.4) {\scriptsize{$h$}};
\node (v11) at (0.7,-1.5) {$\bF_{2p+1}$};
\draw  (v11) edge (v10);
\end{tikzpicture}
\ee
Equating the gauge theory associated to (\ref{suL}) to the gauge theory associated to (\ref{suR}), we obtain the following irreducible duality\footnote{We will display all irreducible dualities inside boxes for the convenience of the reader.} for $m\ge0$, $n\ge3$ and $p\ge0$
\be\label{su1}
\begin{tikzpicture} [scale=1.9]
\node (v1) at (-5.6,0.4) {$\su(2)_0$};
\node at (-5.4,0.9) {$\su(m+n+p+1)_{k_{m,n,p}}+(2n-4)\F+\L^{m+1}+\L^{m+n}$};
\node at (-7.9,0.4) {$=$};
\node (v2) at (-7,0.4) {$\su(m+2)_{m+2}$};
\draw  (v2) edge (v1);
\node (v3) at (-4.6,0.4) {$\su(2)_0$};
\draw  (v1) edge (v3);
\node (v4) at (-3.9,0.4) {$\cdots$};
\draw  (v3) edge (v4);
\node (v5) at (-3.2,0.4) {$\su(2)_0$};
\draw  (v4) edge (v5);
\node (v6) at (-1.9,0.4) {$\su(p+2)_{p+2}$};
\draw  (v5) edge (v6);
\begin{scope}[shift={(-2.7,0.05)}]
\node at (-1.8,-0.1) {$n-2$};
\draw (-3.2,0.15) .. controls (-3.2,0.1) and (-3.2,0.05) .. (-3.1,0.05);
\draw (-3.1,0.05) -- (-1.9,0.05);
\draw (-1.8,0) .. controls (-1.8,0.05) and (-1.85,0.05) .. (-1.9,0.05);
\draw (-1.8,0) .. controls (-1.8,0.05) and (-1.75,0.05) .. (-1.7,0.05);
\draw (-1.7,0.05) -- (-0.35,0.05);
\draw (-0.25,0.15) .. controls (-0.25,0.1) and (-0.25,0.05) .. (-0.35,0.05);
\end{scope}
\draw  (-8.2,1.2) rectangle (-1.2,-0.3);
\end{tikzpicture}
\ee
where the edges between two gauge algebras denote bifundamental matter charged under the two gauge algebras.
\be
\su(m+2)_{m+2}\to\su(2)_\pi
\ee
as $m\to0$ and
\be
\su(p+2)_{p+2}\to\su(2)_\pi
\ee
as $p\to0$. The CS level
\be
k_{m,n,p}=p-m+\half\left(A_{m+1,m+n+p+1}+A_{m+n,m+n+p+1}\right)
\ee
where $\half A_{m,n}$ is the addition to the CS level of an $\su(n)$ theory when a full hyper $f$ transforming in $\L^m$ is integrated into the theory from the direction corresponding to $m_f=-\infty$. The quantity $A_{m,n}$ is also known as the \emph{anomaly coefficient} for the representation $\L^m$ of $\su(n)$ in the literature, and an explicit expression for it is \cite{Aranda:2009wh}
\be
A_{m,n}=\frac{(n-m)(n-m+1)\cdots(n-3)(n-2m)}{(m-1)!}
\ee
for $m\ge 3$,
\be
A_{2,n}=n-4
\ee
and
\be
A_{1,n}=1
\ee
For $n=2$, $m\ge0$ and $p\ge1$, we obtain the irreducible duality
\be\label{su2}
\begin{tikzpicture} [scale=1.9]
\node at (-6.6,0.9) {$\su(m+p+3)_{k_{m,p}}+\L^{m+1}+\L^{m+2}$};
\node at (-4.9,0.9) {$=$};
\node (v2) at (-3.8,0.9) {$\su(m+2)_{-(m+2+\frac p2)}$};
\node (v6) at (-1.8,0.9) {$\su(p+2)_{p+2+\frac m2}$};
\draw  (-8.2,1.2) rectangle (-1,0.6);
\draw  (v2) edge (v6);
\end{tikzpicture}
\ee
with
\be
\su(m+2)_{-(m+2+\frac p2)}\to\su(2)_\pi
\ee
as $m\to0$ and
\be
k_{m,p}=p-m+\half(A_{m+1,m+p+3}+A_{m+2,m+p+3})
\ee
Now the only remaining case is $n=2$ and $m=p=0$. For this case, we obtain the following irreducible duality
\be\label{su3}
\begin{tikzpicture} [scale=1.9]
\node at (-6.6,0.9) {$\su(3)_{0}+2\F$};
\node at (-5.7,0.9) {$=$};
\node (v2) at (-5.05,0.9) {$\su(2)_{\pi}$};
\node (v6) at (-3.8,0.9) {$\su(2)_{\pi}$};
\draw  (-7.4,1.2) rectangle (-3.3,0.6);
\draw  (v2) edge (v6);
\end{tikzpicture}
\ee

We can add matter on both sides of the above irreducible dualities to obtain other dualities. For example, adding $\sum d_i\L^i$ on the left hand side of (\ref{su1}) leads to the following more general reducible dualities
\be
\begin{tikzpicture} [scale=1.9]
\node (v1) at (-5.6,0.4) {$\su(2)_0$};
\node at (-4.7,0.9) {$\su(m+n+p+1)_{k_{m,n,p,\{d_i\}}}+(2n-4)\F+\L^{m+1}+\L^{m+n}+\sum_{i=1}^{m+n+p} d_i\L^i$};
\node at (-8,0.4) {$=$};
\node (v2) at (-7,0.4) {$\su(m+2)_{k_{m,\{d_i\}}}$};
\draw  (v2) edge (v1);
\node (v3) at (-4.6,0.4) {$\su(2)_0$};
\draw  (v1) edge (v3);
\node (v4) at (-3.9,0.4) {$\cdots$};
\draw  (v3) edge (v4);
\node (v5) at (-3.2,0.4) {$\su(2)_0$};
\draw  (v4) edge (v5);
\node (v6) at (-1.8,0.4) {$\su(p+2)_{k_{p,\{d_i\}}}$};
\draw  (v5) edge (v6);
\node (v7) at (-7,-0.2) {$\sum_{i=1}^{m+1}d_i\L^{m+2-i}$};
\draw  (v2) edge (v7);
\node (v8) at (-5.6,-0.2) {$d_{m+2}\F$};
\draw  (v1) edge (v8);
\node (v9) at (-4.6,-0.2) {$d_{m+3}\F$};
\draw  (v3) edge (v9);
\node (v10) at (-3.2,-0.2) {$d_{m+n-1}\F$};
\node at (-3.9,-0.2) {$\cdots$};
\draw  (v5) edge (v10);
\node (v11) at (-1.8,-0.2) {$\sum_{i=1}^{p+1}d_{m+n+i-1}\L^{i}$};
\draw  (v6) edge (v11);
\end{tikzpicture}
\ee
where extra edges attached to gauge algebras denote the extra matter content transforming only under those gauge algebras. For instance, $\su(m+2)$ on the right hand side of the above equation has $d_i$ extra hypers transforming under $\L^{m+2-i}$ of $\su(m+2)$ for $i=1,\cdots,m+1$, and the $\su(2)$ connected to $\su(m+2)$ has extra $d_{m+2}$ hypers transforming in fundamental $\F$ of that $\su(2)$. The theta angles of an $\su(2)$ are relevant only when there are no extra fundamentals, in which case the theta angle must be zero, and this is the reason we have displayed $\su(2)$ as $\su(2)_0$. The CS levels are
\begin{align}
k_{m,n,p,\{d_i\}}&=k_{m,n,p}+\half\sum_{i=1}^{m+n+p}d_iA_{i,m+n+p+1}\\
k_{m,\{d_i\}}&=m+2+\half\sum_{i=1}^{m+1}d_iA_{m+2-i,m+2}\\
k_{p,\{d_i\}}&=p+2+\half\sum_{i=1}^{p+1}d_{m+n+i-1}A_{i,p+2}
\end{align}

To determine the most general reducible dualities associated to an irreducible duality, we simply need to specify how the roots of the algebra on the left side of the duality map to the roots of the algebra on the right side of the duality. This information for irreducible dualities (\ref{su1}), (\ref{su2}) and (\ref{su3}) is provided by the corresponding transition diagram (\ref{suT}). However, for the precise determination of CS levels, we need to also specify\footnote{This convention is already implicitly chosen while converting the geometry to gauge theory while claiming dualities (\ref{su1}), (\ref{su2}) and (\ref{su3}); so we need to be consistent with that convention when adding extra matter.} which end of the Dynkin diagram of an $\su(N)$ algebra is associated to $\F$ and which end of the Dynkin diagram is associated to $\bar\F$. Thus,
\bit
\item For the duality (\ref{su1}), the map of matter content is specified by
\be
\begin{tikzpicture} [scale=1.9]
\draw[red] (-0.2,-0.8) -- (-0.2,-1.4);
\draw[red] (-0.25,-1.35) -- (-0.2,-1.4) -- (-0.15,-1.35);
\draw  (-2.3,-0.2) ellipse (0.1 and 0.1);
\draw (-2.2,-0.2) -- (-2,-0.2);
\node at (-1.8,-0.2) {$\cdots$};
\draw (-1.6,-0.2) -- (-1.4,-0.2);
\draw  (-1.3,-0.2) ellipse (0.1 and 0.1);
\begin{scope}[shift={(1.6,0)}]
\draw  (-2.3,-0.2) ellipse (0.1 and 0.1);
\draw (-2.8,-0.2) -- (-2.4,-0.2);
\end{scope}
\begin{scope}[shift={(1.6,0)}]
\draw (-2.2,-0.2) -- (-2,-0.2);
\node (v1) at (-1.8,-0.2) {$\cdots$};
\draw (-1.6,-0.2) -- (-1.4,-0.2);
\draw  (-1.3,-0.2) ellipse (0.1 and 0.1);
\end{scope}
\begin{scope}[shift={(3.2,0)}]
\draw  (-2.3,-0.2) ellipse (0.1 and 0.1);
\draw (-2.8,-0.2) -- (-2.4,-0.2);
\end{scope}
\begin{scope}[shift={(3.2,0)}]
\draw (-2.2,-0.2) -- (-2,-0.2);
\node at (-1.8,-0.2) {$\cdots$};
\draw (-1.6,-0.2) -- (-1.4,-0.2);
\draw  (-1.3,-0.2) ellipse (0.1 and 0.1);
\end{scope}
\draw[dashed]  (v1) ellipse (0.8 and 0.3);
\node at (-1.8,0.2) {$m$};
\begin{scope}[shift={(0.1,-1.95)}]
\draw  (-2.5,-0.2) ellipse (0.1 and 0.1);
\draw  (-0.6,-0.2) ellipse (0.1 and 0.1);
\draw  (0,-0.2) ellipse (0.1 and 0.1);
\draw (-2.4,-0.2) -- (-2.2,-0.2);
\node at (-2,-0.2) {$\cdots$};
\draw (-1.8,-0.2) -- (-1.6,-0.2);
\draw  (-1.5,-0.2) ellipse (0.1 and 0.1);
\begin{scope}[shift={(1.4,0)}]
\draw  (-2.3,-0.2) ellipse (0.1 and 0.1);
\draw (-2.8,-0.2) -- (-2.4,-0.2);
\end{scope}
\begin{scope}[shift={(1.6,0)}]
\node (v1) at (-1.9,-0.2) {$\cdots$};
\draw  (-1.3,-0.2) ellipse (0.1 and 0.1);
\end{scope}
\begin{scope}[shift={(3.2,0)}]
\draw  (-2.3,-0.2) ellipse (0.1 and 0.1);
\draw (-2.8,-0.2) -- (-2.4,-0.2);
\end{scope}
\begin{scope}[shift={(3.2,0)}]
\draw (-2.2,-0.2) -- (-2,-0.2);
\node at (-1.8,-0.2) {$\cdots$};
\draw (-1.6,-0.2) -- (-1.4,-0.2);
\draw  (-1.3,-0.2) ellipse (0.1 and 0.1);
\end{scope}
\draw[dashed]  (v1) ellipse (0.9 and 0.3);
\node at (-0.3,0.2) {$n$};
\end{scope}
\begin{scope}[shift={(3.2,0)}]
\node at (-1.8,0.2) {$p$};
\draw (-2.4,-0.05) .. controls (-2.4,0) and (-2.4,0.05) .. (-2.3,0.05);
\draw (-2.3,0.05) -- (-1.9,0.05);
\draw (-1.8,0.1) .. controls (-1.8,0.05) and (-1.85,0.05) .. (-1.9,0.05);
\draw (-1.8,0.1) .. controls (-1.8,0.05) and (-1.75,0.05) .. (-1.7,0.05);
\draw (-1.7,0.05) -- (-1.3,0.05);
\draw (-1.2,-0.05) .. controls (-1.2,0) and (-1.2,0.05) .. (-1.3,0.05);
\end{scope}
\begin{scope}[shift={(-0.1,-2)}]
\node at (-1.8,0.2) {$m$};
\draw (-2.4,-0.05) .. controls (-2.4,0) and (-2.4,0.05) .. (-2.3,0.05);
\draw (-2.3,0.05) -- (-1.9,0.05);
\draw (-1.8,0.1) .. controls (-1.8,0.05) and (-1.85,0.05) .. (-1.9,0.05);
\draw (-1.8,0.1) .. controls (-1.8,0.05) and (-1.75,0.05) .. (-1.7,0.05);
\draw (-1.7,0.05) -- (-1.3,0.05);
\draw (-1.2,-0.05) .. controls (-1.2,0) and (-1.2,0.05) .. (-1.3,0.05);
\end{scope}
\begin{scope}[shift={(3.3,-2)}]
\node at (-1.8,0.2) {$p$};
\draw (-2.4,-0.05) .. controls (-2.4,0) and (-2.4,0.05) .. (-2.3,0.05);
\draw (-2.3,0.05) -- (-1.9,0.05);
\draw (-1.8,0.1) .. controls (-1.8,0.05) and (-1.85,0.05) .. (-1.9,0.05);
\draw (-1.8,0.1) .. controls (-1.8,0.05) and (-1.75,0.05) .. (-1.7,0.05);
\draw (-1.7,0.05) -- (-1.3,0.05);
\draw (-1.2,-0.05) .. controls (-1.2,0) and (-1.2,0.05) .. (-1.3,0.05);
\end{scope}
\draw (-2.4,-0.05) .. controls (-2.4,0) and (-2.4,0.05) .. (-2.3,0.05);
\draw (-2.3,0.05) -- (-1.9,0.05);
\draw (-1.8,0.1) .. controls (-1.8,0.05) and (-1.85,0.05) .. (-1.9,0.05);
\draw (-1.8,0.1) .. controls (-1.8,0.05) and (-1.75,0.05) .. (-1.7,0.05);
\draw (-1.7,0.05) -- (-1.3,0.05);
\draw (-1.2,-0.05) .. controls (-1.2,0) and (-1.2,0.05) .. (-1.3,0.05);
\node at (-0.2,0.2) {$n$};
\node at (-2.3,-0.2) {\scriptsize{$\F$}};
\node at (1.9,-0.2) {\scriptsize{$\bar\F$}};
\node at (-2.4,-2.15) {\scriptsize{$\bar\F$}};
\node at (-0.8,-2.15) {\scriptsize{$\F$}};
\node at (0.4,-2.15) {\scriptsize{$\F$}};
\node at (2,-2.15) {\scriptsize{$\bar\F$}};
\end{tikzpicture}
\ee
\item For the duality (\ref{su2}), the map of matter content is specified by
\be
\begin{tikzpicture} [scale=1.9]
\draw[red] (-0.2,-0.8) -- (-0.2,-1.4);
\draw[red] (-0.25,-1.35) -- (-0.2,-1.4) -- (-0.15,-1.35);
\draw  (-2.3,-0.2) ellipse (0.1 and 0.1);
\draw (-2.2,-0.2) -- (-2,-0.2);
\node at (-1.8,-0.2) {$\cdots$};
\draw (-1.6,-0.2) -- (-1.4,-0.2);
\draw  (-1.3,-0.2) ellipse (0.1 and 0.1);
\begin{scope}[shift={(1.6,0)}]
\draw  (-2.3,-0.2) ellipse (0.1 and 0.1);
\draw (-2.8,-0.2) -- (-2.4,-0.2);
\end{scope}
\begin{scope}[shift={(1.6,0)}]
\draw (-2.2,-0.2) -- (-1.4,-0.2);
\node (v1) at (-1.8,-0.2) {};
\draw  (-1.3,-0.2) ellipse (0.1 and 0.1);
\end{scope}
\begin{scope}[shift={(3.2,0)}]
\draw  (-2.3,-0.2) ellipse (0.1 and 0.1);
\draw (-2.8,-0.2) -- (-2.4,-0.2);
\end{scope}
\begin{scope}[shift={(3.2,0)}]
\draw (-2.2,-0.2) -- (-2,-0.2);
\node at (-1.8,-0.2) {$\cdots$};
\draw (-1.6,-0.2) -- (-1.4,-0.2);
\draw  (-1.3,-0.2) ellipse (0.1 and 0.1);
\end{scope}
\draw[dashed]  (v1) ellipse (0.8 and 0.3);
\node at (-1.8,0.2) {$m$};
\begin{scope}[shift={(0.1,-1.95)}]
\draw  (-2.5,-0.2) ellipse (0.1 and 0.1);
\draw (-2.4,-0.2) -- (-2.2,-0.2);
\node at (-2,-0.2) {$\cdots$};
\draw (-1.8,-0.2) -- (-1.6,-0.2);
\draw  (-1.5,-0.2) ellipse (0.1 and 0.1);
\begin{scope}[shift={(1.4,0)}]
\draw  (-2.3,-0.2) ellipse (0.1 and 0.1);
\draw (-2.8,-0.2) -- (-2.4,-0.2);
\end{scope}
\begin{scope}[shift={(1.6,0)}]
\node (v1) at (-1.9,-0.2) {};
\draw  (-1.3,-0.2) ellipse (0.1 and 0.1);
\end{scope}
\begin{scope}[shift={(3.2,0)}]
\draw  (-2.3,-0.2) ellipse (0.1 and 0.1);
\draw (-2.8,-0.2) -- (-2.4,-0.2);
\end{scope}
\begin{scope}[shift={(3.2,0)}]
\draw (-2.2,-0.2) -- (-2,-0.2);
\node at (-1.8,-0.2) {$\cdots$};
\draw (-1.6,-0.2) -- (-1.4,-0.2);
\draw  (-1.3,-0.2) ellipse (0.1 and 0.1);
\end{scope}
\draw[dashed]  (v1) ellipse (0.9 and 0.3);
\end{scope}
\begin{scope}[shift={(3.2,0)}]
\node at (-1.8,0.2) {$p$};
\draw (-2.4,-0.05) .. controls (-2.4,0) and (-2.4,0.05) .. (-2.3,0.05);
\draw (-2.3,0.05) -- (-1.9,0.05);
\draw (-1.8,0.1) .. controls (-1.8,0.05) and (-1.85,0.05) .. (-1.9,0.05);
\draw (-1.8,0.1) .. controls (-1.8,0.05) and (-1.75,0.05) .. (-1.7,0.05);
\draw (-1.7,0.05) -- (-1.3,0.05);
\draw (-1.2,-0.05) .. controls (-1.2,0) and (-1.2,0.05) .. (-1.3,0.05);
\end{scope}
\begin{scope}[shift={(-0.1,-2)}]
\node at (-1.8,0.2) {$m$};
\draw (-2.4,-0.05) .. controls (-2.4,0) and (-2.4,0.05) .. (-2.3,0.05);
\draw (-2.3,0.05) -- (-1.9,0.05);
\draw (-1.8,0.1) .. controls (-1.8,0.05) and (-1.85,0.05) .. (-1.9,0.05);
\draw (-1.8,0.1) .. controls (-1.8,0.05) and (-1.75,0.05) .. (-1.7,0.05);
\draw (-1.7,0.05) -- (-1.3,0.05);
\draw (-1.2,-0.05) .. controls (-1.2,0) and (-1.2,0.05) .. (-1.3,0.05);
\end{scope}
\begin{scope}[shift={(3.3,-2)}]
\node at (-1.8,0.2) {$p$};
\draw (-2.4,-0.05) .. controls (-2.4,0) and (-2.4,0.05) .. (-2.3,0.05);
\draw (-2.3,0.05) -- (-1.9,0.05);
\draw (-1.8,0.1) .. controls (-1.8,0.05) and (-1.85,0.05) .. (-1.9,0.05);
\draw (-1.8,0.1) .. controls (-1.8,0.05) and (-1.75,0.05) .. (-1.7,0.05);
\draw (-1.7,0.05) -- (-1.3,0.05);
\draw (-1.2,-0.05) .. controls (-1.2,0) and (-1.2,0.05) .. (-1.3,0.05);
\end{scope}
\draw (-2.4,-0.05) .. controls (-2.4,0) and (-2.4,0.05) .. (-2.3,0.05);
\draw (-2.3,0.05) -- (-1.9,0.05);
\draw (-1.8,0.1) .. controls (-1.8,0.05) and (-1.85,0.05) .. (-1.9,0.05);
\draw (-1.8,0.1) .. controls (-1.8,0.05) and (-1.75,0.05) .. (-1.7,0.05);
\draw (-1.7,0.05) -- (-1.3,0.05);
\draw (-1.2,-0.05) .. controls (-1.2,0) and (-1.2,0.05) .. (-1.3,0.05);
\node at (-2.3,-0.2) {\scriptsize{$\F$}};
\node at (1.9,-0.2) {\scriptsize{$\bar\F$}};
\node at (-2.4,-2.15) {\scriptsize{$\F$}};
\node at (-0.8,-2.15) {\scriptsize{$\bar\F$}};
\node at (0.4,-2.15) {\scriptsize{$\F$}};
\node at (2,-2.15) {\scriptsize{$\bar\F$}};
\end{tikzpicture}
\ee
\item For the duality (\ref{su3}), the map of matter content can for instance be specified by
\be
\begin{tikzpicture} [scale=1.9]
\draw[red] (-0.2,-0.8) -- (-0.2,-1.4);
\draw[red] (-0.25,-1.35) -- (-0.2,-1.4) -- (-0.15,-1.35);
\draw  (-0.7,-0.2) ellipse (0.1 and 0.1);
\begin{scope}[shift={(1.6,0)}]
\draw (-2.2,-0.2) -- (-1.4,-0.2);
\node (v1) at (-1.8,-0.2) {};
\draw  (-1.3,-0.2) ellipse (0.1 and 0.1);
\end{scope}
\draw[dashed]  (v1) ellipse (0.8 and 0.3);
\begin{scope}[shift={(0.1,-1.85)}]
\draw  (-0.9,-0.2) ellipse (0.1 and 0.1);
\begin{scope}[shift={(1.6,0)}]
\node (v1) at (-1.9,-0.2) {};
\draw  (-1.3,-0.2) ellipse (0.1 and 0.1);
\end{scope}
\draw[dashed]  (v1) ellipse (0.9 and 0.3);
\end{scope}
\node at (0.3,-0.2) {\scriptsize{$\bar\F$}};
\node at (-0.7,-0.2) {\scriptsize{$\F$}};
\end{tikzpicture}
\ee
Actually, the specification of $\F$ and $\bar\F$ does not matter in this case since the $\su(3)$ in the duality (\ref{su3}) has CS level zero, which is left invariant by outer automorphism of $\su(3)$.
\eit
For $n=1$, the transition (\ref{suT}) acts trivially on the gauge algebra. The corresponding irreducible duality for $m,p\ge1$ is generated by the geometry
\be
\begin{tikzpicture} [scale=1.9]
\node (v1) at (-2.8,-0.5) {$\bF_{2m}$};
\node (v3) at (-2.1,-0.5) {$\cdots$};
\node at (-2.5,-0.4) {\scriptsize{$e$}};
\node at (-1.7,-0.4) {\scriptsize{$h$}};
\node (v4) at (-1.4,-0.5) {$\bF_{4}$};
\draw  (v3) edge (v4);
\node (v5) at (-0.4,-0.5) {$\bF_{2}$};
\draw  (v4) edge (v5);
\draw  (v1) edge (v3);
\node (v2) at (0.9,-0.5) {$\bF^2_0$};
\draw  (v5) edge (v2);
\node at (-0.1,-0.4) {\scriptsize{$e$}};
\node (v6) at (2.2,-0.5) {$\bF_2$};
\draw  (v2) edge (v6);
\node at (-1.1,-0.4) {\scriptsize{$e$}};
\node at (-0.7,-0.4) {\scriptsize{$h$}};
\node at (0.4,-0.4) {\scriptsize{$e$+$f$-$x$-$y$}};
\node at (1.4,-0.4) {\scriptsize{$e$+$f$-$x$-$y$}};
\node (v7) at (2.9,-0.5) {$\cdots$};
\draw  (v6) edge (v7);
\node (v8) at (3.6,-0.5) {$\bF_{2p}$};
\draw  (v7) edge (v8);
\node at (3.3,-0.4) {\scriptsize{$e$}};
\node at (2.5,-0.4) {\scriptsize{$h$}};
\node at (1.9,-0.4) {\scriptsize{$e$}};
\begin{scope}[shift={(0,-0.2)}]
\node at (-1.7,0.2) {$m$};
\draw (-3.1,-0.05) .. controls (-3.1,0) and (-3.1,0.05) .. (-3,0.05);
\draw (-3,0.05) -- (-1.8,0.05);
\draw (-1.7,0.1) .. controls (-1.7,0.05) and (-1.75,0.05) .. (-1.8,0.05);
\draw (-1.7,0.1) .. controls (-1.7,0.05) and (-1.65,0.05) .. (-1.6,0.05);
\draw (-1.6,0.05) -- (-0.35,0.05);
\draw (-0.25,-0.05) .. controls (-0.25,0) and (-0.25,0.05) .. (-0.35,0.05);
\end{scope}
\begin{scope}[shift={(4.6,-0.2)}]
\node at (-1.7,0.2) {$p$};
\draw (-2.6,-0.05) .. controls (-2.6,0) and (-2.6,0.05) .. (-2.5,0.05);
\draw (-2.5,0.05) -- (-1.8,0.05);
\draw (-1.7,0.1) .. controls (-1.7,0.05) and (-1.75,0.05) .. (-1.8,0.05);
\draw (-1.7,0.1) .. controls (-1.7,0.05) and (-1.65,0.05) .. (-1.6,0.05);
\draw (-1.6,0.05) -- (-0.9,0.05);
\draw (-0.8,-0.05) .. controls (-0.8,0) and (-0.8,0.05) .. (-0.9,0.05);
\end{scope}
\end{tikzpicture}
\ee
This geometry is left invariant by $\cS$ and produces a self-duality of $\su(m+p+2)_{k_{m,p}}+2\L^{m+1}$ where
\be
k_{m,p}=p-m+A_{m+1,m+p+2}
\ee
This duality only interchanges the perturbative and non-perturbative particles in the theory without modifying wither the gauge algebra or the matter content. Moreover, adding matter on both sides only leads to self-dualities. In this paper, we will ignore such self-dualities, but we will include those irreducible self-dualities which lead to reducible non-self-dualities, that is the nodes of the Dynkin diagram are non-trivially exchanged under the irreducible self-duality. We will discuss such examples later.

\bigskip

Now consider the transition
\be

\ee
For $n=1$, it is regarded as a $\so(2r+1)$ or $\sp(r)$ or $\ff_4$ duality rather than $\su(r+1)$ duality. So we restrict to $n\ge2$. First consider $n\ge3$. The geometry for the right hand side is uniquely fixed to be
\be\label{suR2}
%
\ee
This geometry is flop equivalent to
\be\label{suR2'}
%
\ee
where
\be
q_m=\frac{1+(-1)^m}{2}
\ee
On the other hand, (\ref{suR2}) after the action of $\cS$ and flops is equivalent to the geometry
\be\label{suL2}
%
\ee
Equating the gauge theories arising from (\ref{suL2}) and (\ref{suR2'}), we obtain the following irreducible duality for $m\ge1$, $n\ge3$, $p\ge0$
\be\label{suu4}
%
\ee
with
\be
k_{m,n,p}=p-m-\half+A_{m+1,m+n+p+1}+\half A_{m+n,m+n+p+1}
\ee
and
\be
\su(p+2)_{p+2}\to\su(2)_\pi
\ee
as $p\to0$. The addition of extra matter content to both sides of the above irreducible duality (to produce new reducible dualities) is governed by
\be
%
\ee

For $n=2$, we must have $p=0$ if we want to find a geometry satisfying all of the conditions, listed at the beginning of this section, defining the special class of irreducible dualities we are exploring in this paper. For $n=2$ and $p=0$, the geometry describing right hand side of the transition is
\be
%
\ee
which leads, in a similar way as above, to the irreducible duality for $m\ge1$
\be\label{suu5}
%
\ee
with the addition of extra matter specified by
\be
%
\ee

We can also consider the transition
\be
%
\ee
for which there is a consistent geometry satisfying the required conditions only for $n\ge3$:
\be
%
\ee
which gives rise to the following irreducible duality for $m,p\ge1$ and $n\ge4$
\be
%
\ee
with CS level
\be\label{k}
k_{m,n,p}=p-m+A_{m+1,m+n+p+1}+A_{m+n,m+n+p+1}
\ee
For $n=3$ and $m,p\ge1$, we obtain the irreducible duality
\be
%
\ee
where the CS level $k_{m,3,p}$ can be computed using (\ref{k}). For both of the above dualities, extra matter can be added according to
\be
%
\ee

\subsection{$\so(2r+1)$ dualities}
Let us start with the transition
\be
%
\ee
for $m\ge0$, $n\ge2$ and $p\ge1$. The geometry for the right hand side of the transition is
\be
%
\ee
which can be seen to give rise to the following irreducible duality for $m\ge0$, $n\ge3$ and $p\ge1$
\be
%
\ee
where the extra edge attached to the last node on the right hand side denotes that $\so(2p+5)$ contains $2p$ fundamentals not charged under any other gauge algebra. As always we should remember the replacement
\be
\su(m+2)_{m+2}\to\su(2)_\pi
\ee
as $m\to0$. For $n=2$, $m\ge0$, $p\ge1$ we obtain the following irreducible duality
\be
%
\ee
with
\be
\su(m+2)_{m+p+\frac72}\to\su(2)
\ee
as $m\to0$, as the theta angle of $\su(2)$ is not physically relevant. In both of the above dualities, extra matter is integrated in according to
\be
%
\ee

\bigskip

Now, consider the transition
\be\label{soTT}
%
\ee
for $m\ge0$ and $n\ge2$. The right hand side of this transition can correspond to two geometries satisfying the conditions defining the class of irreducible dualities being studied in this paper. The first geometry is
\be
%
\ee
which leads to the following irreducible duality for $m\ge0$ and $n\ge3$
\be
%
\ee
where the edge between $\su(2)_0$ and $\sp(2)_0$ on the right hand side denotes a full hyper in $\F\otimes\L^2$ of $\su(2)\oplus\sp(2)$ (instead of a bifundamental). As $m\to0$,
\be
\su(m+2)_{m+2}\to\su(2)_\pi
\ee
For $n=2$ and $m\ge0$, we obtain the following irreducible duality
\be
%
\ee
with
\be
\su(m+2)_{m+\frac72}\to\su(2)
\ee
as $m\to0$.

The second geometry associated to this transition is
\be\label{sooR}
%
\ee
which can be seen to give rise to the following irreducible duality for $m\ge0$ and $n\ge3$
\be\label{soo1}
%
\ee
with
\be
\su(m+2)_{m+2}\to\su(2)_\pi
\ee
as $m\to0$. The second geometry (\ref{sooR}) is consistent for $n=2$ only if we have $m=0$, in which case it becomes
\be
%
\ee
and gives rise to the following irreducible duality
\be\label{soo2}
%
\ee
For the above four irreducible dualities, the addition of extra matter is determined by
\be
%
\ee

\bigskip

For the transition
\be\label{soT}
%
\ee
having $m,p\ge1$, a geometry only exists for $n\ge3$ as the reader can construct in a fashion similar to geometries discussed above. This geometry leads to the following irreducible duality for $m\ge1$, $n\ge4$ and $p\ge1$
\be
%
\ee
and the following irreducible duality for $n=3$, $m,p\ge1$
\be
%
\ee
The extra matter for the above two dualities is added according to the transition diagram (\ref{soT}). We don't have to specify anything more since there are no $\su(N\ge3)$ algebras appearing in the above two dualities.

Now consider the transition
\be\label{soT2}
%
\ee
for $m\ge0$ and $n\ge1$. For $n\ge2$, the geometry for right hand side of the transition is
\be
%
\ee
which gives rise to the following irreducible duality for $m\ge0$ and $n\ge3$
\be\label{soo3}
%
\ee
with
\be
\su(m+2)_{m+2}\to\su(2)_\pi
\ee
as $m\to0$. Additional matter content can be added according to
\be
%
\ee
For $n=2$ and $m\ge0$, we obtain the following irreducible duality
\be\label{soo4}
%
\ee
with
\be
\su(m+2)_{-\left(m+\frac 52\right)}\to\su(2)_\pi
\ee
as $m\to0$. The associated reducible dualities are obtained by adding matter according to
\be
%
\ee
For $n=1$, the geometry for the right hand side of the transition (\ref{soT2}) is
\be
%
\ee
which leads us to the following irreducible duality for $m\ge1$
\be\label{soo5}
%
\ee
Additional matter can be incorporated according to
\be
%
\ee

\bigskip

Now consider the transition
\be
%
\ee
for $m\ge1$ and $n\ge2$. A geometry for this transition exists only for $n\ge3$, in which case there are two geometries just as in the case of (\ref{soTT}). The first one leads to the following irreducible duality for $m\ge1$ and $n\ge4$
\be
%
\ee
and the following duality for $n=3$ and $m\ge1$
\be
%
\ee
The second geometry leads to the following irreducible duality for $m\ge1$ and $n\ge4$
\be
%
\ee
and the following irreducible duality for $n=3$ and $m\ge1$
\be
%
\ee

\bigskip

Now consider the transition
\be\label{soTT2}
%
\ee
for which a corresponding geometry (satisfying the conditions detailed at the starting of this section) exists only for $n\ge3$. This geometry leads to the following irreducible duality for $m\ge1$ and $n\ge3$
\be
%
\ee
where the $\su(3)$ is oriented such that extra matter is added according to
\be
%
\ee

The $n=1$ version of the transition (\ref{soTT2}) allows a simple gauge algebra on the right hand side of the transition only for $m=1,2$. The $m=2$ version leads to a duality between $\so(9)$ and $\ff_4$, and is treated in the subsection on $\ff_4$ dualities later. The $m=1$ version leads to a duality between two $\so(7)$ whose Dynkin diagrams are mapped as
\be\label{so7s}
%
\ee
In particular $\F\to\S$ and $\S\to\F$. The geometry describing the right hand side of the above transition is fixed to be
\be
%
\ee
which leads to the following irreducible self-duality
\be\label{soo6}
%
\ee
We emphasize that, in general, the above irreducible self-duality leads to reducible dualities which are not self-dualities since the matter on two sides of the duality is mapped according to (\ref{so7s}).

\bigskip

Now, consider the transition
\be
%
\ee
The geometry for the right hand side of the transition for $n\ge3$ and $m\ge0$ is uniquely determined to be
\be
%
\ee
which implies the following irreducible duality for $m\ge0$ and $n\ge3$
\be
%
\ee
where
\be
\su(m+2)_{m+2}\to\su(2)_\pi
\ee
as $m\to 0$. Extra matter is added according to
\be
%
\ee
For $n=2$, only $m=0$ admits a consistent geometry satisfying all the imposed conditions. This geometry is
\be
%
\ee
and leads to the following irreducible duality
\be
%
\ee

\bigskip

The final transition we need to consider in this subsection is
\be
%
\ee
for $m\ge1$. This transition has a geometric representation satisfying all the required properties only for $n\ge3$, leading to the following irreducible duality for $m\ge1$ and $n\ge4$
\be
%
\ee
and the following irreducible duality for $n=3$ and $m\ge1$
\be
%
\ee

\subsection{$\sp(r)$ dualities}
Let us start with the transition
\be
%
\ee
where $m,p\ge0$ and $n\ge2$. The geometry associated to the right hand side of the transition must be
\be
%
\ee
which leads to the following irreducible duality for $m\ge0$, $n\ge3$ and $p\ge0$
\be
%
\ee
with
\be
\su(m+2)_{m+2}\to\su(2)_\pi
\ee
as $m\to 0$. For $n=2$ and $m,p\ge0$, we obtain the following irreducible duality
\be\label{spp1}
%
\ee
with
\be
\su(m+2)_{m+p+3}\to\su(2)_\pi
\ee
as $m\to 0$. Extra matter in the above two dualities is added according to the diagram
\be
%
\ee

\bigskip

Now let us consider the transition
\be
%
\ee
We require $n\ge3$ for the existence of a geometry satisfying all the requirements, and obtain the following irreducible duality for $m\ge1$, $n\ge3$ and $p\ge0$
\be
%
\ee

\bigskip

Finally, let us consider the transition
\be
%
\ee
for $m\ge1$. The geometry corresponding to the left hand side of the transition is uniquely determined to be the geometry (\ref{spd}) for the pure gauge theory. This leads to the following irreducible duality for $m\ge1$
\be\label{spsu}
%
\ee
generalizing the irreducible duality (\ref{exS}) discussed for illustration in Section \ref{S}. Extra matter is added according to
\be
%
\ee

\subsection{$\so(2r)$ dualities}
Let us start by considering the transition
\be
%
\ee
for $p\ge1$ and $n\ge2$. The right hand side of the above transition is governed by the geometry
\be
%
\ee
which leads to the following irreducible duality for $m\ge0$, $n\ge3$ and $p\ge1$
\be
%
\ee
with
\be
\su(m+2)_{m+2}\to\su(2)_\pi
\ee
as $m\to0$. For $n=2$, $m\ge0$ and $p\ge1$, we obtain the following irreducible duality
\be
%
\ee
with
\be
\su(m+2)_{m+p+4}\to\su(2)_\pi
\ee
as $m\to0$. Extra matter is added in the above two dualities according to
\be
%
\ee

\bigskip

Now, consider the transition
\be\label{soeT}
%
\ee
with $n\ge2$. There are two consistent geometries that can describe the right hand side of this transition. The first geometry is valid for all $n\ge2$ and is shown below
\be
%
\ee
which leads to the following irreducible duality for $m\ge0$ and $n\ge3$
\be
%
\ee
with
\be
\su(m+2)_{m+2}\to\su(2)_\pi
\ee
as $m\to0$. For $n=2$ and $m\ge0$, the above geometry gives rise to the following irreducible duality
\be
%
\ee
with
\be
\su(m+2)_{m+4}\to\su(2)_\pi
\ee
as $m\to0$. The second geometry is valid for $n\ge3,m\ge0$ and $n=2,m=0$. It is presented below
\be
%
\ee
For $m\ge0$ and $n\ge3$, it gives rise to the following irreducible duality
\be\label{soe}
%
\ee
with
\be
\su(m+2)_{m+2}\to\su(2)_\pi
\ee
as $m\to0$. Extra matter for the above three dualities is added according to 
\be
%
\ee
and gives rise to the following irreducible duality
\be\label{soe2}
%
\ee
Extra matter is added according to
\be
%
\ee
That is, the fundamental of $\su(2)$ on the right side is chosen to correspond to $\F$ of $\so(8)$ on the left side and thus partially fix a triality frame of $\so(8)$.

\bigskip

Now, let us move on to consider the transition
\be
%
\ee
for $n\ge3$. The geometry describing the right hand side of the above transition is
\be
%
\ee
which leads to the following irreducible duality for $m\ge0$ and $n\ge4$
\be\label{soe3}
%
\ee
with
\be
\su(m+2)_{m+2}\to\su(2)_\pi
\ee
as $m\to0$. Extra matter is added according to
\be
%
\ee
For $\so(2m+8)=\so(4k+2)$, $\S$ and $\C$ are complex conjugates of each other\footnote{Geometrically, this means that partially integrating in $\S$ is flop equivalent to partially integrating in $\C$.}, so we can choose whether integrating in a hyper in spinor of $\so(2m+8)=\so(4k+2)$ is equivalent to integrating a hyper in $\F$ of $\su(2)_\pi$ or $\F$ of $\su(3)_{\frac m2+3}$ with negative contribution to CS level of $\su(3)$. Similar comments apply to many of the following cases.

For $n=3$ and $m\ge0$, we obtain the following irreducible duality
\be\label{soe4}
%
\ee
with
\be
\su(m+2)_{-m-\frac52}\to\su(2)_\pi
\ee
as $m\to0$. Extra matter is added according to
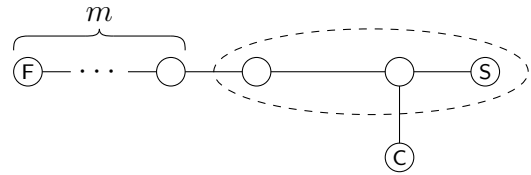
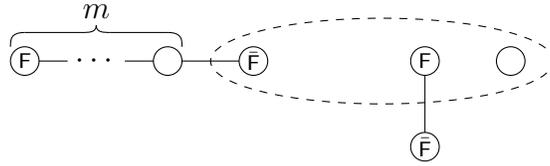
\be
%
\ee

\bigskip

Let us now consider the following transition
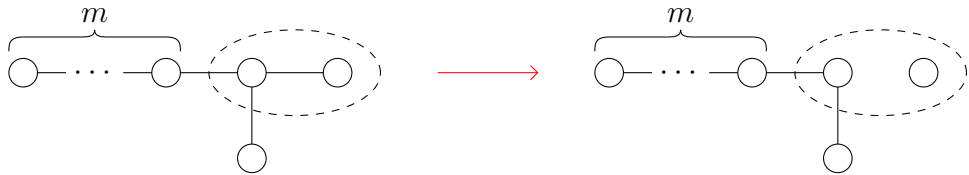
\be
%
\ee
for which we can restrict $m\ge2$ to avoid overcounting. There are two geometries that can describe the right hand side of this transition. The first one is
\be
%
\ee
and leads to the following irreducible duality for $m\ge2$
\be
%
\ee
and additional matter content can be added according to
\be\label{soeM}
%
\ee
The second geometry is
\be
%
\ee
and leads to the following irreducible duality for $m\ge2$
\be\label{soe5}
%
\ee
into which additional matter is added according to (\ref{soeM}).

\bigskip

Now consider the transition
\be
%
\ee
for $n\ge4$. The corresponding geometry for the right hand side of the transition is
\be
%
\ee
which leads to the following irreducible duality for $m\ge0$ and $n\ge4$
\be\label{soe6}
%
\ee
with
\be
\su(m+2)_{m+2}\to\su(2)_\pi
\ee
as $m\to0$. Extra matter is added according to
\be
%
\ee

\bigskip

For the transition
\be
%
\ee
we can take $m\ge2$ without loss of generality. The right hand side of the transition is described by the geometry
\be
%
\ee
leading to the following irreducible duality for $m\ge2$
\be\label{soe7}
%
\ee
Extra matter is added according to
\be
%
\ee

\bigskip

Now, we start considering transitions which admit more than one edge between some pairs of nodes. The first such transition we consider is
\be
%
\ee
for $m,p\ge1$ and $n\ge2$. A geometry satisfying all the criteria exists only for $n\ge3$ and leads to the following irreducible duality for $m\ge1$, $n\ge3$ and $p\ge1$
\be
%
\ee

Now, consider the transition
\be
%
\ee
which admits two geometric descriptions just like (\ref{soeT}). The first one leads to the following irreducible duality for $m\ge1$ and $n\ge3$
\be
%
\ee
The second one leads to the following irreducible duality for $m\ge1$ and $n\ge4$
\be
%
\ee
and the following irreducible duality for $n=3$ and $m\ge1$
\be
%
\ee

\bigskip

Now, consider the transition
\be
%
\ee
whose right hand side corresponds to the following geometry
\be
%
\ee
which makes sense for $n\ge3$. Thus, it leads to the following irreducible duality valid for $m\ge0$ and $n\ge3$
\be
%
\ee
with
\be
\su(m+2)_{m+2}\to\su(2)_\pi
\ee
as $m\to0$. For $n=2$, we can find the following consistent geometry if $m=0$
\be
%
\ee
leading to the irreducible duality
\be
%
\ee
where the edge between $\su(2)$ and $\so(7)$ with an $\S$ on top of it near $\so(7)$ denotes a full-hyper charged under $\F\otimes\S$ of $\su(2)\oplus\so(7)$. Extra matter is added in the above two dualities according to
\be
%
\ee
leads to the following irreducible duality for $m\ge1$ and $n\ge4$
\be
%
\ee
There is no corresponding geometry for $n=2$. Extra matter in the above two dualities is added according to
\be
%
\ee
admits a corresponding geometry only for $n\ge4$ and gives rise to the following irreducible duality for $m\ge1$ and $n\ge4$
\be
%
\ee
into which extra matter can be added according to
\be
%
\ee

\bigskip

Now, consider the transition
\be
%
\ee
for $n\ge3$. For $n\ge4$, the geometry describing the right hand side of the above transition is
\be
%
\ee
which leads to the following irreducible duality for $m\ge0$ and $n\ge4$
\be\label{soe8}
%
\ee
For $n=3$, a geometry for the right hand side of the transition satisfying all the requirements exists only for $m=0$, and the corresponding irreducible duality is
\be\label{soe9}
%
\ee
Extra matter in the above two dualities is added according to
\be
%
\ee

For the transition
\be
%
\ee
a consistent geometry exists only for $n\ge4$. For $m\ge1$ and $n\ge5$, we obtain the following irreducible duality
\be
%
\ee
For $n=4$ and $m\ge1$, we obtain the following irreducible duality
\be
%
\ee
Extra matter in the above two dualities is added according to
\be
%
\ee

\bigskip

Now, consider the transition
\be
%
\ee
Its right hand side is described by the geometry
\be
%
\ee
which leads to the following irreducible duality
\be
%
\ee
Notice that adding an $\F$ of $\su(2)$ or $\A$ of $\ff_4$ on the right side both are equivalent to adding $\S$ on the left side of the above duality.

\bigskip

Now, consider the transition
\be
%
\ee
This leads to the following irreducible duality for $m\ge2$
\be
%
\ee
and additional matter can be added according to
\be
%
\ee

Now, consider the following transition
\be
%
\ee
for $n\ge4$. This leads to the following irreducible duality for $m\ge1$ and $n\ge5$
\be
%
\ee
For $n=4$ and $m\ge1$ we obtain the following irreducible duality
\be
%
\ee

\bigskip

Finally, consider the transition
\be
%
\ee
which leads to the following irreducible duality for $m\ge1$
\be\label{soe10}
%
\ee

\subsection{$\fe_r$ dualities}
Let us start by considering the transition
\be
%
\ee
The geometry associated to the right hand side of the above transition is uniquely determined to be
\be
%
\ee
which leads to the following irreducible duality
\be
%
\ee

\bigskip

In a similar fashion, the transition
\be
%
\ee
 for $n\ge2$ and $m+n\le3$ leads to the following irreducible dualities
\be
%
\ee
It is clear how to add extra matter to the first two dualities, while extra matter for the last duality is added according to
\be
%
\ee
for $n\ge2$ and $m+n\le4$. For $m=0$ and $2\le n\le4$, it leads to the following irreducible dualities
\be
%
\ee
For $m=1,2$, we obtain the following irreducible dualities
\be
%
\ee
Extra matter in dualities associated to this transition is added according to
\be
%
\ee

Now, consider the transition
\be
%
\ee
for $n\ge3$ and $m+n\le5$. As for (\ref{soeT}), there are two geometries which can describe the right hand side of the above transition. The first one leads to the following irreducible duality for $m=0$ and $3\le n\le5$
\be
%
\ee
and the following set of irreducible dualities for $m=1,2$
\be
%
\ee
The second geometry leads to the following irreducible duality for $m=0$ and $3\le n\le5$
\be
%
\ee
and the following set of irreducible dualities for $m=1,2$
\be
%
\ee
Extra matter can be added into the dualities associated to this transition according to
\be
%
\ee

\bigskip

Now consider the transition
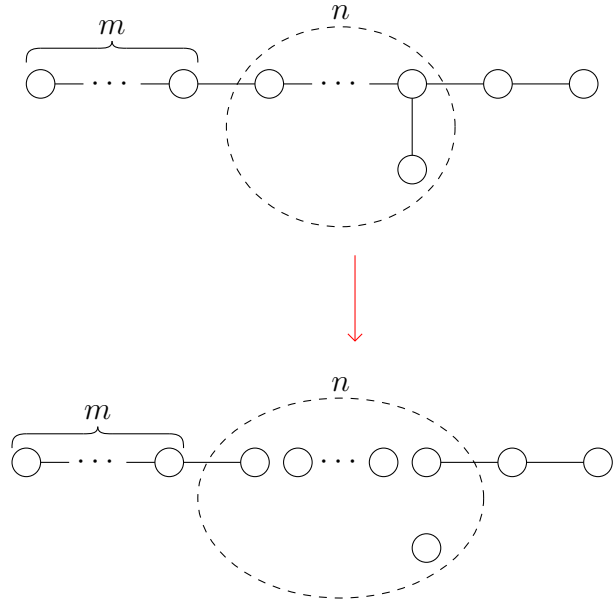
\be
%
\ee
This leads to the following irreducible duality for $1\le m\le3$
\be
%
\ee
Extra matter can be added according to
\be
%
\ee
with $n\ge4$ and $m+n\le6$ and leading to the following irreducible duality
\be
%
\ee
with
\be
\su(m+2)_{m+2}\to\su(2)_\pi
\ee
as $m\to0$. Extra matter can be added according to
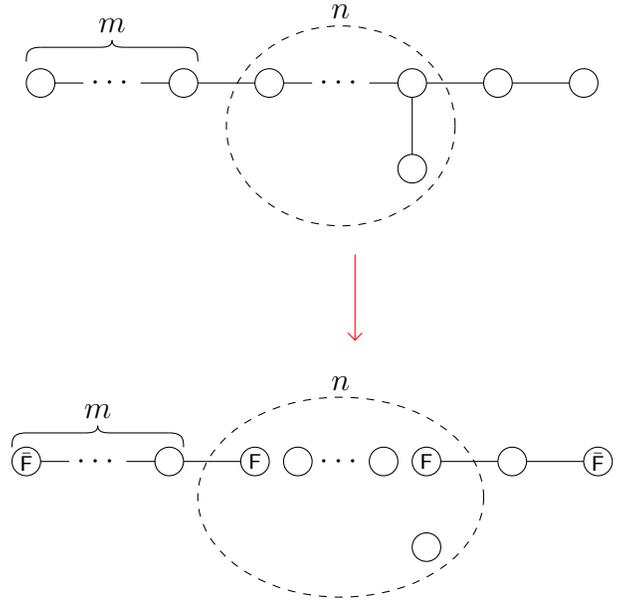
\be
%
\ee

\bigskip

Now, let us move to the transition
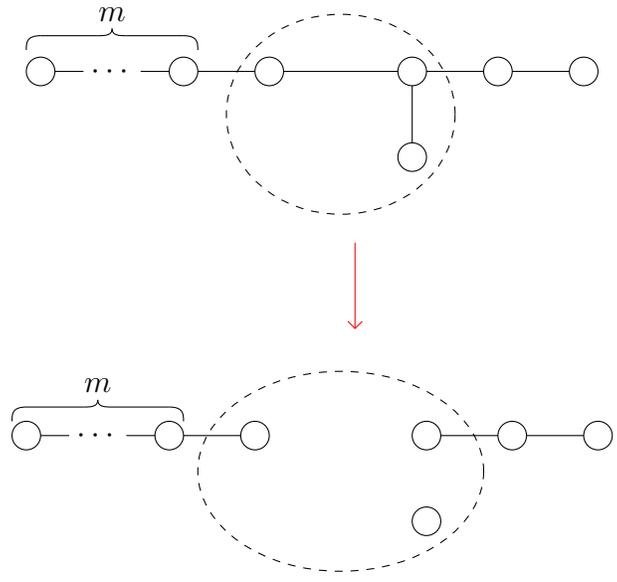
\be
%
\ee
which give rise to the following irreducible duality for $1\le m\le3$
\be
%
\ee
Additional matter content is integrated in according to
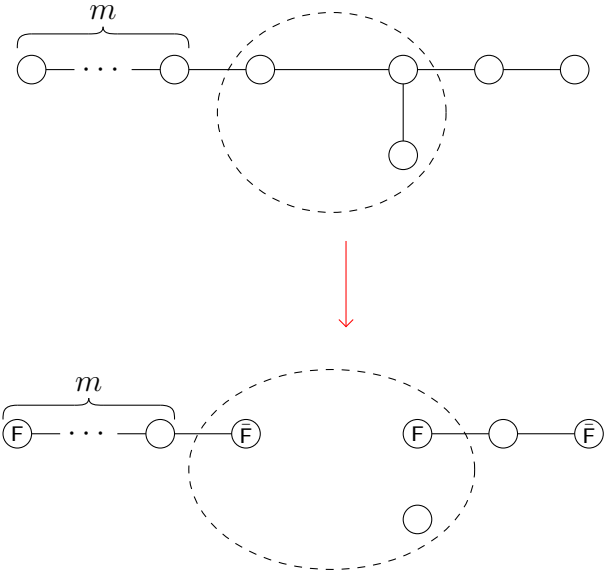
\be
%
\ee

\bigskip

For the transition
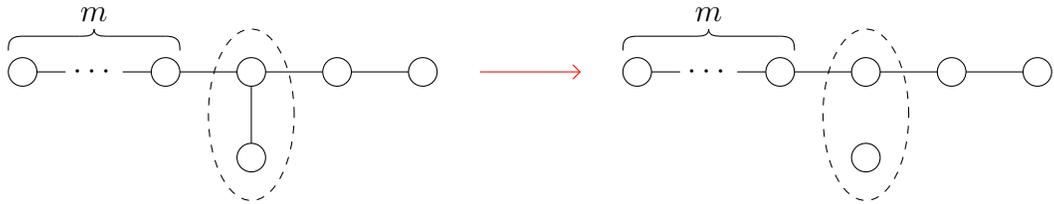
\be
%
\ee
with $2\le m\le4$, there are two possible geometries. The first one gives rise to the following irreducible duality
\be
%
\ee
with
\be
k_m=\frac{(3m+7)(m-2)}{4}
\ee
The second geometry gives rise to the following irreducible duality
\be
%
\ee
where
\be
k_m=\frac{(m+3)(m-2)}{2}
\ee
Additional matter into the above two dualities is incorporated using
\be
%
\ee
This leads to the following irreducible duality for $n\ge4$ and $5\le m+n\le7$
\be
%
\ee
with
\be
\su(m+2)_{m+2}\to\su(2)_\pi
\ee
as $m\to0$, and additional matter content is added according to
\be
%
\ee

\bigskip

Consider now the transition
\be
%
\ee
which leads to the following irreducible duality for $m=3,4$
\be
%
\ee
Extra matter is incorporated according to
\be
%
\ee

\bigskip

Consider now the transition
\be
%
\ee
It leads to the following irreducible duality for $n\ge5$ and $6\le m+n\le 8$
\be
%
\ee
with
\be
\su(m+2)_{m+2}\to\su(2)_\pi
\ee
as $m\to0$. Additional matter content is added according to
\be
%
\ee
leading to the following irreducible duality for $m=3,4$
\be
%
\ee
into which matter is added according to the rule
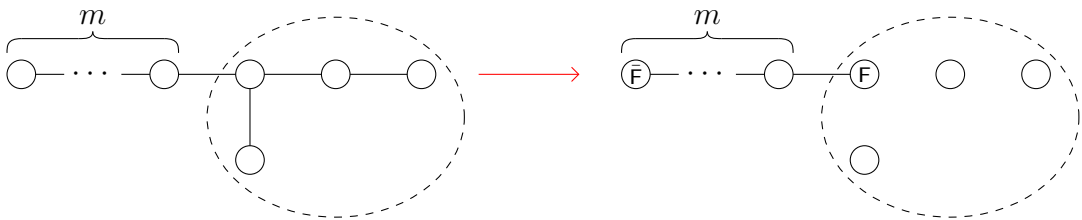
\be
%
\ee

\bigskip

Now consider performing the transition
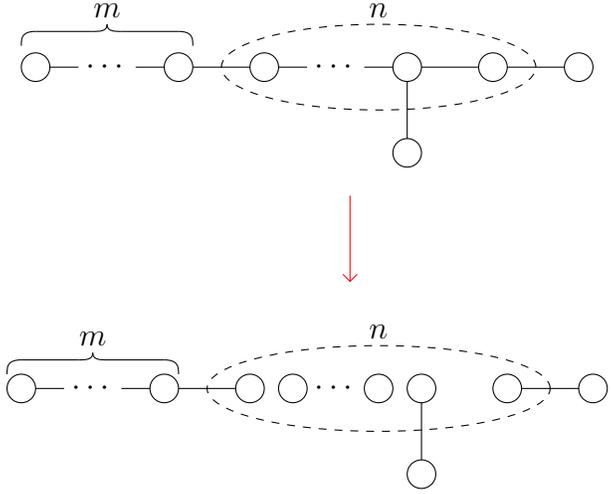
\be
%
\ee
This transition gives rise to the following irreducible duality for $n\ge4$ and $m+n\le6$
\be
%
\ee
with
\be
\su(m+2)_{m+2}\to\su(2)_\pi
\ee
as $m\to0$. Extra matter is added according to
\be
%
\ee
which gives rise to the following irreducible duality for $1\le m\le3$
\be
%
\ee
Extra matter is added according to
\be
%
\ee
which leads to the following irreducible duality for $m=3,4$
\be
%
\ee
where
\be
k_m=-\left(\frac{m^2+5m+14}{4}\right)
\ee
Extra matter is added according to
\be
%
\ee
which leads to the following irreducible duality for $n\ge5$ and $m+n\le7$
\be
%
\ee

\bigskip

Consider now the transition
\be
%
\ee
which leads to the following irreducible duality for $m=2,3$
\be
%
\ee
with extra matter content being added according to the following specification
\be
%
\ee
has two geometric solutions. The first one leads to the following irreducible duality for $m=3,4$
\be
%
\ee
and the second one leads to the following irreducible duality for $m=3,4$
\be
%
\ee
In both of the above dualities, extra matter is added according to
\be
%
\ee
This leads to the following irreducible duality for $m=3,4$
\be
%
\ee
where extra matter is added according to
\be
%
\ee
which leads to the irreducible duality 
\be
%
\ee
There are two geometries which can describe the right hand side of the above transition. The first one leads to the following set of irreducible dualities
\be
%
\ee
The second geometry leads to the following set of irreducible dualities
\be
%
\ee
In all the dualities associated to this transition, extra matter is added according to
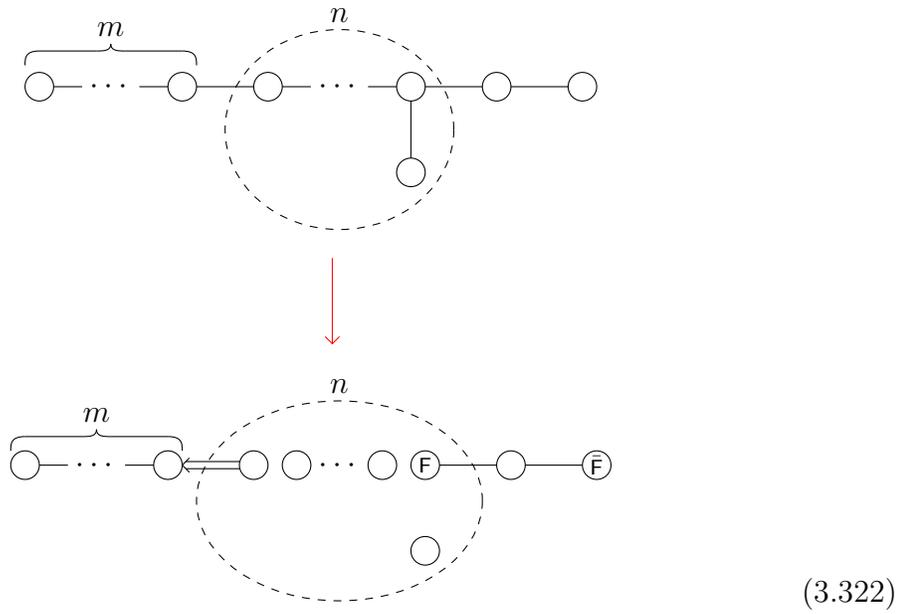
\be
%
\ee

\bigskip

Now, let us consider the transition
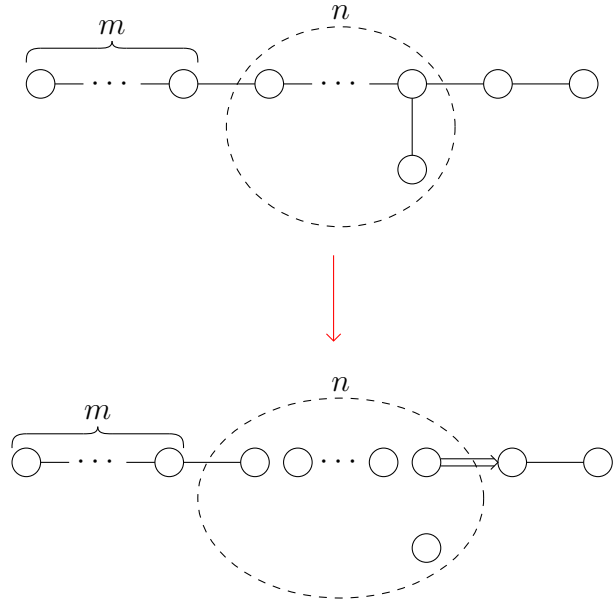
\be
%
\ee
which leads to the following irreducible duality for $n\ge3$ and $m+n\le5$
\be
%
\ee
which leads to the following set of irreducible dualities
\be
%
\ee
which leads to the following irreducible duality for $n\ge3$ and $m+n\le5$
\be
%
\ee
which leads to the following set of irreducible dualities
\be
%
\ee

\bigskip

Now, consider the following transition
\be
%
\ee
which leads to the following set of irreducible dualities
\be
%
\ee
from which we obtain the following irreducible dualities for $n\ge4$ and $m+n\le6$
\be
%
\ee
This leads to the following irreducible duality for $n\ge5$, $m\ge1$ and $m+n\le7$
\be
%
\ee
and the following irreducible duality for $n=4$ and $1\le m\le3$
\be
%
\ee
Extra matter in the above two dualities is added according to
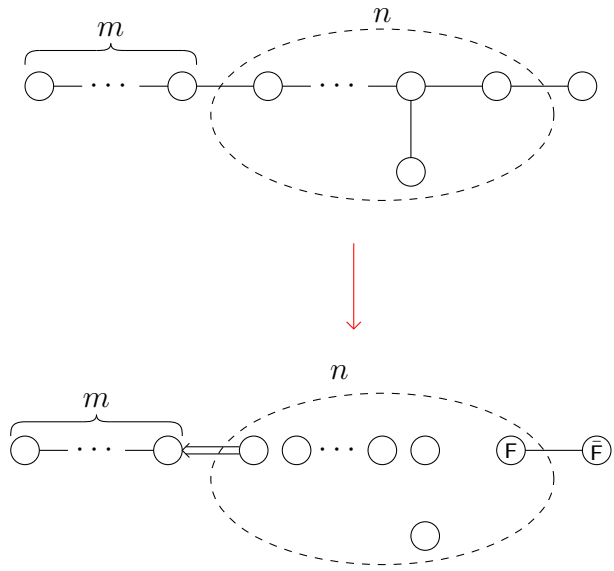
\be
%
\ee
which leads to the following irreducible duality for $n\ge4$ and $5\le m+n\le7$
\be
%
\ee
with
\be
\su(m+2)_{m+2}\to\su(2)_\pi
\ee
as $m\to0$, and additional matter content is added according to
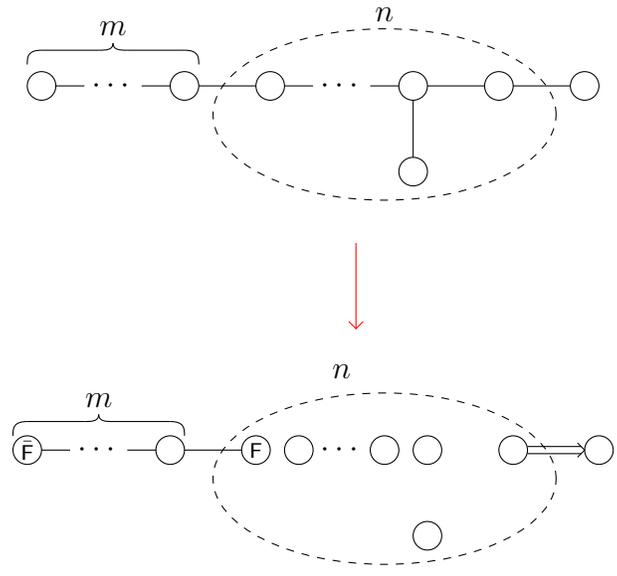
\be
%
\ee

\bigskip

Now consider the transition
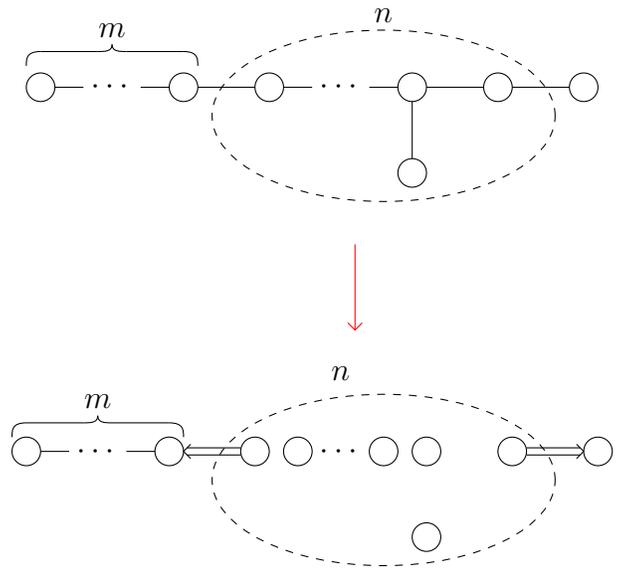
\be
%
\ee
This leads to the following irreducible duality for $n\ge4$, $m\ge1$ and $m+n\le7$
\be
%
\ee
It leads to the following irreducible duality for $n\ge6$, $m\ge1$ and $m+n\le 8$
\be
%
\ee
and the following irreducible duality for $n=5$ and $1\le m\le3$
\be
%
\ee
leading to the following irreducible duality for $m=3,4$
\be
%
\ee

\bigskip

Now consider performing the transition
\be
%
\ee
This transition gives rise to the following irreducible duality for $n\ge4$, $m\ge1$ and $m+n\le6$
\be
%
\ee
which leads to the following irreducible duality for $n\ge5$, $m\ge1$ and $m+n\le7$
\be
%
\ee
This leads to the following set of irreducible dualities
\be
%
\ee

\bigskip

Finally, consider the transition
\be
%
\ee

\subsection{$\ff_4$ dualities}
Let us start by considering the transition
\be
%
\ee
The geometry for the right hand of the transition is uniquely fixed to be
\be
%
\ee
which leads to the following irreducible duality
\be
%
\ee

\bigskip

Now, let us consider the transition
\be
%
\ee
The right hand side of this transition can correspond to two geometries satisfying the conditions listed at the beginning of this section. The first geometry is
\be
%
\ee
and leads to the following irreducible duality
\be
%
\ee
where $\A$ denotes the adjoint representation of $\ff_4$, and the edge between $\sp(3)$ and $\su(2)$ with $\L^3$ on top of it denotes a full hyper in $\L^3\otimes\F$ of $\sp(3)\oplus\su(2)$. The second geometry is
\be
%
\ee
The right hand side of the transition is described uniquely by the geometry
\be
%
\ee
giving rise to the following irreducible duality
\be
%
\ee
whose right hand side is described by
\be
%
\ee
which gives rise to the following irreducible duality
\be\label{f4}
%
\ee
Extra matter can be added on both sides according to the following diagram
\be
%
\ee

\bigskip

Finally, consider the following transition
\be
%
\ee
The right hand side of the above transition is described by the geometry
\be
%
\ee
giving rise to the following irreducible duality
\be
%
\ee

\subsection{$\fg_2$ dualities}
There are two irreducible dualities in which $\fg_2$ participates. The first one arises via the transition
\be
%
\ee
whose right hand side is described by the geometry
\be
%
\ee
which leads us to the following irreducible duality
\be\label{g2sp2}
%
\ee

\bigskip

The second irreducible duality arises via the transition
\be
%
\ee
whose right hand side is described by the geometry
\be
%
\ee
giving rise to the following irreducible duality
\be\label{g2su}
%
\ee
Extra matter can be added into the above duality according to
\be
%
\ee

\section{Further generalizations}\label{FG}
In this section, we discuss some ways in which the considerations of this paper can be generalized.
\subsection{Semi-simple to semi-simple}
We can consider transitions with semi-simple algebras on both sides. For example, we can consider the transition
\be
%
\ee
whose left hand side can be described by the geometry
\be
%
\ee
This geometry leads to the following irreducible duality
\be
%
\ee
Extra matter content can be added according to
\be
%
\ee

\subsection{Allowing extra edges}
We can allow more edges to appear in the geometry as compared to the Dynkin diagram in the phase where $\S$ transformations are applied. For example, consider the geometry
\be
%
\ee
which can describe the left hand side of the transition
\be
%
\ee
The intersection matrix of the above geometry describes $\su(4)$ gauge algebra but the number of edges in the geometry are more than the number of edges in the Dynkin diagram of $\su(4)$. The above geometry leads to the irreducible duality
\be
%
\ee
Extra matter can be added according to
\be
%
\ee

\subsection{Mixing $\cS$ with other isomorphisms}
We can combine the isomorphisms $\cI_n$ with $\cS$ to implement the dualities. For example, we claim that the geometry
\be\label{GG1}
%
\ee
can describe the right hand side of the transition
\be
%
\ee
Even though the geometry does lead to intersection matrix of $\su(3)\oplus\su(2)$, applying $\cS$ naively to all the nodes does not lead to the intersection matrix of $\su(4)$. Let us instead first perform $\cS$ on the lower node to rewrite the above geometry as
\be
%
\ee
Now we perform $\cI_0$ on the lower node using a blowup $y$ not equal to $x$. We obtain
\be
%
\ee
Applying $\cI^{-1}_0$ on the lower node using a blowup $z$ not equal to $x,y$, we obtain
\be
%
\ee
Now applying $\cI_0$ on the lower node using a blowup $w$ not equal to $x,y,z$, we obtain
\be
%
\ee
Performing $\cI_0^{-1}$ on the lower node using the blowup $x$, we obtain
\be
%
\ee
Now we perform $\cS$ on all the nodes to obtain
\be\label{GG2}
%
\ee
whose intersection matrix indeed leads to $\su(4)$. The precise irreducible duality obtained by matching (\ref{GG2}) with (\ref{GG1}) is
\be
%
\ee

\subsection{Reducing irreducible dualities}
It is possible to obtain irreducible dualities by adding matter to other irreducible dualities and composing them. For example, consider the irreducible duality
\be
\sp(2)_\pi=\su(3)_{-5}
\ee
following from (\ref{spsu}). As noted there, we can add matter to the above duality to obtain the following duality
\be\label{spsuM}
\sp(2)_\pi+2\L^2=\su(3)_{-6}+2\F
\ee
Now, note the irreducible duality (\ref{g2su})
\be
\fg_2=\su(3)_{-7}
\ee
which can be extended to the duality
\be\label{g2suM}
\fg_2+2\F=\su(3)_{-6}+2\F
\ee
Combining (\ref{spsuM}) and (\ref{g2suM}), we find that
\be
\fg_2+2\F=\sp(2)_\pi+2\L^2
\ee
which is precisely the irreducible duality (\ref{g2sp2})! 

Thus, the irreducible dualities can be ``reduced'' further into more primitive dualities. The notion of irreducible dualities used in this paper is defined when the gauge algebras on the two sides of the duality are kept fixed while removing matter, but as we have just seen, if one changes gauge algebra by composing with other dualities, then it is possible sometimes to further remove matter. It will be interesting to incorporate this and other new notions of reducibility in future works and find the truly primitive $5d$ dualities from which all the other $5d$ dualities can be derived by performing some physical operations.

\section*{Acknowledgements}
The author thanks Hee-Cheol Kim and especially Gabi Zafrir for useful discussions. This work is supported by NSF grant PHY-1719924.\\
The author thanks Sakura Schafer-Nameki and Brian Willett for pointing out typos in a previous version of this paper.

\bibliographystyle{ytphys}
\let\bbb\bibitem\def\bibitem{\itemsep4pt\bbb}
\bibliography{ref}

\end{document}